\documentclass{aa}  
\usepackage{graphicx}
\usepackage{txfonts}
\usepackage{linenoaa}
\usepackage{newtxtext,newtxmath}
\usepackage{dcolumn}
\usepackage{placeins}
\usepackage{placeins}
\usepackage{hyperref}
\hypersetup{colorlinks,allcolors=[rgb]{0,0,0.8}}

\newcommand{\diff}{\mathrm{d}}
\renewcommand{\vec}[1]{\mbox{\boldmath$#1$}}

\DeclareRobustCommand{\VAN}[3]{#2}
\let\VANthebibliography\thebibliography
\def\thebibliography{\DeclareRobustCommand{\VAN}[3]{##3}\VANthebibliography}
\defcitealias{bocquet22b}{B22-I}
\defcitealias{gatti21}{Gatti \& Sheldon et al. 2021}
\defcitealias{myles21}{Myles \& Alarcon et al. 2021}

\usepackage{float}
\usepackage{graphicx}   
\usepackage{amsmath}
\usepackage{tabularx}
\usepackage{booktabs}

\newcommand{\azeta}{\ensuremath{\ln{\zeta_0}}}
\newcommand{\bzeta}{\ensuremath{{\zeta_\mathrm{M}}}}
\newcommand{\czeta}{\ensuremath{{\zeta_\mathrm{z}}}}
\newcommand{\dzeta}{\ensuremath{{\sigma_{\ln\zeta}}}}
\newcommand{\alambda}{\ensuremath{{\lambda_0}}}
\newcommand{\blambda}{\ensuremath{{\lambda_\mathrm{M}}}}
\newcommand{\clambda}{\ensuremath{{\lambda_\mathrm{z}}}}
\newcommand{\dlambda}{\ensuremath{{\sigma_{\ln\lambda}}}}
\newcommand{\aWL}{\ensuremath{{\ln M_{\mathrm{WL}_0}}}}
\newcommand{\bWL}{\ensuremath{{M_{\mathrm{WL}_\mathrm{M}}}}}
\newcommand{\cWL}{\ensuremath{{M_{\mathrm{WL}_\mathrm{z}}}}}
\newcommand{\asigmaWL}{\ensuremath{{\ln\sigma^2_{\ln\mathrm{WL}_0}}}}
\newcommand{\bsigmaWL}{\ensuremath{{\sigma^2_{\ln\mathrm{WL}_\mathrm{M}}}}}
\newcommand{\csigmaWL}{\ensuremath{{\sigma^2_{\ln\mathrm{WL}_\mathrm{z}}}}}

\newcommand{\hzeta}{\ensuremath{{\hat{\zeta}}}}
\newcommand{\hlambda}{\ensuremath{{\hat{\lambda}}}}

\usepackage{eso-pic}

\AddToShipoutPictureBG*{%
  \AtPageUpperLeft{%
    \hspace{0.712\paperwidth}%
    \raisebox{-1.5\baselineskip}{%
      \makebox[0pt][l]{\textnormal{DES-2024-0844}}
}}}%

\AddToShipoutPictureBG*{%
  \AtPageUpperLeft{%
    \hspace{0.712\paperwidth}%
    \raisebox{-2.5\baselineskip}{%
      \makebox[0pt][l]{\textnormal{FERMILAB-PUB-24-0363-PPD}}
}}}%

\begin{document}

    \title{Galaxy cluster matter profiles\\ I. Self-similarity, mass calibration, and observable-mass relation validation employing cluster mass posteriors 
    }
   
    \titlerunning{Galaxy cluster matter profiles and mass posteriors}
        \author{A.~Singh
        \inst{1,2}\fnmsep\thanks{aditya.singh@physik.lmu.de} \and
        J.~J.~Mohr \inst{1,2}\and
        C.~T.~Davies \inst{1}\and
        S.~Bocquet \inst{1}\and
        S.~Grandis \inst{1,3}\and
        M.~Klein \inst{1}\and 
        J.~L.~Marshall \inst{32}\and\\
        M.~Aguena \inst{4}\and
        S.~S.~Allam \inst{5}\and
        O.~Alves \inst{6}\and
        F.~Andrade-Oliveira \inst{4,7}\and
        D.~Bacon \inst{8}\and
        S.~Bhargava \inst{9}\and
        D.~Brooks \inst{10}\and
        A.~Carnero~Rosell \inst{11,4}\and
        J.~Carretero \inst{12}\and
        M.~Costanzi \inst{13,14,15}\and
        L.~N.~da Costa \inst{4}\and
        M.~E.~S.~Pereira \inst{16}\and
        S.~Desai \inst{17}\and
        H.~T.~Diehl \inst{5}\and
        P.~Doel \inst{10}\and
        S.~Everett \inst{18}\and
        B.~Flaugher \inst{5}\and
        J.~Frieman \inst{5,19}\and
        J.~Garc\'ia-Bellido \inst{20}\and
        E.~Gaztanaga \inst{21,8,22}\and
        R.~A.~Gruendl \inst{23,24}\and
        G.~Gutierrez \inst{5}\and
        D.~L.~Hollowood \inst{25}\and
        K.~Honscheid \inst{26,27}\and
        D.~J.~James \inst{28}\and
        K.~Kuehn \inst{29,30}\and
        M.~Lima \inst{31,4}\and
        J. Mena-Fern{\'a}ndez \inst{33}\and
        F.~Menanteau \inst{23,24}\and
        R.~Miquel \inst{34,12}\and
        J.~Myles \inst{35}\and
        A.~Pieres \inst{4,36}\and
        A.~K.~Romer \inst{9}\and
        S.~Samuroff \inst{37}\and
        E.~Sanchez \inst{38}\and
        D.~Sanchez Cid \inst{38}\and
        I.~Sevilla-Noarbe \inst{38}\and
        M.~Smith \inst{39}\and
        E.~Suchyta \inst{40}\and
        M.~E.~C.~Swanson \inst{23}\and
        G.~Tarle \inst{6}\and
        C.~To \inst{26}\and
        D.~L.~Tucker \inst{5}\and
        V.~Vikram \inst{41}\and
        N.~Weaverdyck \inst{42,43}\and
        P.~Wiseman \inst{39} \\
        (the DES and SPT Collaborations)
        }

        \institute{Affiliations at the end of the paper}
   
   \date{Received---; accepted ---}

    \abstract
    {We present a study of the weak lensing inferred matter profiles $\Delta\Sigma(R)$ of 698 South Pole Telescope (SPT) thermal Sunyaev-Zel'dovich effect (tSZE) selected and MCMF optically confirmed galaxy clusters in the redshift range $0.25 <z< 0.94$ that have associated weak gravitational lensing shear profiles from the Dark Energy Survey (DES). Rescaling these profiles to account for the mass dependent size and the redshift dependent density produces average rescaled matter profiles $\Delta\Sigma(R/R_\mathrm{200c})/(\rho_\mathrm{crit}R_\mathrm{200c})$ with a lower dispersion than the unscaled $\Delta\Sigma(R)$ versions, indicating a significant degree of self-similarity. Galaxy clusters from hydrodynamical simulations also exhibit matter profiles that suggest a high degree of self-similarity, with RMS variation among the average rescaled matter profiles with redshift and mass falling by a factor of approximately six and 23, respectively, compared to the unscaled average matter profiles. We employed this regularity in a new Bayesian method for weak lensing mass calibration that employs the so-called cluster mass posterior $P(M_{\mathrm{200c}}|\hzeta, \hlambda, z)$, which describes the individual cluster masses given their tSZE ($\hzeta$) and optical ($\hlambda$, $z$) observables. This method enables simultaneous constraints on richness $\lambda$-mass and tSZE detection significance $\zeta$-mass relations using average rescaled cluster matter profiles. We validated the method using realistic mock datasets and present observable-mass relation constraints for the SPT$\times$DES sample, where we constrained the amplitude, mass trend, redshift trend, and intrinsic scatter. Our observable-mass relation results are in agreement with the mass calibration derived from the recent cosmological analysis of the SPT$\times$DES data based on a cluster-by-cluster lensing calibration. Our new mass calibration technique offers a higher efficiency when compared to the single cluster calibration technique. We present new validation tests of the observable-mass relation that indicate the underlying power-law form and scatter are adequate to describe the real cluster sample but that also suggest a redshift variation in the intrinsic scatter of the $\lambda$-mass relation may offer a better description. In addition, the average rescaled matter profiles offer high signal-to-noise ratio (S/N) constraints on the shape of real cluster matter profiles, which are in good agreement with available hydrodynamical $\Lambda$CDM simulations. This high S/N profile contains information about baryon feedback, the collisional nature of dark matter, and potential deviations from general relativity.}
    
    
    \maketitle



\section{Introduction}

Galaxy clusters constitute the most massive collapsed halos in the Universe. Studying their abundance as a function of redshift and mass provides insights into structure formation history and therefore serves as a powerful tool for constraining cosmological models  \citep[e.g.,][]{White1993MNRAS.262.1023W,Haiman2001,Vikhlinin2009ApJ...692.1060V,Mantz2010MNRAS.406.1759M,PlanckCollaboration2016A&A...594A..13P,Chiu23,Bocquet2024IIPhRvD.110h3510B}. The ability to accurately measure cluster masses plays an important role in cluster cosmological studies, as it enables constraints on the rate of cosmic structure growth, the dark energy equation of state, and other cosmological parameters, such as the amplitude of matter fluctuations and the matter density parameter. The development of robust weak lensing (WL) and cosmic microwave background (CMB) lensing has informed mass calibration techniques \citep{Becker11,vonderlinden2014MNRAS.439....2V,Dietrich2019MNRAS.483.2871D,Zubeldia2019MNRAS.489..401Z,Grandis21,BocquetI2024PhRvD.110h3509B}, and the availability of associated high quality WL datasets from, for example, the Hyper Suprime-Cam Subaru Strategic Program (HSC-SSP), the Dark Energy Survey (DES), and the Kilo-Degree Survey (KiDS) have set the stage for progress in constraining the standard $\Lambda$CDM and wCDM parameters \citep{Costanzi+19,Abbott+20,Costanzi+21,To+21,Chiu23,Bocquet2024IIPhRvD.110h3510B,ghirardini+24} as well as model extensions, including the modification of general relativity and interacting dark matter \citep[e.g.,][]{Mantz2014,Cataneo+15,Vogt2024arXiv240913556V,Mazoun2024arXiv241119911M}.  

These same WL datasets can be employed to study the matter distribution within galaxy clusters.  A challenge is that in the existing WL datasets based on large photometric surveys, the matter profiles of individual clusters often have a low signal-to-noise ratio (S/N). Combining WL matter profiles from multiple galaxy clusters provides a way to improve the S/N and reduce the intrinsic variations in the matter distribution from cluster to cluster that arise from their different formation histories. Previous works have employed WL measurements of multiple clusters to constrain cluster masses by combining tangential shear profiles or projected matter profiles of clusters \citep{Oguri+11, Umetsu:2015baa, McClintock_2018, Bellagamba+19, Giocoli+21, Lesci+22}. A challenge in this approach is that there are systematic variations in the projected matter profiles of galaxy clusters with cluster mass and redshift.  An average WL matter profile therefore reflects the characteristics of the cluster sample, and it depends sensitively on the distribution of the sample in mass and redshift. 
This approach also requires careful modeling of the spatial distribution and masking of the WL source galaxies on a cluster-by-cluster basis to enable accurate modeling of the multi-cluster matter profile.

If the systematic variations of the matter profiles with mass and redshift can be accurately characterized, then they can also be scaled out, enabling average matter profiles of high S/N that are largely independent of the characteristics of the cluster sample from which they are constructed.  In particular, if cluster matter profiles are approximately self-similar in nature -- that is, they exhibit similar shapes that vary systematically with mass and redshift -- then these systematic trends can be easily removed. In this limit, the need to accurately model the spatial distribution and masking of WL source galaxies is also no longer required.

Approximate self-similarity is a generic prediction of gravitational structure formation \citep{1986MNRAS.222..323K}.  In N-body simulations, cluster halos are well described by so-called Navarro, Frenk, and White (NFW) models \citep{Navarro_1997} that exhibit weak trends in concentration or shape with mass and redshift. In hydrodynamical simulations, self-similar behavior has been seen in cluster gas profiles \citep{Lau_2015} and pressure profiles \citep{Nelson_2014}. Observationally, approximate self-similarity has been demonstrated in the intracluster medium (ICM) density, pressure profiles, and temperature profiles \citep{Vikhlinin_2006,Arnaud_2010,Baldi+12,McDonald2014ApJ...794...67M}, whereas WL studies of cluster matter profiles have tended to focus on whether the NFW model is a good description of the real matter profiles \citep[e.g.,][]{Umetsu_2014,Niikura+2015}. In particular, \cite{Umetsu_2014} studied average matter profiles and found the NFW profile to be an excellent description of real clusters. In contrast, \cite{Niikura+2015} analyzed scaled average matter profiles by rescaling with the scale radius of the NFW profile and with the critical density of the Universe and found evidence of a single underlying or universal cluster matter profile.

In this analysis, we use hydrodynamic structure formation simulations and direct WL observations of cluster samples to examine cluster matter profiles, which reveal the remarkable consistency and approximate self-similarity of the simulated and real matter profiles. We exploit this self-similarity to study high S/N cluster matter profiles and then employ them to perform a calibration of observable-mass relations. This mass calibration approach offers a computationally more efficient technique to analyze large cluster WL datasets compared to a cluster-by-cluster approach \citep{Bocquet_2019,BocquetI2024PhRvD.110h3509B} without loss of information.

The paper is organized as follows. We present the simulated and observed dataset in Sect.~\ref{sec:data}. The self-similarity of galaxy cluster matter profiles is explored in  Sect.~\ref{sec:self-similarity}. The mass calibration method along with the likelihood calculation and the hydrodynamical model is discussed in Sect.~\ref{sec:method}. In Sect.~\ref{sec:results}, we validate the analysis method using mock data and present the results using the South Pole Telescope (SPT) clusters and DES-WL data. 
We conclude with a summary and outlook in Sect.~\ref{sec:summary}. 
Throughout the paper, we employ a flat $\Lambda$CDM cosmology with parameters $\Omega_{\mathrm{m}}$ = 0.3 and $h$ = 0.7.  All uncertainties are quoted at the 68\% credible interval unless otherwise specified.


\section{Data}
\label{sec:data}
In this section, we first describe the data used in our work: 1) the SPT cluster catalogs and 2) DES (Year ~3, hereafter Y3) WL and photometric redshift measurements. Then we summarize the simulation datasets from \textit{Magneticum} and IllustrisTNG, which are used to explore the impact of baryons on cluster matter profiles.

\begin{figure}
\centering
\includegraphics[width=\columnwidth]{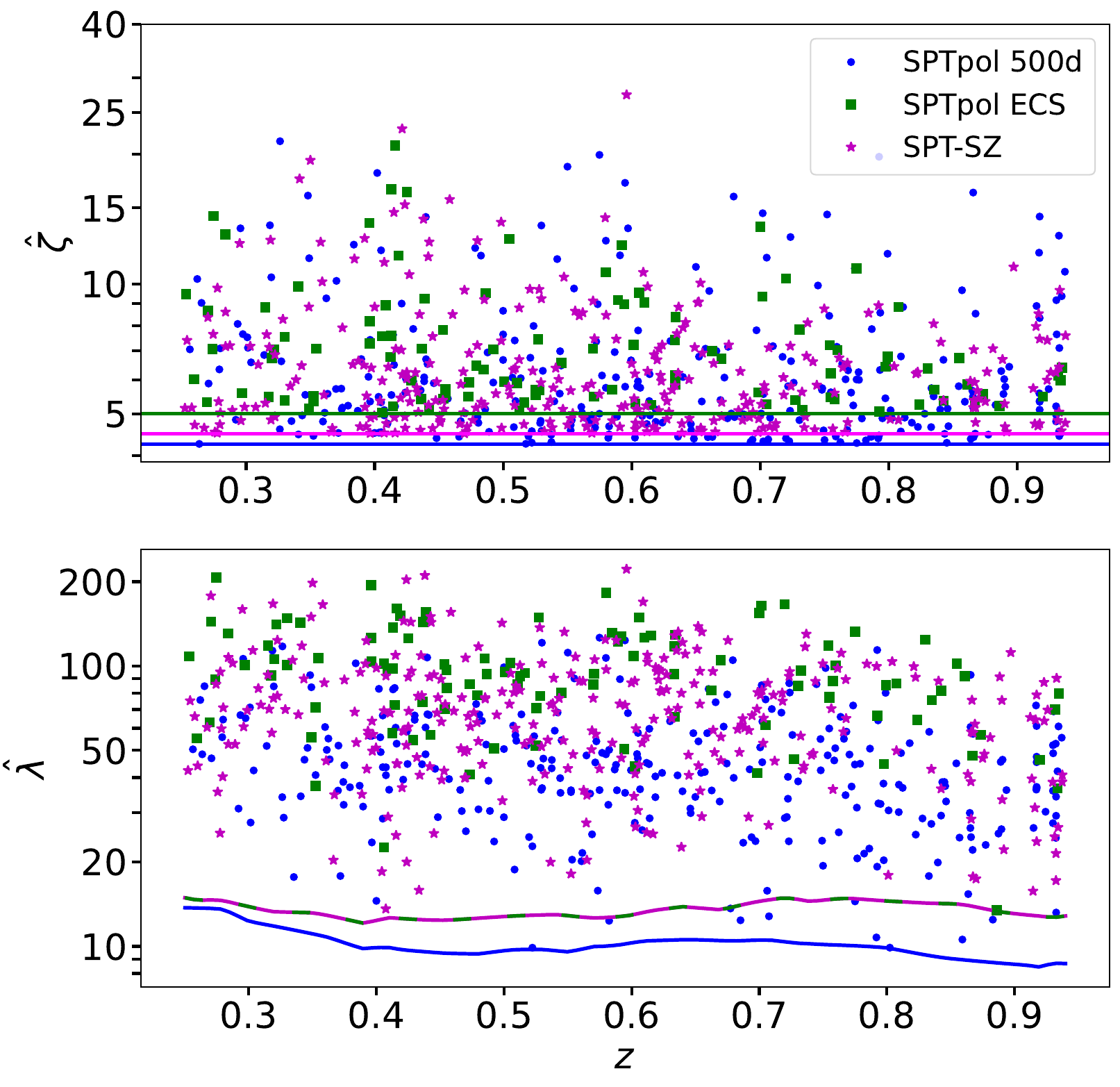}
  \vskip-0.10in
  \caption{Distribution of observed tSZE detection significance $\hat\zeta$ and richness $\hat\lambda$ as a function of redshift for the 698 galaxy clusters in the SPT sample that overlap the DES region. The solid line in the top figure shows the detection threshold for the three survey regions. In the bottom figure, the colored lines correspond to the $\hlambda_\mathrm{min}(z)$ detection threshold for each survey.}
  \label{fig:zeta_lambda_z}
\end{figure}

\subsection{South Pole Telescope cluster catalogs}
We use a combination of three thermal Sunyaev-Zel'dovich effect (tSZE) selected cluster catalogs that have been extracted from surveys carried out by the South Pole Telescope \citep{carlstrom11} collaboration: SPT-SZ \citep{Bleem_2015,Klein2023arXiv230909908K}, SPTpol ECS \citep{Bleem_2020}
and SPTpol-500d \citep{Bleem2023-500d}.
The SPT-SZ survey covers 2,500\,deg$^2$, and the SPTpol ECS survey spans 2,770\,deg$^2$ in the southern sky, while the SPTpol-500d survey pushes to a greater depth within a 500\,deg$^2$ patch inside the SPT-SZ survey. 
Galaxy cluster candidates are selected from the mm-wave maps at 90 and 150\,GHz using a matched filter technique \citep{Melin06}, which employs galaxy cluster tSZE models with a range of angular scales \citep{Vanderlinde10}.
Only cluster candidates at redshifts $z>0.25$ are considered, because clusters at lower redshifts are larger in angular extent and therefore more strongly impacted by the matched filtering, which is designed to remove atmospheric noise as well as increased noise contributions from the primary CMB.  At low redshift the angular scales filtered out overlap with the scales important for the galaxy cluster tSZE, strongly impacting the candidate detection significance and thereby complicating its use as a cluster halo mass proxy. Additionally, we only analyze clusters with a redshift $z<0.95$ due to the depth and systematics of the DES WL sample described below.

These cluster candidates are then studied  using the Multi-Component Matched Filter cluster confirmation tool \citep[MCMF;][]{klein18}.  This processing results in a cluster catalog that includes measurements of optical richnesses $\hat\lambda$, sky positions and redshifts.  
The measured optical richness allows for efficient removal of contaminants from the tSZE candidate list by evaluating the likelihood of each candidate being a random superposition of a physically unassociated optical system with a tSZE noise fluctuation \citep{Klein2023arXiv230909908K}.
The exclusion threshold corresponds to an observed richness threshold that varies with redshift $\hat\lambda_\mathrm{min}(z)$ \citep{Klein2019MNRAS.488..739K} and has been determined by analyzing the richness distributions along random lines of sight within the survey.  The final MCMF-confirmed cluster catalogs have a constant contamination fraction at all redshifts.  

The selection threshold in the tSZE detection significance is $\hat\zeta>4.25$ for SPTpol-500d, $\hat\zeta>4.5$ for SPT-SZ and $\hat\zeta>5$ for SPTpol ECS, while the MCMF selection threshold $\hat\lambda_\mathrm{min}(z)$ is adjusted to maintain a contamination fraction of $<2$\,\% in the final MCMF-confirmed cluster lists from both surveys. Figure~\ref{fig:zeta_lambda_z} shows the distribution of observed richness and tSZE detection significance as a function of redshift for the MCMF confirmed SPT sample we study here.

\subsection{DES~Y3 lensing}
\label{sec:DES-WL}
The Dark Energy Survey is a photometric survey in five broadband filters ($grizY$) which covers an area of $\sim$5,000\,deg$^2$ in the southern sky. The survey was conducted using the Dark Energy Camera \citep[DECam;][]{flaugher15} at the 4m Blanco telescope at the Cerro Tololo Inter-American Observatory (CTIO) in Chile.
In this work, we use WL data from the first three years of DES observations (DES~Y3), which cover the entire 5,000\,deg$^2$ survey footprint.

The DES~Y3 shape catalog (\citetalias{gatti21}) is constructed from the $r,i,z$-bands using the \textsc{Metacalibration}
pipeline \citep{huff&mandelbaum17, sheldon&huff17}.
Other DES~Y3 works contain detailed information on the Point-Spread Function modeling \citep{jarvis21}, the photometric dataset \citep{sevilla-noarbe21}, and image simulations \citep{maccrann22}. After all source selection cuts, the shear catalog consists of roughly 100~million galaxies over an area of 4,143~deg$^2$. The typical source density is 5 to 6~arcmin$^{-2}$, depending on the selection choices of a specific analysis.

\begin{figure}
\centering
\includegraphics[width=\columnwidth]{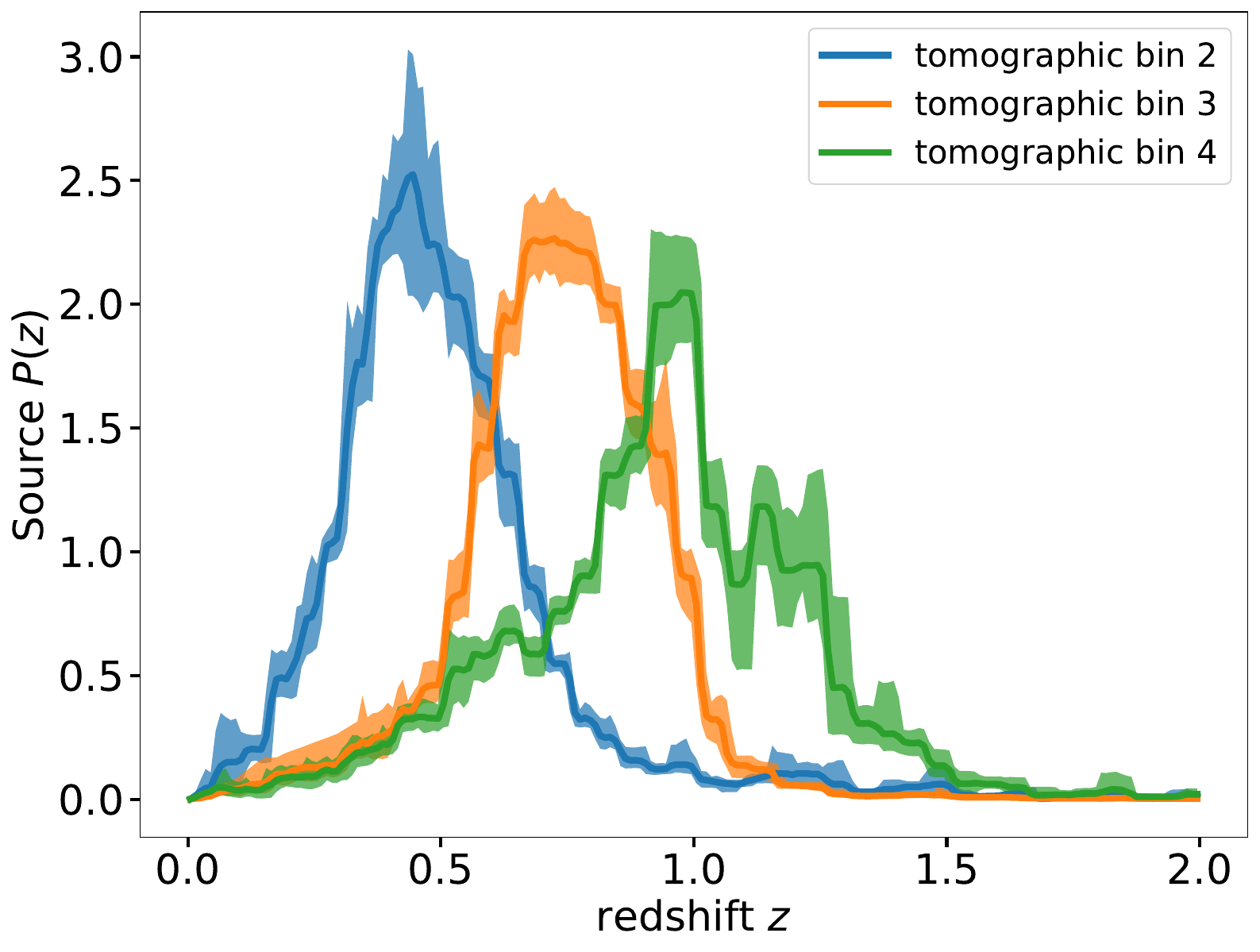}
  \vskip-0.10in
  \caption{Dark Energy Survey~Y3 lensing source redshift distribution for tomographic bins 2 through 4 that are used in this analysis. The solid line represents the mean and the shaded region depicts the 2$\sigma$ uncertainties on the redshift distributions.
  }
  \label{fig:source_distribution}
\end{figure}

Our work follows the selection of lensing source galaxies in four tomographic bins (Fig.~\ref{fig:source_distribution} shows the three tomographic bins used in our analysis) as employed in the DES 3x2pt analysis \citep{DES_Y3_3x2pt}. The selection is defined and calibrated in \cite{myles21} and \cite{gatti22}, where source redshifts are estimated using Self-Organizing Maps Photo-z (SOMPZ). The final calibration accounts for the (potentially correlated) systematic uncertainties in source redshifts and shear measurements. For each tomographic source bin, the mean redshift distribution is provided, and the systematic uncertainties are captured using 1,000 realizations of the source redshift distribution. The amplitude of the source redshift distribution is scaled by a factor $1+m$ to account for the multiplicative shear bias $m$. In addition to the tomographic bins and SOMPZ, we used Directional Neighbourhood Fitting \citep[\textsc{dnf};][]{devicente16} galaxy photo-z estimates when determining the expected fraction of the lensing source galaxy population in each tomographic bin that is contributed by member galaxies from a particular cluster of interest-- the so-called cluster member contamination.

\subsection{Hydrodynamical simulations}
\label{sec:sim_data}
In this work, we use the \textit{Magneticum} Pathfinder suite of cosmological hydrodynamical simulations \citep{hirschmann14, teklu15, beck16, Bocquet2016MNRAS.456.2361B, dolag17}.  We use Box1, which has a box size of 896\,$h^{-1}$\,Mpc on a side with $2\times1526^3$ particles and the particle mass $1.3\times10^{10}\,h^{-1}M_\odot$ for dark matter particles, and $2.6\times10^9\,h^{-1}M_\odot$ for gas particles. The simulation is run with cosmological parameters ($\Omega_m=0.272$, $\Omega_b=0.0457$, $H_0=70.4$, $n_s=0.963$, $\sigma_8=0.809$), which correspond to the WMAP7 constraints for a spatially flat $\Lambda$CDM model \citep{Komatsu2011ApJS..192...18K}. From this simulation, we use snapshots at five redshifts $z_\text{snap} = 0.01, 0.25, 0.47, 0.78, 0.96$.

In addition, we also use the data from IllustrisTNG300-1 \citep{pillepich18, marinacci18, springel18, nelson18, naiman18, nelson19}. These include $2\times2500^3$ resolution elements for a box size of 205\,$h^{-1}$\,Mpc on a side. The cosmology corresponds to the Planck2015 constraints for a spatially flat $\Lambda$CDM cosmology \citep{PlanckCollaboration2016A&A...594A..13P}: $\Omega_m=0.3089$, $\Omega_b=0.0486$, $\sigma_8=0.8159$, $n_s=0.9667$, and $h=0.6774$. We use snapshots correponding to redshift $z_\text{snap}\in \{0.01, 0.24,\,0.42,\,0.64,\,0.95\}$. 
From these simulation snapshots, we then extract halos with $M_\text{200c}> 3\times 10^{13}\,h^{-1} $ M$_\odot$. Shear maps are generated following \citet{Grandis21} in a cylinder with a projection depth of $20~h^{-1}$\,Mpc. 


\section{Self-similarity in cluster matter profiles}
\label{sec:self-similarity}
Gravitational lensing is the phenomenon through which photon geodesics are perturbed by gravitational potentials. For a distant galaxy, this causes a distortion in the observed image relative to its true shape \citep{Schneider_2006}. In this work, we are interested in WL, where distortions in source galaxy images induced by intervening matter along the line of sight are small. In this regime, the WL signal must be extracted through statistical correlations of source galaxies.
The observable of interest in this context is the reduced shear, which is defined as
\begin{equation}
    g = \frac{\gamma}{1-\kappa},
        \label{eq:red_shear}
\end{equation}
where $\gamma$ is the WL shear and $\kappa$ is the WL convergence  \citep[for detailed discussion see][]{Schneider_2006}. 
The ensemble averaged source ellipticity, $e$, and the shear response, $R_{\gamma}$, are related to the reduced shear as
\begin{equation}
    \langle g \rangle = \langle R_{\gamma} \rangle^{-1} \langle e \rangle.
        \label{eq:shear_response}
\end{equation}
The term $R_{\gamma}$ is the average response of the measured ellipticity to a shear. Due to instrumental and atmospheric effects and noise this shear response typically is less than one. The tangential reduced shear profile induced by an object with a projected mass distribution $\Sigma(R)$ is related to the critical surface mass density $\Sigma_{\mathrm{crit}}$ by 
\begin{equation}    
    \Delta\Sigma(R)=\Sigma_{\mathrm{crit}}\,\gamma_\mathrm{t}(R) = \langle \Sigma(<R) \rangle - \Sigma(R), 
        \label{eq:gamma_delta_sigma}
\end{equation}
where $\Sigma_{\mathrm{crit}}$ depends on the geometry of the source-lens system and is defined as
\begin{equation}
    \Sigma_{\mathrm{crit}}(\mathrm{z_s},\mathrm{z_l}) = \frac{c^2}{4\pi G} \frac{D_\mathrm{s}}{D_\mathrm{l}D_{\mathrm{ls}}},
        \label{eq:sigma_crit}
\end{equation}
where $z_\mathrm{s}$ and $z_\mathrm{l}$ are the source and lens redshifts, respectively, and $D_\mathrm{s}, D_\mathrm{l}, D_{\mathrm{ls}}$ are the angular diameter distances to the source, lens, and between the source-lens pair. When $z_{\mathrm{s}} \leq z_{\mathrm{l}}$, $\Sigma_{\mathrm{crit}}^{-1}$ is defined to be zero.

Analogous to Eq.~\ref{eq:gamma_delta_sigma}, we introduce an observed quantity \begin{equation}    
    \Delta\Sigma_{\mathrm{reduced}}(R)=\Sigma_{\mathrm{crit}}\,g_\mathrm{t}(R) 
        \label{eq:gt_delta_sigma}.
\end{equation}
In this paper, we use the excess surface mass density as defined in the above equation but refer to it as $\Delta\Sigma(R)$. We note that, although $\Sigma(R)$ is the more fundamental quantity, we use $\Delta\Sigma(R)$ (calculated using eq.~\ref{eq:gt_delta_sigma}) in our work because it can be directly measured through WL.  For convenience we refer to $\Delta\Sigma(R)$ simply as the cluster matter profile.

For a given source redshift distribution $P(\mathrm{z_s})$, we can compute the average lensing efficiency for a given lens as
\begin{equation}
    \Sigma_{\mathrm{crit}}^{-1}(\mathrm{z_l}) = \int \mathrm{dz_s} P(\mathrm{z_s}) \Sigma_{\mathrm{crit}}^{-1}(\mathrm{z_s},\mathrm{z_l}).
        \label{eq:sigma_crit_integral}
\end{equation}

From Eq. \ref{eq:gamma_delta_sigma} we can see that the differential surface mass density at a given projected radius $R$ can be expressed as the difference between the mean enclosed surface mass density and the surface mass density $\Sigma$ at that projected radius, which is expressed as follows
\begin{eqnarray}
    \Sigma(R) = \int_{-\infty}^{\infty} \mathrm{d\chi 
    }
    \rho \bigg( \sqrt{R^2 + \mathrm{\chi}^2} \bigg),\\
    \langle \Sigma(<R) \rangle = \frac{2}{R^2} \int_{0}^{R} \mathrm{d}R' R' \Sigma(R').
\end{eqnarray}
 Where $\rho(r)$ is the density distribution of the halo and 
 $\chi$ is the comoving distance along the line of sight. 
 For the shear signal induced by a halo of mass $M$, the average excess three-dimensional matter density is given by 
\begin{equation}
    \rho(r) = \rho_{\mathrm{m}} \xi_{\mathrm{hm}}(r|M),
\end{equation}
where $\rho_{\mathrm{m}}  = \Omega_{{\mathrm{m,0}}}\rho_{{\mathrm{crit,0}}}(1+z)^3$ is the mean matter density of the universe and $\xi_{\mathrm{hm}}(r|M)$ is the halo-matter correlation function at the halo redshift.  
At a small radius, $\xi_{\mathrm{hm}}(r|M)$ is dominated by the cluster density profile, and this region is called the ``one-halo'' region. At a larger radius, most of the contribution comes from correlated structures around the halo, and it is therefore referred to as the  ``two-halo'' region. In this work, we examine both regions but focus on the one-halo region for the cluster mass calibration.  

There is a strong theoretical expectation that, barring the impact of baryonic effects, the one-halo region of a halo should be described by the NFW model \cite{Navarro_1996,Navarro_1997}.  In this model, the cluster matter profile  within the radius $r_\mathrm{200c}$, which encloses a region with a mean density that is 200 times the critical density $\rho_\mathrm{crit}$ is well described as 
\begin{equation}
\label{eq:NFW}
\rho(r)=\delta_\mathrm{s}\rho_\mathrm{crit}\left[\frac{r}{cr_\mathrm{200c}}\left(1+\left(\frac{r}{cr_\mathrm{200c}}\right)^2\right)\right]^{-1},
\end{equation}
where $\delta_\mathrm{s}$ is a characteristic overdensity depending on $c$, which is the halo concentration parameter. Such a halo characterized by $r_\mathrm{200c}$ has a mass that can be expressed as
\begin{equation}
\label{eq:mass_definition}
    M_\mathrm{200c} = 200\rho_\mathrm{crit}(z)\frac{4\pi}{3}  r_\mathrm{200c}^3.
\end{equation}
This underlying density profile implies a particular projected $\Delta\Sigma$ matter profile \citep{Bartelmann1996A&A...313..697B}, whose amplitude scales with the extent of the cluster along the line of sight (i.e., $r_\mathrm{200c}$) and depends on the critical density, which varies with redshift as $\rho_\mathrm{crit}=3H^2(z)/8\pi G$, where $H(z)$ is the Hubble parameter at redshift $z$ and $G$ is the Gravitational constant.  For the projected profiles discussed below, we rename $r_\mathrm{200c}$ to be $R_\mathrm{200c}$, corresponding to the projected distance equal to the 3D radius that encloses the mass $M_\mathrm{200c}$.

Within simulations the halo shapes vary from cluster to cluster due to formation history differences, and systematic trends in concentration with mass and redshift have been identified \citep{Bhattacharya:2011vr,Covone:2014nla}.  But in the limit that the systematic variation in $c$ with mass and redshift is small, the average projected matter profiles would have the same shape in the space of $R/R_\mathrm{200c}$.  The amplitudes of these projected matter profiles would scale as $R_\mathrm{200c}\rho_\mathrm{crit}$.  This suggests a rescaled projected matter profile, $\widetilde{\Delta\Sigma}$, that would allow for easy exploration of departures from self-similarity:
\begin{equation}
    \widetilde{\Delta\Sigma}
    \left(\frac{R}{R_\mathrm{200c}},z\right)=
    \frac{\Delta\Sigma
    \left(\frac{R}{R_\mathrm{200c}},z\right)}
    {R_\mathrm{200c}\rho_\mathrm{crit}(z)}.
    \label{eq:scaledDeltaSigma}
\end{equation}
We note here that for any given $R_\Delta$, where $\Delta$ represents the overdensity with respect to the critical or mean background densities, one can create a rescaled profile $\widetilde{\Delta\Sigma}$.  If one chooses mean background density $\left<\rho(z)\right>$, which corresponds to a scale radius of $R_{\mathrm{200m}}$ rather than $R_\mathrm{200c}$, the rescaling in amplitude would have to be adjusted to follow the correct redshift evolution of the mean background density $\left<\rho(z)\right>=\Omega_\mathrm{m}\rho_\mathrm{crit}(z=0)\left(1+z\right)^3$.

Finally, because observationally determined cluster centers are not perfect tracers of the true halo center, one must consider also the impact that this mis-centering will have on the observed matter profile.  When considering a mis-centering radius $R_{\mathrm{mis}}$ the azimuthal average of the surface mass density can be expressed as
\begin{equation}
    \Sigma(R,R_{\mathrm{mis}}) = \frac{1}{2\pi} \int_{0}^{2\pi} \mathrm{d}\theta \, \Sigma \bigg( \sqrt{R^2 + R_{\mathrm{mis}}^2 - 2RR_{\mathrm{mis}}\mathrm{cos}\theta} \bigg).
    \label{eq:mis-centered}
\end{equation}
Generally speaking, the mis-centering effects when using optical centers \citep[MCMF adopts the brightest cluster galaxy (BCG) position or the center of mass of the red galaxy distribution if the BCG is significantly offset from that;][]{Klein2023arXiv230909908K} or X-ray or even tSZE centers, the impact of mis-centering has only a minor impact outside the inner core region of the cluster.  We return to this issue in Sect.~\ref{sec:mis-centering}.

\begin{figure*}
        \includegraphics[width=\columnwidth]{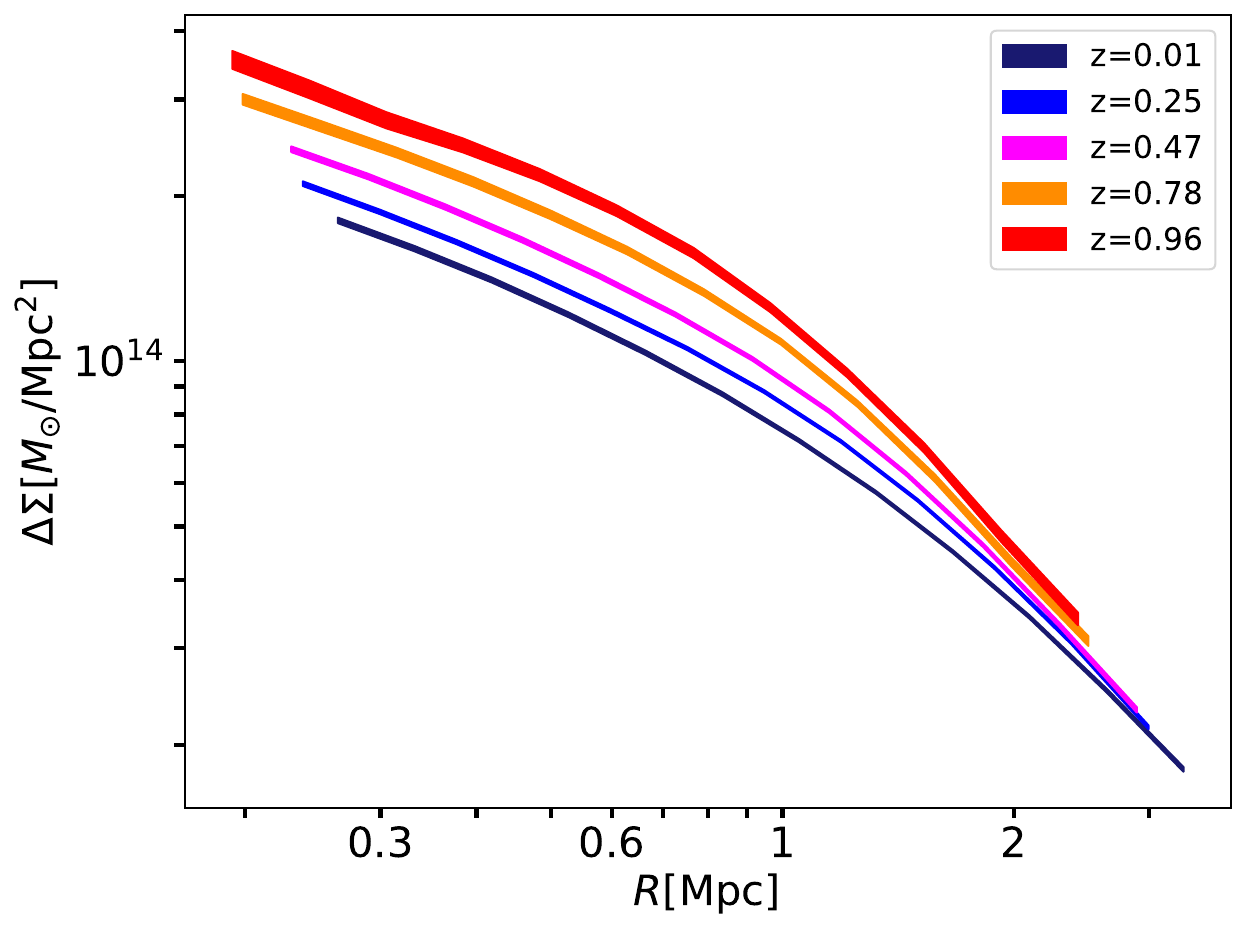}
        \includegraphics[width=\columnwidth]{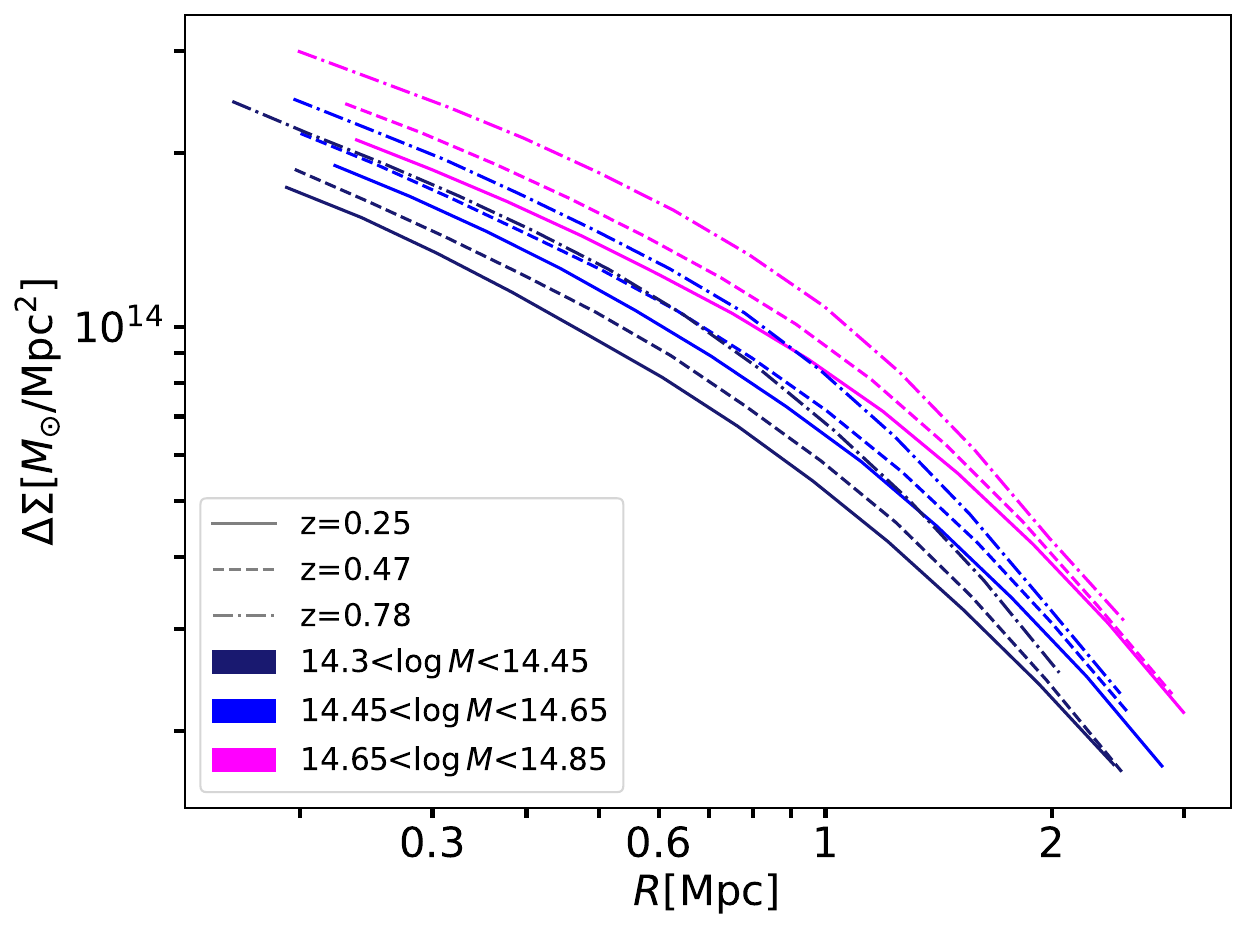}
        \vskip-0.10in
    \caption{Average cluster matter profiles $\Delta\Sigma(R)$ in the \textit{Magneticum} simulation at five redshifts (color-coded on left) for the same halo-mass bin, $14.65<\log (M_{200\mathrm{c}}/M_{\odot})<14.85$, and for three mass and redshift bins (color and line-type coded on right). The dependence of the matter profiles on cluster mass and redshift is clearly visible.  The thickness of the lines represents the 68\% credible region in the average matter profiles.}
    \label{fig:sims_DeltaSigma}
\end{figure*}

\begin{figure*}
     \includegraphics[width=\columnwidth]{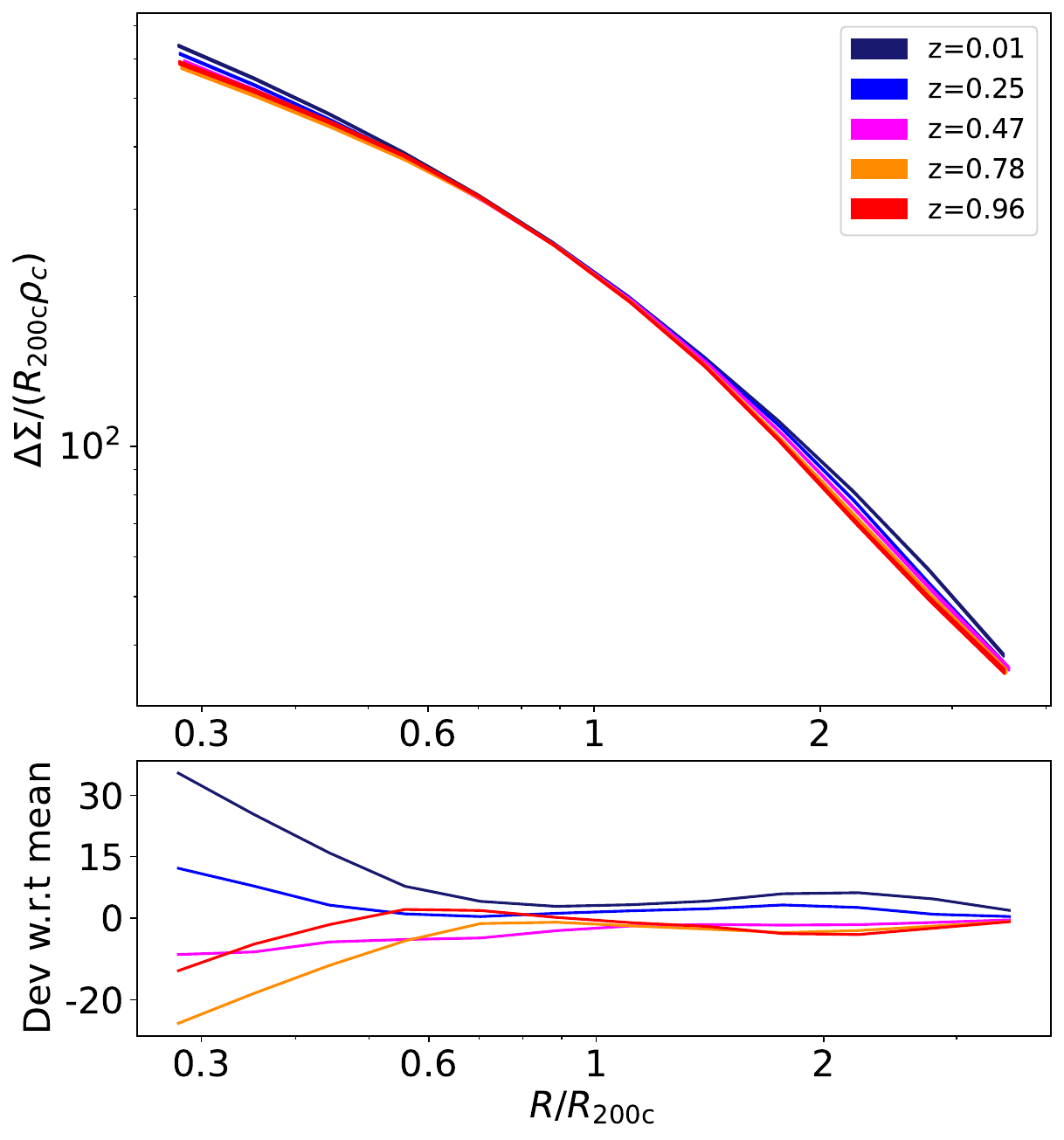}
     \includegraphics[width=\columnwidth]{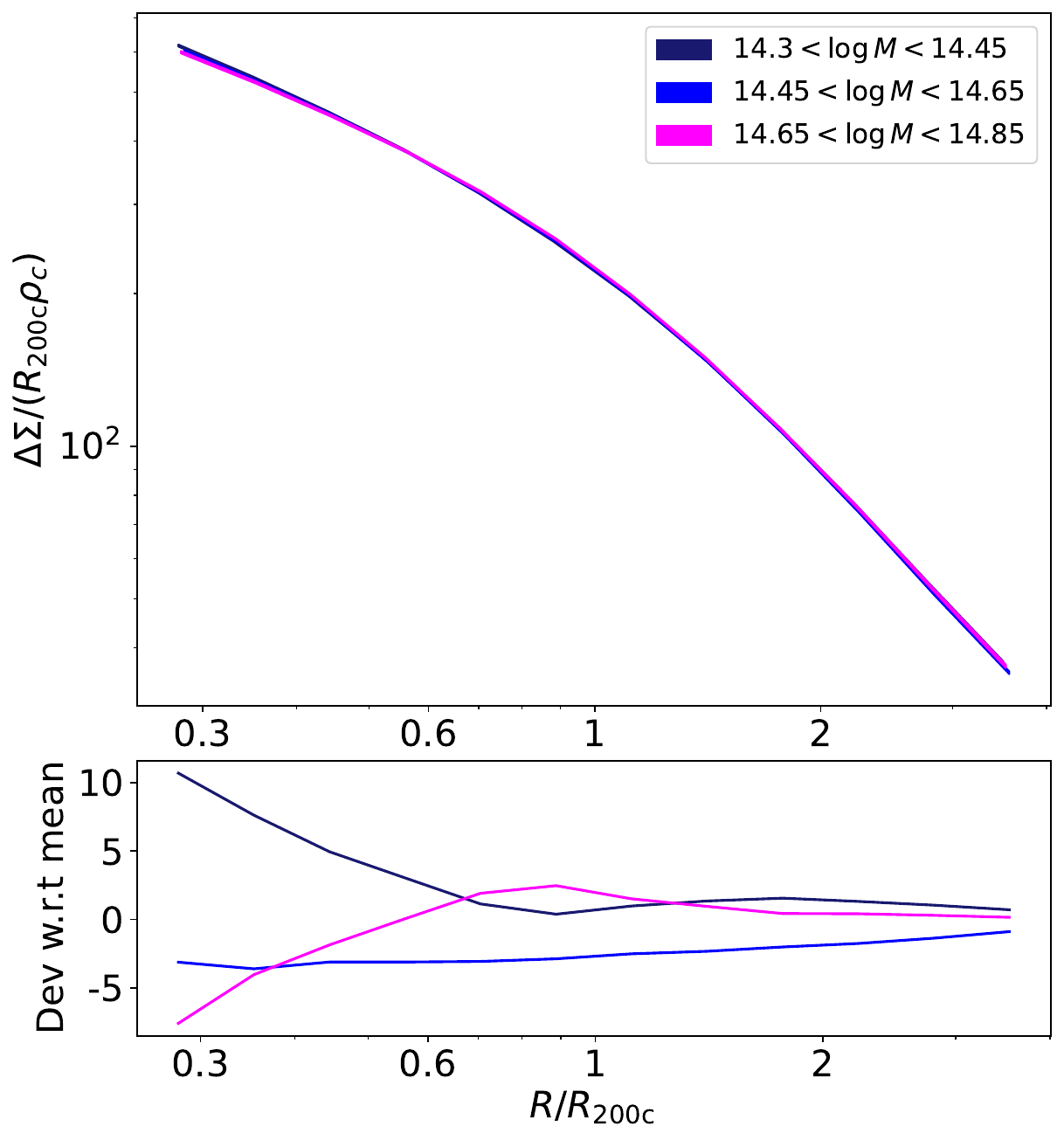}
        \vskip-0.10in
    \caption{Average cluster matter profiles in the \textit{Magneticum} simulation rescaled as in Eq.~\eqref{eq:scaledDeltaSigma} to $\widetilde{\Delta\Sigma}(R/R_\mathrm{200c})$. In the left plot, we average all the clusters ($14.3<\log (M_{200\mathrm{c}}/M_{\odot})<15$) for a given redshift and analyze the redshift trend. In the right plot, we average clusters over all redshift ranges ($0.01\leq z\leq 0.96$) within a mass bin and analyze the mass trends. In both panels, the bottom plot shows the deviation of average profiles with respect to the mean of all the profiles. The rescaling dramatically reduces systematic trends in mass and redshift, highlighting the degree of self-similarity in the matter profiles even when baryonic components are included. The thickness of the lines in the upper panels represents the 68\% credible region in the average matter profiles.}
    \label{fig:sims_scaledDeltaSigma}
\end{figure*}

\subsection{Average matter profiles: Hydrodynamical simulations}
\label{sec:sims_results}
To enable our study of the average cluster matter profiles, we extracted 
cluster $\Delta\Sigma$ profiles following the method described in \cite{Grandis21} for each redshift for both the IllustrisTNG and \textit{Magneticum} simulations. In total, we extracted 903, 852, 780, 684, and 528 cluster matter profiles at redshifts of 0.01, 0.25, 0.47, 0.78, and 0.96 respectively. In the absence of measurement uncertainties, we constructed average $\Delta\Sigma$ profiles for further study using
\begin{equation}
    \Delta\Sigma(R_j) = \frac{1 }{ N }\sum_i \Delta\Sigma_i(R_j),
    \label{eq:delta_sigma_stack_sims}
\end{equation}
where the sum is over the $N$ clusters $i$ in the sample and $R_j$ is the radial binning adopted for the cluster matter profiles.

\begin{figure*}
        \includegraphics[width=0.99\columnwidth]{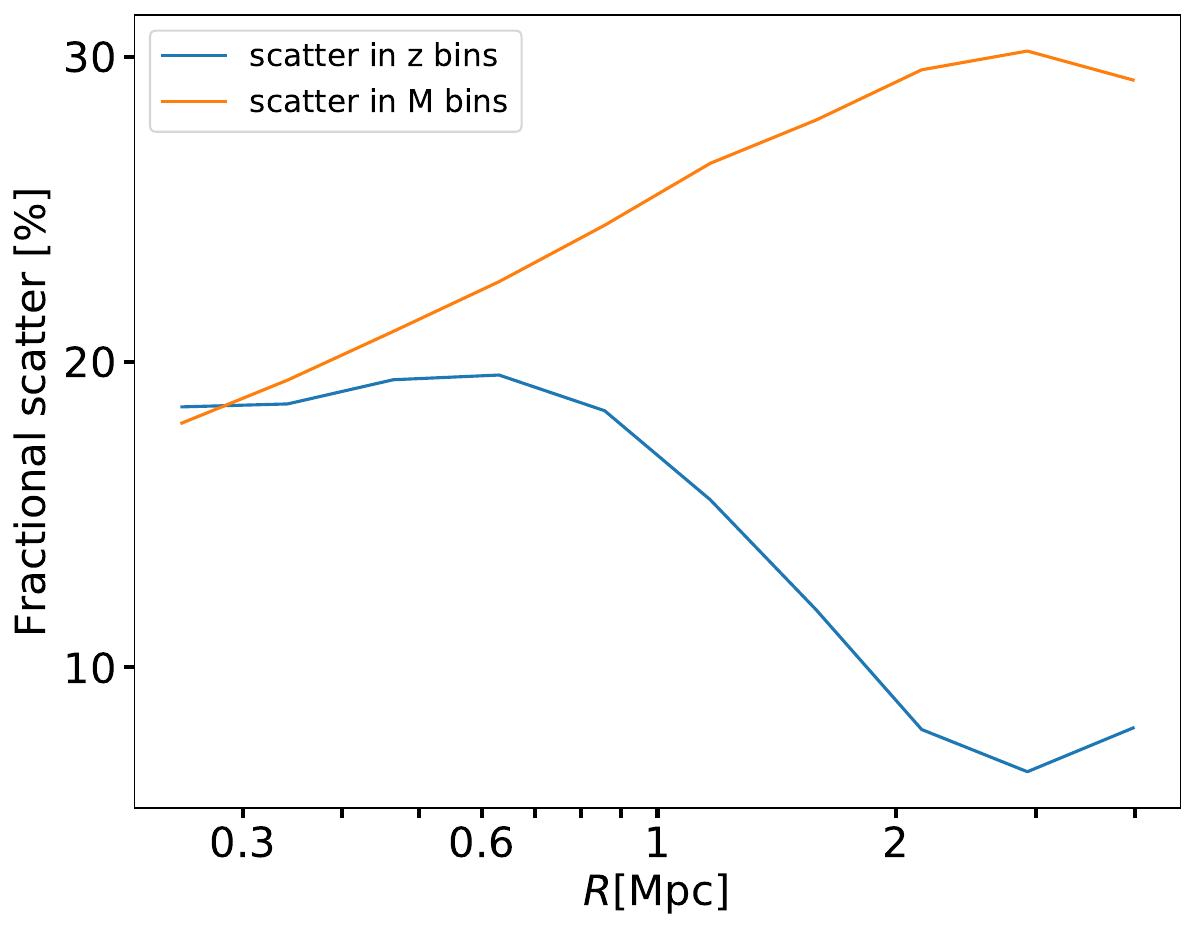}
        \includegraphics[width=\columnwidth]{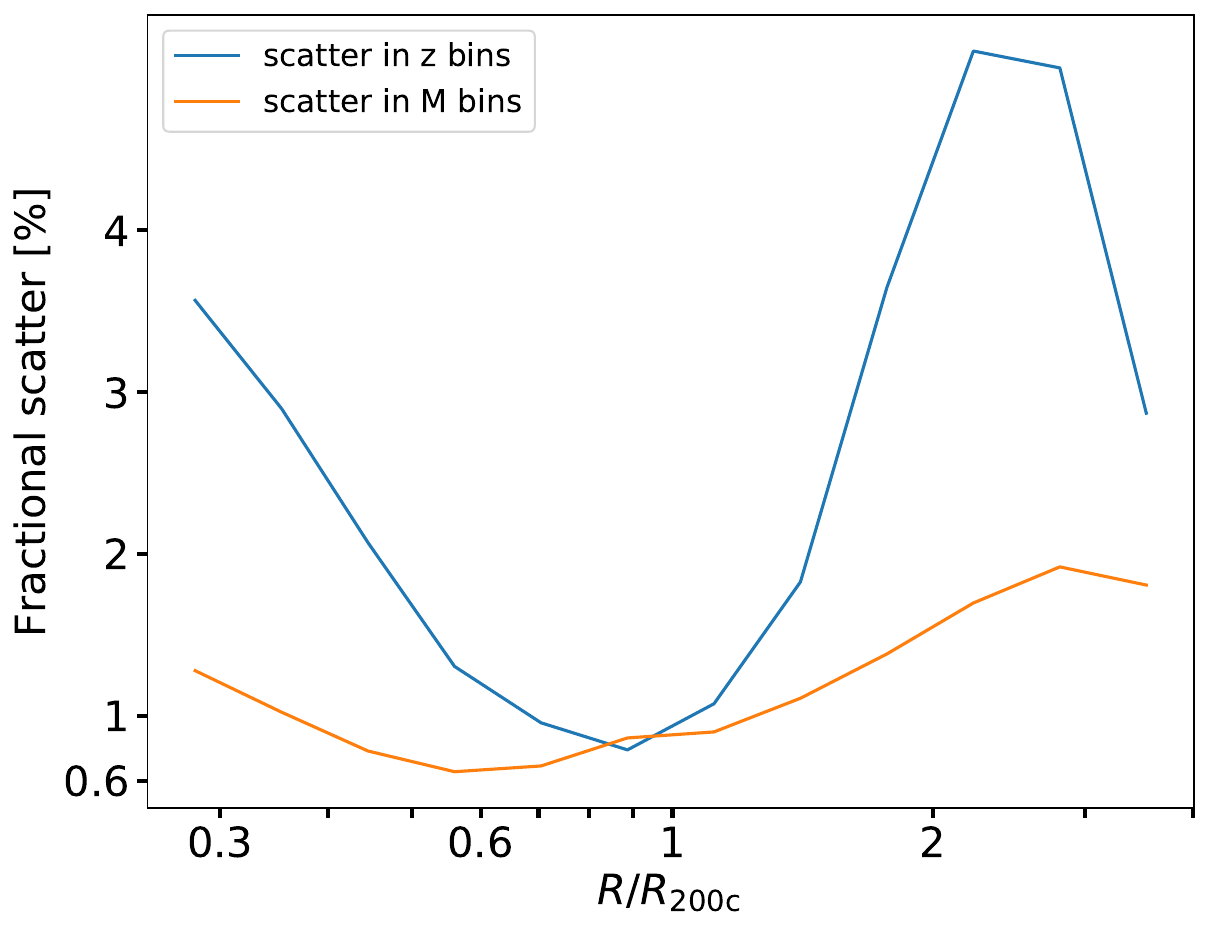}
        \vskip-0.10in
    \caption{Fractional variation in the average matter profiles versus radius shown for $\Delta\Sigma(R)$ on the left and for average rescaled matter profiles $\widetilde{\Delta\Sigma}(R/R_\mathrm{200c})$ on the right. We track the scatter due to redshift variations in blue and mass variations in orange. The average rescaled matter profiles exhibit approximately six and 23 times less variation on average with redshift and mass, respectively, within the one-halo region.}
    \label{fig:sims_fractional_scatter}
\end{figure*}

\subsubsection{$\Delta\Sigma(R)$ dependence on mass and redshift}

To start, we compute the average matter profiles within each of the five redshifts.
We compare the average matter profiles $\Delta\Sigma(R)$ at different redshifts using the \textit{Magneticum} sample in the left panel of Fig.~\ref{fig:sims_DeltaSigma}. To ensure we are seeing only trends with redshift, we average a narrow range of cluster mass ($14.65<\log (M/M_{\odot})<14.85$) for all of the redshifts. The average cluster matter profiles show a significant dependence on redshift.

In the right panel of Fig.~\ref{fig:sims_DeltaSigma} we plot average matter profiles $\Delta\Sigma(R)$ at three redshifts of 0.25, 0.47, and 0.78. Each redshift bin is divided into three mass bins ($14.30<\log (M/M_{\odot})<14.45$, $14.45<\log (M/M_{\odot})<14.65$ and $14.65<\log (M/M_{\odot})<14.85$). Profiles show a significant dependence on halo mass for all three redshifts. Higher mass clusters have higher amplitude $\Delta\Sigma(R)$ when compared to the low mass clusters for a given redshift, due chiefly to the increased extent of the cluster along the line of sight. Figure~\ref{fig:sims_DeltaSigma} makes clear that averaging cluster matter profiles in this way will therefore lead to results that are sensitive to the distribution of the cluster sample in redshift and mass (as well as the spatial distribution and masking of the source galaxies), which complicates the interpretation and characterization of such mean matter profiles.

\subsubsection{Evidence for self-similarity in mass and redshift}
Motivated by the results from the previous section and the behavior of NFW profiles derived from N-body simulations, we used the same simulated clusters to explore a rescaled matter profile $\widetilde{\Delta\Sigma}(R/R_\mathrm{200c})$ (see Eq.~\ref{eq:scaledDeltaSigma}). This profile would be identical for all samples of clusters if the population were truly self-similar.

To determine the average $\widetilde{\Delta\Sigma}$ profiles we combine the individual cluster matter profiles $\widetilde{\Delta\Sigma}_i$ as
\begin{equation}
\widetilde{\Delta\Sigma}\left(\frac{R}{R_\mathrm{200c}}\right)_j = \frac{1}{N}\sum_{i} \widetilde{\Delta\Sigma_i}\left(\frac{R}{R_\mathrm{200c}}\right)_j,
\label{eq:sims_averagescaledDeltaSigma}
\end{equation}
where the summation $i$ is over the N clusters in the sample, and $j$ denotes the radial bin in units of $R/R_\mathrm{200c}$.

In the left panel of Fig.~\ref{fig:sims_scaledDeltaSigma} we show average rescaled cluster matter profiles $\widetilde{\Delta\Sigma}\left(R/R_\mathrm{200c}\right)$ at five redshifts: 0.01, 0.25, 0.47, 0.78 and 0.96. We average all the clusters for a given redshift and then analyze the redshift trend. The profiles show very small variations and little change in amplitude with redshift as seen in the bottom left panel. This is in contrast to the behavior observed in Fig.~\ref{fig:sims_DeltaSigma} where we show $\Delta\Sigma(R)$. The average matter profiles line up for all redshifts from $R/R_\mathrm{200c} \approx 0.6$ to $R/R_\mathrm{200c} \approx 1$ with some small, remaining redshift trend at low and high $R/R_\mathrm{200c}$.

Similarly, in the right panel of Fig.~\ref{fig:sims_scaledDeltaSigma} we combine all the redshift samples and divide them into three mass bins ($14.30<\log (M/M_{\odot})<14.45$, $14.45<\log (M/M_{\odot})<14.65$ and $14.65<\log (M/M_{\odot})<14.85$) to study the mass trends in average $\widetilde{\Delta\Sigma}\left(R/R_\mathrm{200c}\right)$.  The profiles show remarkably small variation.  This is an indication that even when hydrodynamical effects are included, simulated galaxy clusters over this range of mass and redshift have matter profiles that exhibit strikingly similar shape.  As discussed in Sect.~\ref{sec:self-similarity}, the lack of variation in shape is an indication of the self-similarity of cluster matter profiles along dimensions of mass and redshift.

\begin{figure*}
        \includegraphics[width=\columnwidth]{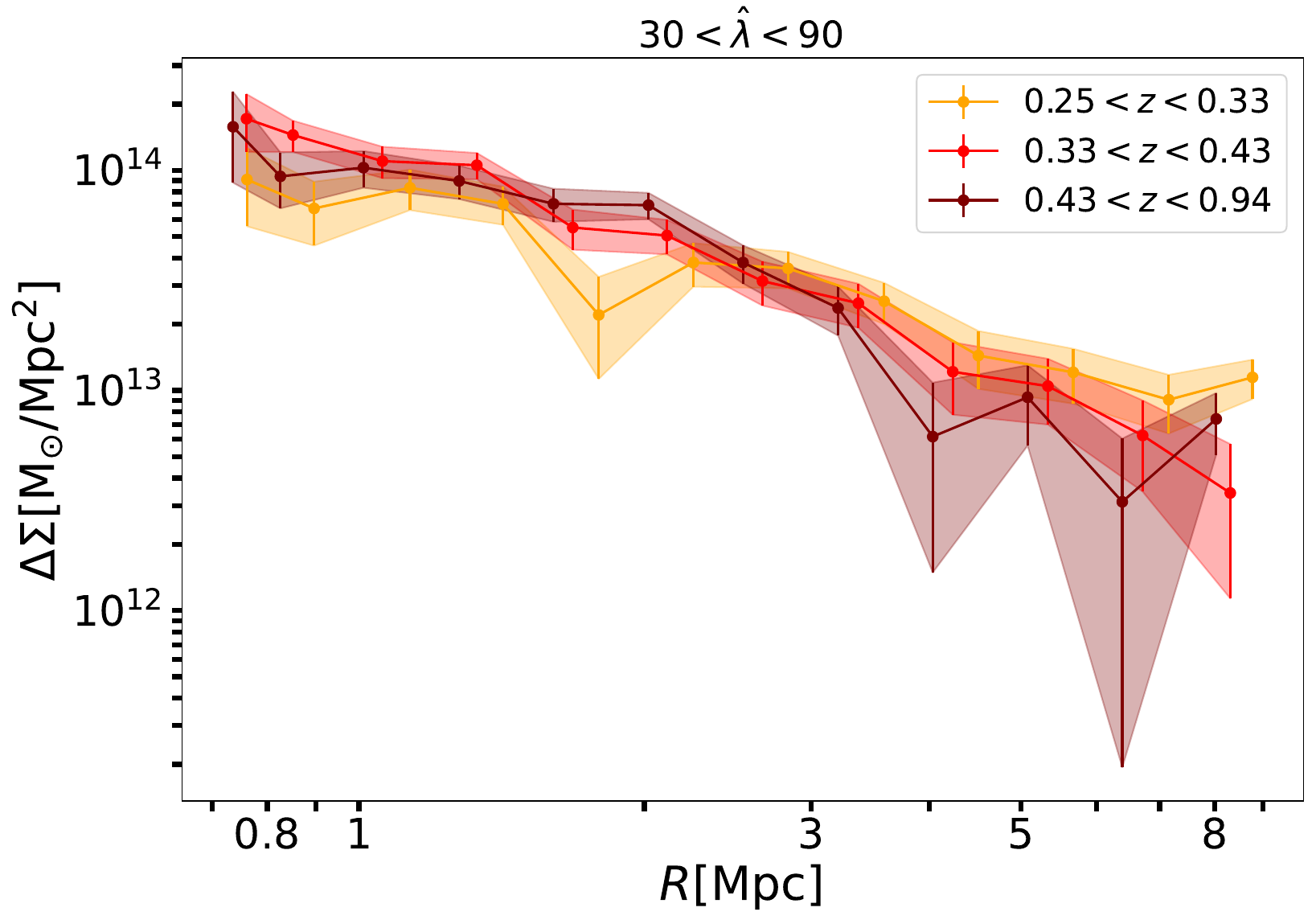}
        \includegraphics[width=\columnwidth]{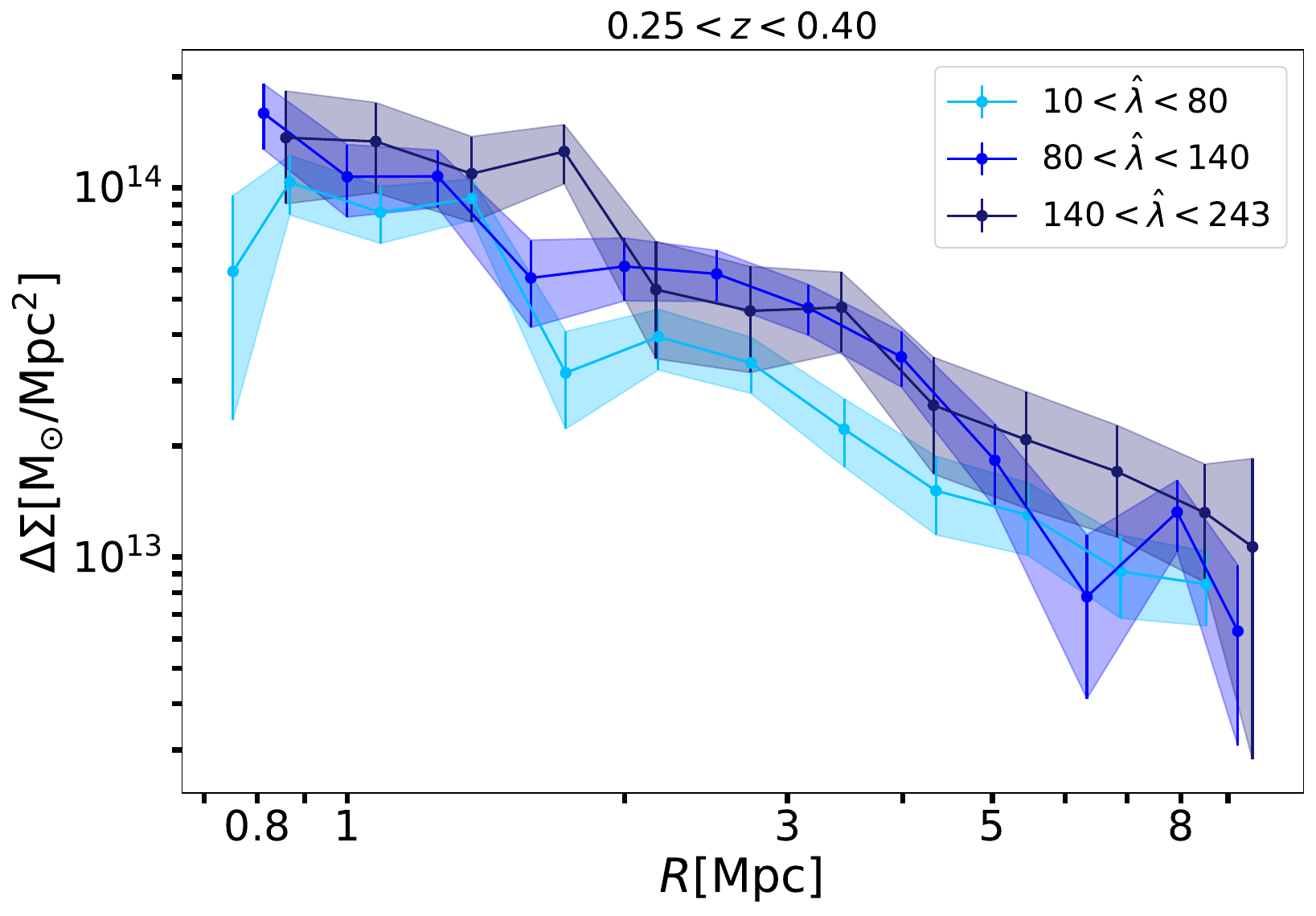}
        \vskip-0.10in
    \caption{SPT cluster average matter profiles $\Delta\Sigma(R)$ for three redshift bins in a given richness bin. In the right panel we show the average matter profiles $\Delta\Sigma(R)$ for different richness bins in a given redshift bin.  The color bands encode the 68\% credible region for each profile.  The profiles show variation with redshift and richness that is consistent with that shown in Fig.~\ref{fig:sims_DeltaSigma} for the simulated clusters in redshift and mass.} \label{fig:SPT_DeltaSigma}
\end{figure*}

\begin{figure*}
        \includegraphics[width=\columnwidth]{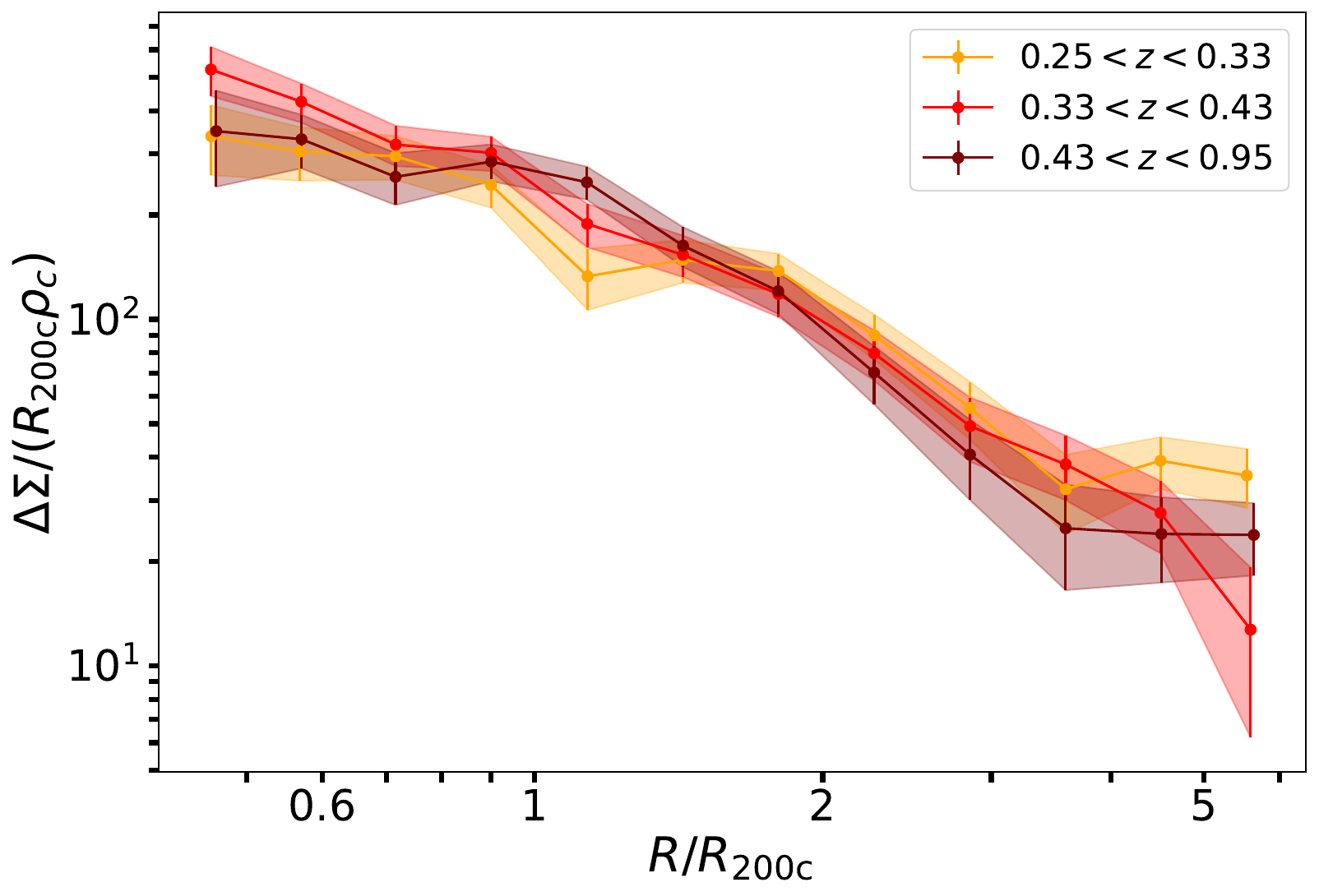}
        \includegraphics[width=\columnwidth]{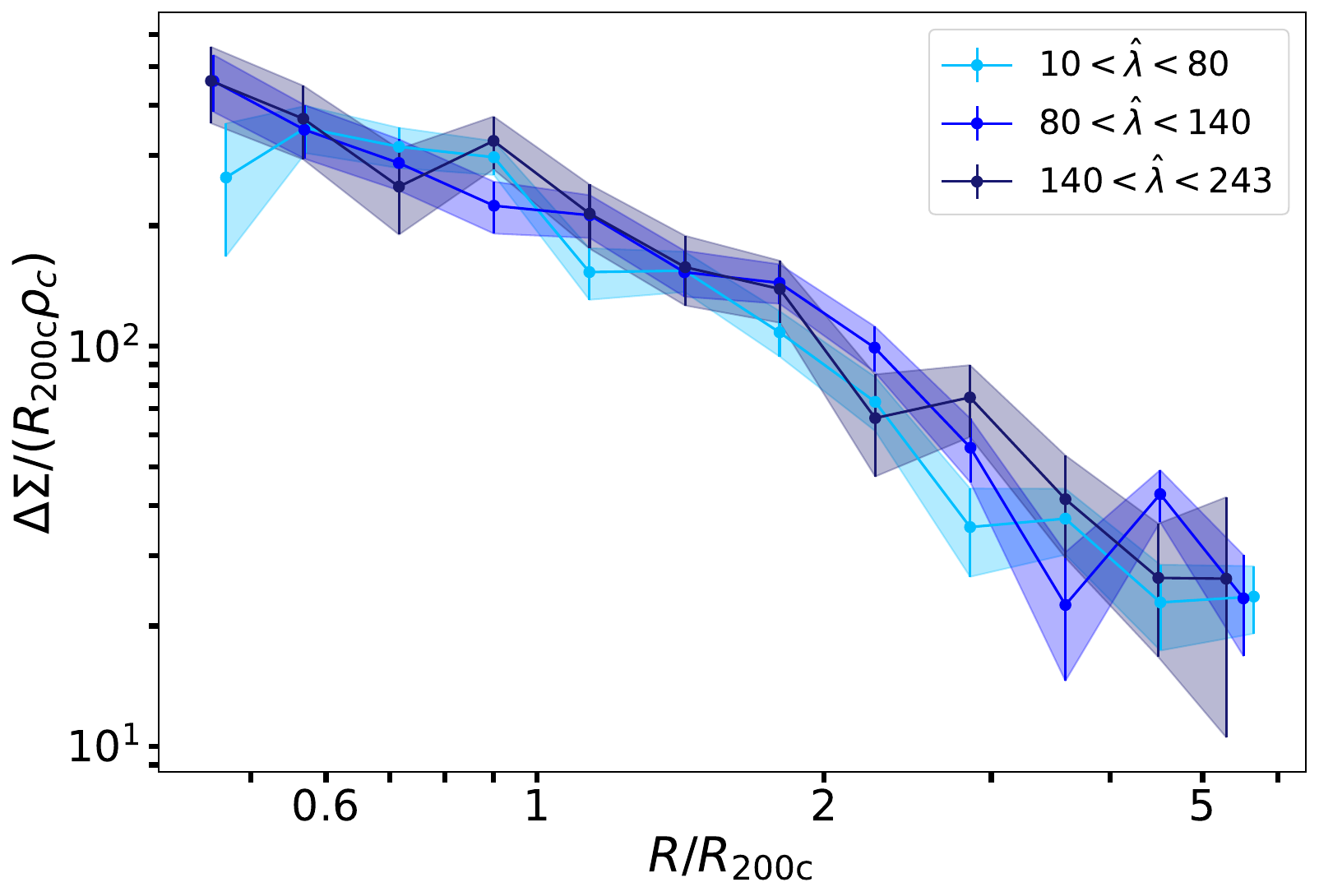}
        \vskip-0.10in
    \caption{Average rescaled SPT cluster matter profiles $\widetilde{\Delta\Sigma}\left(R/R_\mathrm{200c}\right)$ constructed using the mass calibration posteriors from \protect\cite{Bocquet2024IIPhRvD.110h3510B}. In the left panel, we show $\widetilde{\Delta\Sigma}\left(R/R_\mathrm{200c}\right)$ profiles for three redshift bins, and in the right panel we show $\widetilde{\Delta\Sigma}\left(R/R_\mathrm{200c}\right)$ profiles for different richness bins.  The average rescaled matter profiles show regularity similar to that seen in Fig.~\ref{fig:sims_scaledDeltaSigma} for the cluster simulations.}
    \label{fig:SPT_scaledDeltaSigma}
\end{figure*}

\subsubsection{Variations in average matter profiles}
\label{sec:re-scaled vs. non-re-scaled space}
In this section we aim to quantify the degree of variation we see in cluster matter profiles. For this, we calculate the fractional scatter in the average matter profiles as a function of redshift and mass. We also compare this to the fractional scatter values obtained when averaging profiles in physical space. The fractional scatter is given by $\sigma_{\Delta\Sigma}/<\Delta\Sigma>$, where $\sigma_{\Delta\Sigma}$ is the standard deviation of the sample and $<\Delta\Sigma>$ is the mean of the sample.

We used the \textit{Magneticum} simulation, which has a larger volume and therefore contains many more halos in comparison to IllustrisTNG. This allows us to measure a higher SNR through averaging profiles, which in turn enables us to better quantify the intrinsic scatter in the average matter profiles. We ignore the uncertainty on the average matter profiles when calculating the fractional scatter with redshift and mass, because the uncertainty is much smaller than the scatter.

Starting with the $\Delta\Sigma(R)$ profiles, we first divide each redshift bin into four mass bins. We calculate the fractional scatter as a function of redshift in a given mass bin and then report the average value of fractional scatter as a function of redshift of four mass bins as a function of radius in the left panel (blue curve) of Fig.~\ref{fig:sims_fractional_scatter}. The scatter ranges from $\approx 7\%$ to $\approx 19\%$ with an average of $\approx 17\%$. Similarly, when analyzing the fractional scatter as a function of mass, we compute the scatter at each redshift bin and report the mean value for five redshift bins. The orange curve in the left panel of Fig.~\ref{fig:sims_fractional_scatter} shows the mean scatter values as a function of radius $R$ with an average of $\approx 26\%$. 

Moving to the average rescaled matter profiles $\widetilde{\Delta\Sigma}\left(R/R_\mathrm{200c}\right)$, we average all of the clusters for a given redshift and then report the fractional scatter among the five redshift bins. The blue curve in the right panel of Fig.~\ref{fig:sims_fractional_scatter} shows the fractional scatter as a function of redshift. The value varies from $\approx 0.8\%$ to $\approx 5\%$ with an average value of $\approx 2.6\%$ and the minimum value is achieved around $\approx 0.9 R/R_\mathrm{200c}$. The scatter in $\widetilde{\Delta\Sigma}\left(R/R_\mathrm{200c}\right)$ is reduced by a factor of 6 in comparison to the scatter obtained in $\Delta\Sigma(R)$. Similarly, when analyzing the fractional scatter as a function of mass, we divide each redshift into four mass bins and stack all of the clusters with different redshifts in a given mass bin. The orange curve in Fig.~\ref{fig:sims_fractional_scatter} shows the trend of the fractional scatter as a function of mass with scaled radius with an average value of $\approx 1.1\%$, which is 23 times smaller relative to the scatter with mass in $\Delta\Sigma(R)$.

\subsection{Average matter profiles: Observations}
\label{sec:ObservedProfiles}
In this section we examine the matter profiles of the 698 SPT tSZE-selected clusters using the WL data from DES in physical and rescaled space.    
Given the cosmological parameters $\vec p$ (Table~\ref{tab:parameter_priors}), the average $\Delta\Sigma$ estimator for the cluster ensemble in a radial bin $R_j$ is a triple sum over clusters $k$, lensing source galaxy tomographic bins $b$, and individual lensing source galaxies $i$ as
\begin{equation}
   \Delta\Sigma(R_j|\vec p) = \frac{\sum_{k,b,i} \frac{\Sigma_{\mathrm{crit},k,b} w_{k,b} 
   \mathcal{W}_{k,b,i}^{\mathrm{s}} e_{\mathrm{t},k,b,i} }{(1-f_{\mathrm{cl},k,b})}}{\sum_{k,b,i} w_{k,b}
   \mathcal{W}_{k,b,i}^{\mathrm{s}} (R_{\gamma_{\mathrm{t},i}} + R_{\mathrm{sel}})}.
    \label{eq:obs_averageDeltaSigma}
\end{equation}
Here, $\mathcal{W}^{\mathrm{s}}_{k,b,i}$ is the scaled source weight, and $w_{k,b}$ is the tomographic bin weight
\begin{eqnarray}
    \mathcal{W}^{\mathrm{s}}_{k,b,i} = & w_i^{\mathrm{s}} \left(\frac{1-f_{\mathrm{cl},k,b}}{\Sigma_{\mathrm{crit},k,b}}\right)^2\nonumber\\
    w_{k,b} = & \Sigma_{\mathrm{crit},k,b}^{-1},
\end{eqnarray}
where the $w^{\mathrm{s}}_i$ represent individual source weights (defined as the inverse variance in the measured ellipticity).
Following \cite{BocquetI2024PhRvD.110h3509B}, we employed only the tomographic bins 2 to 4 in this analysis. Additionally, we note that we only use the tomographic bins for which the median source redshift is larger than the cluster redshift. $\Sigma_{\mathrm{crit},k,b}$ is the critical surface density, which depends on cluster redshift and source galaxy redshift distribution, calculated as in Eq.~\ref{eq:sigma_crit_integral}. The ellipticity of a source galaxy $i$ from a tomographic bin $b$ and lying in the background of a cluster $k$ is $e_{\mathrm{t},k,b,i}$, and $R_{\gamma_{\mathrm{t},i}}$ is the shear response for galaxy $i$, which is needed to scale the ellipticity to the reduced shear. Additionally, the selection response $R_\mathrm{sel}$ accounts for the fact that lensing sources are selected based on their (intrinsically) sheared observations. We use $R_\mathrm{sel} = -0.0026$ for optical centers as measured previously for this sample \citep{BocquetI2024PhRvD.110h3509B}. We also scale the ellipticity with a factor of $1/(1-f_{\mathrm{cl},k,b})$ to correct the profiles for the cluster member contamination, which is measured separately for each tomographic bin (see the discussion in Sect.~\ref{sec:systematics}). 

The corresponding uncertainty in $\Delta\Sigma$  for a radial bin $R_j$ in the average cluster matter profile is calculated as
\begin{equation}
    \sigma_{\Delta\Sigma}^2(R_j|\vec p) = \frac{\sum_{k,b,i} \left(\frac{\Sigma_{\mathrm{crit},k,b} w_{k,b} \mathcal{W}_{k,b,i}^{\mathrm{s}} \sigma_{\mathrm{eff}, b}}{(1-f_{\mathrm{cl},k,b})}\right)^2  }{ (\sum_{k,b,i} w_{k,b} 
    \mathcal{W}_{k,b,i}^{\mathrm{s}})^2}.
    \label{eq:obs_averageDeltaSigma_uncertainty}
\end{equation}
Here $\sigma^2_{\mathrm{eff}, b}$ is the effective shape variance for sources in a given tomographic bin, and all other elements are as described in Eq.~\ref{eq:obs_averageDeltaSigma}.

Similarly, after accounting for the rescaling described in Eq.~\ref{eq:scaledDeltaSigma} and given the cosmological parameters $\vec p$ and the cluster mass $M_\mathrm{200c}$ or equivalently the cluster radius $R_\mathrm{200c}$, the inverse variance weighted average $\widetilde{\Delta\Sigma}$ estimator in a radial bin $(R/R_\mathrm{200c})_j$ is given by 
\begin{equation}
    \widetilde{\Delta\Sigma}\left(\frac{R}{R_\mathrm{200c}}\bigg|\vec{M_\mathrm{200c}}, \vec p\right)_j = \frac{\sum_{k,b,i} \frac{\Sigma_{\mathrm{crit},k,b} w_{k,b} 
   \widetilde{\mathcal{W}}_{k,b,i}^{\mathrm{s}} e_{\mathrm{t},k,b,i}}{\rho_{\mathrm{crit}_k}R_{\mathrm{200c}_k}(1-f_{\mathrm{cl},k,b})}}{\sum_{k,b,i} w_{k,b}
   \widetilde{\mathcal{W}}_{k,b,i}^{\mathrm{s}} (R_{\gamma_{\mathrm{t}, i}} + R_{\mathrm{sel}})},
   \label{eq:obs_rescaledDeltaSigma}
\end{equation}
where $\widetilde{\mathcal{W}}_{k,b,i}^{\mathrm{s}}$ is the re-scaled source weight corresponding to $\widetilde{\Delta\Sigma}$, which is defined as 
\begin{equation}
    \widetilde{\mathcal{W}}_{k,b,i}^{\mathrm{s}} = w_i^{\mathrm{s}} \left(\frac{\rho_{\mathrm{crit}_{k} R_{\mathrm{200c}_{k}}(1-f_{\mathrm{cl},k,b})}}{ \Sigma_{\mathrm{crit},k,b}}\right)^2,
\end{equation}
and the corresponding uncertainty is given by
\begin{equation}
    \sigma_{\widetilde{\Delta\Sigma}}^2\left(\frac{R}{R_\mathrm{200c}}\bigg|\vec{M_\mathrm{200c}}, \vec p\right)_j = \frac{\sum_{k,b,i} \left(\frac{\Sigma_{\mathrm{crit},k,b} w_{k,b} \widetilde{\mathcal{W}}_{k,b,i}^{\mathrm{s}} \sigma_{\mathrm{eff}_{b}}}{\rho_{\mathrm{crit},k}R_{\mathrm{200c}_{k}(1-f_{\mathrm{cl},k,b})}}\right)^2} {(\sum_{k,b,i} w_{k,b} 
    \widetilde{\mathcal{W}}_{k,b,i}^{\mathrm{s}})^2}.
    \label{eq:obs_rescaledDeltaSigma_uncertainty}
\end{equation}
The mean estimated scaled radius of a given radial bin $j$ is calculated from the equation
\begin{equation}
    \left( \frac{R}{R_\mathrm{200c}}\bigg|\vec{M_\mathrm{200c}}, \vec p \right)_j = \frac{\sum_{k,b,i} w_{k,b} \widetilde{\mathcal{W}}_{k,b,i}^{\mathrm{s}} R_{k,b,i}/R_{\mathrm{200c}_{k}} }{ \sum_{k,b,i} w_{k,b} \widetilde{\mathcal{W}}_{k,b,i}^{\mathrm{s}} },
\end{equation}
where $R_{k,b,i}$ is the projected separation of ellpiticity $i$ in tomographic bin $b$ from the center of cluster $k$.  A similar expression for the mean radius $R_j$ within a bin pertains, but without the $1/R_\mathrm{200c}$ scaling.

In the left panel of Fig.~\ref{fig:SPT_DeltaSigma}, we show the average cluster matter profile $\Delta\Sigma(R)$, in three redshift bins (for a richness bin, $30<\hlambda<90$ containing 426 clusters), and in the right panel, we show the average cluster matter profiles in three richness bins (for a redshift bin, $0.25<z<0.40$ containing 123 clusters).
While the measurement uncertainties are significant in this sample (color bands represent 68\% credible regions), it is still possible to discern variations among the presented average matter profiles.

To study the average $\widetilde{\Delta\Sigma}\left(R/R_\mathrm{200c}\right)$ profiles, we need a robust value of $R_\mathrm{200c}$ for each cluster, which then implies we need good mass constraints for each system (Eq.~\ref{eq:mass_definition}). 
For this purpose, we adopt the observable-mass relation ($\zeta$-mass and $\lambda$-mass-- see discussion in Sect.~\ref{sec:observable-mass}) posteriors from \cite{Bocquet2024IIPhRvD.110h3510B}, where the mass calibration constraints were obtained using the same DES WL data as part of the cosmological cluster abundance analysis of the sample \citep{BocquetI2024PhRvD.110h3509B}. 
We used the full sample while analyzing clusters in redshift or richness bins. The average rescaled cluster matter profiles exhibit less evidence for variation than the $\Delta\Sigma(R)$ profiles in Fig.~\ref{fig:SPT_scaledDeltaSigma}.  In other words, the tSZE selected clusters show indications of self-similarity with redshift and richness similar to those presented above for the clusters from hydrodynamical simulations with redshift and mass.

This simplicity in the average rescaled matter profiles $\widetilde{\Delta\Sigma}\left(R/R_\mathrm{200c}\right)$ of the simulated and observed galaxy cluster population offers some advantages. It allows us to combine large ensembles of clusters with a wide halo mass and redshift range, creating higher SNR cluster matter profiles, which can be used to test different models of structure formation. Moreover, the modeling of average cluster matter profiles $\widetilde{\Delta\Sigma}\left(R/R_\mathrm{200c}\right)$ for observed cluster samples becomes more straightforward, because the final average profile is insensitive to the redshift and mass distribution of the cluster sample (and to the spatial distribution and masking of the source galaxies). We employ this simplicity in matter profiles in Sect.~\ref{sec:mass_calib}, where we present a new method of galaxy cluster mass calibration that exploits the approximate self-similarity of galaxy clusters.


\section{Mass calibration method}
\label{sec:method}
Measurements of the WL signal induced by foreground galaxy clusters can be used to robustly estimate the cluster mass. However, because the S/N of the WL signal is low for individual clusters, it is practical to perform mass calibration using the lensing signal averaged over many clusters \citep[e.g.,][]{Umetsu_2014, Umetsu:2015baa, Okabe_2010, Okabe_2013}. Because the average rescaled matter profiles $\widetilde{\Delta\Sigma}\left(R/R_\mathrm{200c}\right)$ are particularly simple to model, they offer the possibility to improve upon previous average matter profile based WL mass calibration methods.

The method presented below involves 1) building an ensemble of average rescaled matter profiles-- one for each bin in cluster observable, 2) extracting a likelihood of these matter profiles given a model profile and then 3) iterating with a Markov Chain Monte Carlo method to characterize the posteriors of the model parameters that describe the galaxy cluster observable-mass relations discussed in Sect.~\ref{sec:observable-mass}.  The mean posteriors describe the parameters for which the observed and model average matter profiles are consistent over the full range of cluster observables.  The average matter profile model is discussed in Sect.~\ref{sec:hydro-model}, and the mass calibration likelihood is presented in Sect.~\ref{sec:likelihood_model}.  A discussion of the systematic effects and their correction then appears in Sect.~\ref{sec:systematics}.

\label{sec:mass_calib}

\subsection{Observable-mass relations}
\label{sec:observable-mass}
In our analysis, each confirmed cluster has four associated observable quantities.  These include the tSZE detection significance $\hat\zeta$, the MCMF obtained richness $\hat\lambda$ and redshift $z$ and the WL mass $M_\mathrm{WL}$ that is derived using the WL shear and photometric redshift measurements of the background, lensed source galaxies. The WL masses are measured using average matter profiles $\widetilde{\Delta\Sigma}\left(R/R_\mathrm{200c}\right)$, and these masses are used to constrain the so-called cluster observable--mass relations \citep[e.g.,][]{Mohr1997, Mohr1999, Finoguenov2001A&A...368..749F, Chiu16, Chiu18, Bulbul2019ApJ...871...50B} that describe the redshift dependent statistical binding between the observables (i.e., detection significance, richness and WL mass) and the underlying halo mass, which for this analysis we take to be $M_\mathrm{200c}$.  

\subsubsection{tSZE detection significance $\hat\zeta$}
\label{tSZE detection significance}
As described in an early SPT analysis \citep{Vanderlinde10}, the tSZE detection significance or signal-to-noise ratio $\hat{\zeta}$ is related to the unbiased significance $\zeta$ as
\begin{equation} \label{eq:xizeta}
P(\hat{\zeta}|\zeta) = \mathcal N\left(\sqrt{\zeta^2 + 3}, 1\right),
\end{equation}
where $\mathcal N$ denotes a Gaussian distribution. This relationship accounts for the maximization bias in $\hat{\zeta}$ caused during the cluster matched filter candidate selection \citep{Melin06}, which has three free parameters (cluster sky location and cluster model filter scale).
The normal distribution models the impact of the unit noise in the appropriately rescaled mm-wave maps.
The mean unbiased detection significance is modeled as a power-law relation in mass and redshift: 
\begin{equation}
    \langle\ln\zeta|M_\mathrm{200c},z\rangle = \azeta + \bzeta \ln\left(\frac{M_\mathrm{200c}}{M_{\mathrm{piv}}}\right)  + \czeta \ln\left(\frac{E(z)}{E(z_{\mathrm{piv}})}\right),
    \label{eq:zetaM}
\end{equation}
where \azeta\ is the normalization, \bzeta\ is the mass trend, \czeta\ is the redshift trend, $E(z)$ is the dimensionless Hubble parameter, $M_\mathrm{piv}=3\times10^{14}h^{-1}$M$_\odot$ is the pivot mass and $z_\mathrm{piv}=0.6$ is the pivot redshift, which are chosen to reflect the median mass and redshift of our confirmed cluster sample. To account for the variable depth of the SPT survey fields, we rescaled the amplitude $\zeta_0$ on a field-by-field basis
\begin{equation}
\zeta_{0,i} = \gamma_i \zeta_0, 
\label{eq:aszfield}
\end{equation}
where $\gamma_i$ is obtained from simulated maps \citep{Bleem_2015,Bleem_2020,Bleem2023-500d}. 
This approach allows us to combine the full SPT cluster sample when empirically modeling the $\zeta$-mass relation.
We model the intrinsic scatter in $\zeta$ at fixed mass and redshift as log-normal $\sigma_{\mathrm{ln}\zeta}$.  This single scatter parameter has been shown to be sufficient to model the SPT tSZE-selected cluster sample $\zeta$-mass relation \citep{Bocquet_2019,Bocquet2024IIPhRvD.110h3510B}.  We return to this question with new validation tools in Sect.~\ref{sec:goodness_fit}.

\subsubsection{Cluster richness $\hat\lambda$}
\label{sec:cluster richness}
The observed cluster richness $\hat\lambda$ is related to the intrinsic richness $\lambda$ as
\begin{equation}
P(\hat\lambda|\lambda) = \mathcal N\left(\lambda, \sqrt{\lambda}\right),
\end{equation}
which models the Poisson sampling noise in the limit of a normal distribution where the dispersion is $\sigma=\sqrt{\lambda}$.
The mean intrinsic richness is modeled as a power law in mass and redshift 
\begin{equation}\label{eq:lambdamass}
    \left< \ln \lambda | M_\mathrm{200c},z\right> = \ln \alambda  + \blambda \ln\left(\frac{M_\mathrm{200c}}{M_\mathrm{piv}}\right) + \clambda \ln\left(\frac{1+z}{1+z_\mathrm{piv}}\right),
\end{equation}
\noindent
where \alambda\ is the normalization, \blambda\ is the mass trend, \clambda\ is the redshift trend and, as above, $M_\mathrm{piv}=3\times10^{14}h^{-1}$M$_\odot$ and $z_\mathrm{piv}=0.6$. The intrinsic scatter of the intrinsic richness $\lambda$ at fixed mass and redshift is modeled as a log-normal distribution with the parameter \dlambda.  This scatter is the same for all redshifts and masses, which has been shown to be adequate for modeling the $\lambda$-mass relation of the SPT selected cluster sample \citep{Bocquet2024IIPhRvD.110h3510B}.  We return to this question also with new validation tools in Sect.~\ref{sec:goodness_fit}.

\subsubsection{Weak lensing mass $M_\mathrm{WL}$}
\label{sec:weak lensing mass}
In addition to the $\zeta$-mass and $\lambda$-mass relations, we also include a mapping between the so-called WL mass $M_\mathrm{WL}$,  
which is the mass one would infer by fitting a model profile to an individual cluster matter profile, and the halo mass $M_\mathrm{200c}$. This follows the approach adopted in previous work \citep{Becker11,Dietrich2019MNRAS.483.2871D,Grandis21,BocquetI2024PhRvD.110h3509B} and is a mechanism for incorporating corrections for systematic biases that may arise from the interpretation of the average matter profiles and for marginalizing over the remaining systematic uncertainties in those bias corrections.  For example, the uncertainties associated with hydrodynamical simulations and the subgrid physics they incorporate can be modeled with this $M_\mathrm{WL}$-mass relation, incorporating an effective systematic floor in the accuracy of the final, calibrated masses.  We characterize this relation as
\begin{equation}
        \label{eq:WL_mass_rel}
        \begin{split}
         \left\langle \mathrm{ln} \left( \frac{M_{\mathrm{WL}}}{M_\mathrm{piv}}  \right)  \right\rangle = &
         \aWL + 
        \bWL \mathrm{ln} \left( \frac{M_\mathrm{200c}}{M_\mathrm{piv}} \right) \\
         &+  \cWL \mathrm{ln} \left( \frac{1+z}{1+z_\mathrm{piv}} \right) \, ,
        \end{split}
\end{equation}
where \aWL\ is the logarithmic bias at $M_\mathrm{200c}=M_\mathrm{piv}$ and \bWL\ and \cWL\ are the mass and redshift trends, respectively, of this bias.  For symmetry with the other observable-mass relations, we explicitly include the redshift trend parametrization, whereas in previous analyses  \citep[see][]{BocquetI2024PhRvD.110h3509B} the relation has been defined at specific redshifts where the required simulation outputs are available.  

The WL mass $M_{\mathrm{WL}}$ estimated from individual clusters exhibits a mass dependent log-normal scatter $\sigma_{\ln\mathrm{WL}}$ about the mean relation, which we model as
    \begin{equation}
    \begin{split}
        \label{eq:WL_mass_var}
      \ln\sigma^2_{\ln\mathrm{WL}} = & \asigmaWL 
      + \bsigmaWL  \ln\left( \frac{M_\mathrm{200c}}{M_\mathrm{piv}} \right) \\
      & + \csigmaWL \mathrm{ln} \left( \frac{1+z}{1+z_\mathrm{piv}} \right) \, ,
    \end{split}
    \end{equation}
where \asigmaWL\ is the logarithm of the variance of $M_\mathrm{WL}$ around $M_\mathrm{200c}$ at $M_\mathrm{piv}$ and $z_\mathrm{piv}$ and \bsigmaWL\  and \csigmaWL\ are the mass and redshift trends, respectively, of this variance.  For an average rescaled matter profile that is produced using $N$ clusters, the effective scatter of the extracted $M_\mathrm{WL}$ about the true mass would scale down as $1/\sqrt{N}$, reducing the stochasticity associated with the estimate of the underlying halo mass and reducing the importance of possible correlations between the scatter in $M_\mathrm{WL}$ and other observables.

The parameter posteriors on these relations are extracted through a $M_\mathrm{WL}$ calibration exercise carried out on hydrodynamical simulations of clusters output over a range of redshifts \citep{Grandis21}. This calibration exercise employs a model profile or set of model profiles as discussed in the next section and characterizes the biases and scatter associated with that model. In addition, systematic uncertainties on photometric redshifts, the multiplicative shear bias, the cluster member contamination model and the cluster mis-centering model are also incorporated into the posteriors on these parameters, making it straightforward to marginalize over all critical systematic uncertainties in the mass calibration analysis.  We return to this in Sect.~\ref{sec:systematics}.

\subsection{Average rescaled matter profile model $\widetilde{\Delta\Sigma}_\mathrm{mod}$}
\label{sec:hydro-model}

We used the IllustrisTNG and the \textit{Magneticum} simulations at five redshifts between 0 to 1 (as described in the Sect. \ref{sec:sim_data}) to create an average WL model for use in mass calibration.  Because there is little variation in the average rescaled matter profiles $\widetilde{\Delta\Sigma}\left(R/R_\mathrm{200c}\right)$ with mass and redshift, we could adopt a single average matter profile at all redshifts and masses for the model used in mass calibration.  We could then correct for the small biases introduced by this assumption of perfect self-similarity using the $M_\mathrm{WL}$-mass relations (Eq.~\ref{eq:WL_mass_rel} and \ref{eq:WL_mass_var}).  However, an examination of the average matter profiles presented in Fig.~\ref{fig:sims_scaledDeltaSigma} provides clear evidence for small departures from self-similarity with redshift, while showing no convincing evidence of departures from self-similarity in mass.  Therefore, we adopted a model $\widetilde{\Delta\Sigma}\left(R/R_\mathrm{200c}\right)$ profile that varies with redshift, while assuming perfect self-similarity in mass.  This approach sets the mean logarithmic bias \aWL\ in the $M_\mathrm{WL}$-mass relation to zero over all masses and redshifts.

To construct an average WL model that represents the typical behavior across both sets of simulations, we select the same number of simulated clusters from both the \textit{Magneticum} and IllustrisTNG simulations. Since the subgrid physics differ between the two, the average matter profiles may differ. Given that IllustrisTNG has fewer halos, we randomly choose an equal number of halos from \textit{Magneticum}. Specifically, we select 301, 284, 260, 228, and 176 halos corresponding to the redshifts 0.01, 0.25, 0.47, 0.78, and 0.96, respectively.

For each halo, we extracted three mis-centered cluster matter profiles (using the method described in \cite{Grandis21}) and average all of the clusters at a given redshift from both simulations. We follow the mis-centering distribution model as described in Sect. \ref{sec:mis-centering}. Because mis-centering depends on the richness of the cluster, we assign each halo a richness value using the richness-mass relation (Eq.~\ref{eq:lambdamass}) using the parameters obtained in a previous cluster cosmology analysis \citep{Chiu23}. 
These individual $\Delta\Sigma(R)$ profiles are then rescaled into $\widetilde{\Delta\Sigma}\left(R/R_\mathrm{200c}\right)$ profiles and averaged following Eq.~\ref{eq:sims_averagescaledDeltaSigma}.

Given that the two sets of simulations do not have outputs at exactly the same redshifts (e.g., 0.42 versus 0.47 and 0.64 versus 0.78), we quantified the differences in the average $\widetilde{\Delta\Sigma}\left(R/R_\mathrm{200c}\right)$ profiles at these redshifts, verifying that this induces a negligible uncertainty. This is achieved by interpolating the profiles from both simulations as a function of redshift separately and comparing for each simulation the differences in the profiles at both the redshifts, and finding them to be very small (percentage error of $\approx$ 0.4 \%, see Fig.~\ref{fig:Mag_TNG_redshift_differences}). We therefore adopted the mean value of the redshift (in case the redshifts are different) while combining the profiles from both simulations. The combined profiles are then interpolated as a function of redshift to capture the slight differences we see with redshift in the average $\widetilde{\Delta\Sigma}\left(R/R_\mathrm{200c}\right)$ profiles.

Once an average rescaled matter profile model has been chosen, it is used to characterize the bias and scatter in the $M_\mathrm{WL}$ estimates with respect to the true underlying halo masses, determining posteriors of the parameters in Eqs.~\ref{eq:WL_mass_rel} and \ref{eq:WL_mass_var}.  As part of this calibration process, uncertainties on the other crucial systematics (uncorrelated large-scale structure covariance, cluster mis-centering, cluster member contamination of the source galaxy sample, and hydrodynamical uncertainties on the model) are also included. We note that our model uses Eq.~\ref{eq:gamma_delta_sigma} to calculate $\Delta\Sigma$, while we use Eq.~\ref{eq:gt_delta_sigma} to calculate the observed profiles. This introduces a ~1\% bias in our modeling; however, given that our analysis is dominated by statistical uncertainties, this has no significant impact on our conclusions.

\subsection{Mass calibration likelihood}
\label{sec:likelihood_model}
In this section we present the mass calibration likelihood that we employ with the average rescaled matter profile $\widetilde{\Delta\Sigma}\left(R/R_\mathrm{200c}\right)$.  The lowest level observational constraint from weak gravitational lensing is a tangential reduced shear profile (Eq.~\ref{eq:red_shear}) constructed for each of a series of tomographic bins within which the shear galaxy sample is organized.  A complication with using the tangential shear profiles \citep[see, e.g.,][]{BocquetI2024PhRvD.110h3509B}, is that the profiles from the different bins have amplitudes that depend on $\Sigma_\mathrm{crit}$, which in turn depends on the redshift distributions of the background galaxies (Eq.~\ref{eq:sigma_crit}).  The matter profile $\Delta\Sigma(R)$ (Eq.~\ref{eq:gamma_delta_sigma}) is simpler in that the profiles for each tomographic bin are all estimators of the same underlying projected matter density of the cluster \citep[see, e.g.,][]{McClintock_2018}.  The observable we adopt here $\widetilde{\Delta\Sigma}\left(R/R_\mathrm{200c}\right)$ (Eq.~\ref{eq:scaledDeltaSigma}) offers additional simplicity, because this profile is approximately the same for all clusters, independent of their mass and redshift.  
However, the matter profile $\Delta\Sigma(R)$ and rescaled matter profile $\widetilde{\Delta\Sigma}\left(R/R_\mathrm{200c}\right)$ are no longer pure observables.  They both have dependences on cosmological parameters that impact the distance-redshift relation, and the rescaled matter profile also has dependences on the masses and redshifts of the constituent clusters. This dependence has to be considered within the likelihood, as outlined in the next subsection.

\subsubsection{Likelihood of the rescaled matter profile}
The lensing likelihood for an average rescaled matter profile is given by a product of the independent Gaussian probabilities of obtaining the observed matter profile given the model within each radial bin. 
Because the rescaled matter profile $\widetilde{\Delta\Sigma}\left(R/R_\mathrm{200c}\right)$ depends on cosmological parameters and the masses and radii of the constituent clusters, the likelihood has to be altered to account for these dependencies.
The likelihood transformation for a data vector $t_{\theta}$ which is some function of a data vector $y$ (independent of model parameters) and model parameters $\theta$ is given by \citep{0da8c2ff-3f64-3424-9f60-e5cf9a7c5cb1}
\begin{equation}
    P(t_{\theta}|\theta)d\theta \longrightarrow P(t_{\theta}|\theta) \bigg| \frac{\partial t_{\theta}}{\partial \mu} \bigg| \mathrm{d} \mu.
\end{equation}
Here, $\mu$ denotes a function of data $y$ that has the same dimension as $t_{\theta}$ and is independent of the model parameters $\theta$. In the case where $\mu$ cannot be expressed with the same dimension as $t_{\theta}$, the likelihood transformation is then given by 
\begin{equation}
    P(t_{\theta}|\theta)d\theta \longrightarrow P(t_{\theta}|\theta) \bigg| \frac{\partial t_{\theta}}{\partial \mu} \bigg(\frac{\partial t_{\theta}}{\partial \mu} \bigg)^{\mathrm{T}} \bigg|^{1/2} \mathrm{d} \mu.
    \label{eq:LT_dim}
\end{equation}
Using Eq.~\ref{eq:LT_dim}, we can write the transformed Gaussian likelihood (lensing likelihood) for a rescaled matter profile $\widetilde{\Delta\Sigma}\left(\vec{M_\mathrm{200c}}, \vec p\right)$ (given by Eq.~\ref{eq:obs_rescaledDeltaSigma}) with $j$ radial bins as
\begin{equation}
    P(\widetilde{\Delta\Sigma}\left(\vec{M_\mathrm{200c}},\vec{z}, \vec p\right)|\widetilde{\Delta\Sigma}_{\mathrm{mod}},\vec z) =  \prod_j P_{\mathrm{G},j}\,
    \bigg| \frac{\partial \widetilde{\Delta\Sigma}_j}{\partial \vec e_{t,j}} \bigg(\frac{\partial \widetilde{\Delta\Sigma}_j}{\partial \vec e_{t,j}} \bigg)^{\mathrm{T}} \bigg|^{1/2},
\end{equation}
where $\widetilde{\Delta\Sigma}_\mathrm{mod}$ is the model profile and the second factor in the above equation is the transformation calculation for all the radial bins, $\vec e_{t,j} = [ e^1_{t},e^2_{t},e^3_{t}...e^n_{t}]$ is the vector containing the $n$ source ellipticities in a given radial bin $j$ and $P_{\mathrm{G},j}$ is the Gaussian likelihood for that bin
\begin{equation}
\begin{split}
    P_{\mathrm{G},j} =
    \bigg(\sqrt{2\pi}\sigma_{\widetilde{\Delta\Sigma},j} \bigg)^{-1} \exp{\Bigg [-\frac{1}{2} \bigg( \frac{\widetilde{\Delta\Sigma}_j-\widetilde{\Delta\Sigma}_{\mathrm{mod,j}}}{\sigma_{\widetilde{\Delta\Sigma},j}} \bigg)^2 \Bigg]},
\end{split}
\end{equation}
where $\sigma_{\widetilde{\Delta\Sigma},j}$ is the rescaled shape noise as described in Eq.~\ref{eq:obs_rescaledDeltaSigma_uncertainty}.
The likelihood transformation is a one-dimensional partial derivative matrix with length $n$ and is given by
 
\begin{equation}
    \frac{\partial \widetilde{\Delta\Sigma}_j}{\partial \vec e_{t,j}}  = \Bigg[ \frac{\partial \widetilde{\Delta\Sigma}_j}{\partial e^1_{t,j}},  \frac{\partial \widetilde{\Delta\Sigma}_j}{\partial e^2_{t,j}},..., \frac{\partial \widetilde{\Delta\Sigma}_j}{\partial e^n_{t,j}} \Bigg]. 
\end{equation}
The transformation factor can then be expressed as
\begin{eqnarray}
     \bigg| \frac{\partial \widetilde{\Delta\Sigma}_j}{\partial \vec e_{t,j}} \bigg(\frac{\partial \widetilde{\Delta\Sigma}_j}{\partial \vec e_{t,j}} \bigg)^{T} \bigg|^{1/2} = & \Bigg( \left(\frac{\partial \widetilde{\Delta\Sigma}_j}{\partial e^1_{t,j}}\right)^2 +  \left(\frac{\partial \widetilde{\Delta\Sigma}_j}{\partial e^2_{t,j}}\right)^2 +... \nonumber\\ 
     &
     + \left(\frac{\partial \widetilde{\Delta\Sigma}_j}{\partial e^n_{t,j}}\right)^2 \Bigg)^{1/2}, 
\end{eqnarray}
where we just compute the derivative of Eq.~\ref{eq:obs_rescaledDeltaSigma} with respect to the measured ellipticities.

\subsubsection{Likelihood of single cluster rescaled matter profile}

The intrinsic scatter in the observable mass relations, the measurement noise on the observables, and the posteriors of the observable-mass relation parameters and the cosmological parameters all contribute to create the posterior mass distribution $P(M|\hzeta, \hlambda, z, \vec p)$ for a given cluster.  Even in the limit of perfect knowledge of the observable-mass relation and cosmological parameters, this posterior distribution has some characteristic width determined by the mass trends in each observable together with the sources of scatter mentioned above.  Moreover, even in the case of a perfect match between the observed and model rescaled matter profiles, the resulting WL mass estimate is a biased and scattered estimator of the true halo mass $M_\mathrm{200c}$ as described by the $M_\mathrm{WL}$-halo mass relations (Eq.~\ref{eq:WL_mass_rel} and \ref{eq:WL_mass_var}).

The cluster mass uncertainty represented by this mass posterior and any biases and scatter in the WL mass estimate have to be accounted for. Therefore, the single cluster lensing likelihood for cluster $k$ with the WL mass posterior $P_k(M_\mathrm{WL}|\hzeta, \hlambda, z, \vec p)$ and rescaled matter profile $\widetilde{\Delta\Sigma}_k$ is written as
\begin{equation}
    \begin{split}
    P(\widetilde{\Delta\Sigma}_k|\widetilde{\Delta\Sigma}_{\mathrm{mod}},z) = \int \diff M_\mathrm{WL} & P_k(M_\mathrm{WL}|\hzeta, \hlambda, z, \vec p) \\  
    & P(\widetilde{\Delta\Sigma}_k(M_\mathrm{WL})|\widetilde{\Delta\Sigma}_{\mathrm{mod}},z),\\
    \end{split}
\label{eq:lensing_average_likelihood}     
\end{equation}
where we are explicit with subscript $k$ to emphasize that this expression represents a weighted likelihood for a single cluster.  We note that the $M_\mathrm{WL}$ dependence  of $\widetilde{\Delta\Sigma}_k$ is due to the cluster radius (Eq.~\ref{eq:scaledDeltaSigma}), which is mass dependent as in Eq.~\ref{eq:mass_definition}.

We calculate the mass posterior using three observables $\hat\zeta$, $\hat\lambda$ and $z$ (we neglect the cluster photometric redshift uncertainty because it is too small relative to other sources of scatter to be important). According to Bayes' theorem, the expression for the mass posterior of a cluster that accounts for intrinsic and measurement scatter in the observables is
\begin{equation}
\begin{split}
     & P(M_\mathrm{WL}|\hzeta, \hlambda, z, \vec p) = \\
     & \ \   
     \frac{\iiint \diff M \diff \lambda \diff \zeta P(\hlambda|\lambda) P(\hzeta|\zeta) P(\zeta, \lambda, M_\mathrm{WL}|M, z, \vec p) P(M|z, \vec p)}
     {P(\hlambda,\hat\zeta|z, \vec p)},
\end{split}
\label{eq: mass_cal},     
\end{equation}
where the measurement noise is represented by $P(\hlambda|\lambda)$ and $P(\hzeta|\zeta)$, the intrinsic scatter and any bias in the observable about mass by $P(\zeta, \lambda, M_\mathrm{WL}|M, z, \vec p)$,
$P(M|z, \vec p)$ is the halo mass function factor which allows us to account for Eddington bias due to the selection, and $P(\hlambda,\hat\zeta|z, \vec p)$ is just the numerator integrated over $M_\mathrm{WL}$.

In the context of multiple observables, the single cluster mass calibration likelihood $\mathcal{L_{\mathrm{single}}}$ can be written (assuming $\widetilde{\Delta\Sigma}$ is uncorrelated with other observables) as
a product of the single cluster lensing likelihood and the likelihood of the observables

\begin{equation}
    \mathcal{L_{\mathrm{single}}} = P(\widetilde{\Delta\Sigma}_k|\widetilde{\Delta\Sigma}_{\mathrm{mod}},z)  P(\hlambda|\hzeta, z, p),
    \label{eq:single_cluster_likelihood}
\end{equation}
where the second component is the likelihood of the observed richness $\hlambda$ given the observed tSZE detection significance $\hzeta$ and redshift $z$. It can be calculated using Bayes' theorem accounting for intrinsic scatter in the observables
\begin{equation}
\begin{split}
   & P(\hlambda|\hzeta, z, \vec p) = \\
   & \ \ \ \ \ \ \ 
   \frac{\iiint \diff M \diff \lambda \diff \zeta  P(\hlambda|\lambda) P(\hzeta|\zeta) P(\zeta, \lambda|M, z, \vec p) P(M|z, \vec p) }{P(\hzeta|z,\vec p)},
    \end{split}
\label{eq:lambda_given_xi}    
\end{equation}
where $P(\hzeta|z,\vec p)$ is just the normalization that comes from integrating the numerator over all $\hat\lambda$, including importantly the $\hat\lambda_\mathrm{min}(z)$ selection threshold, which is crucial for accounting for Malmquist bias. Additionally, in this study, we assume that there is no correlated scatter between $\zeta$ and $\lambda$, so $P(\zeta, \lambda|M, z, \vec p)$ can be further simplified as $P(\zeta|M, z, \vec p) P(\lambda|M, z, \vec p)$.

\subsubsection{Likelihood of multi-cluster average rescaled matter profile }
\label{sec:stacked_likelihood}
For a given $\hzeta-\hlambda-z$ bin containing $n$ clusters, the rescaled matter profile $\widetilde{\Delta\Sigma}$ is notionally calculated as in Eq.~\ref{eq:obs_rescaledDeltaSigma}.  However, as discussed above for the single cluster rescaled matter profile, the mass posteriors of the clusters must be included.  Rather than extracting an average likelihood by marginalizing over the WL mass posterior $P(M_\mathrm{WL}|\hzeta, \hlambda, z, \vec p)$ as in the single cluster case (Eq.~\ref{eq:lensing_average_likelihood}), in the multi-cluster case we adopt a Monte Carlo integration approach that allows us to efficiently marginalize over the mass posteriors of all $n$ clusters simultaneously.  In effect, we rebuild the average matter profile $\widetilde{\Delta\Sigma}$ for the cluster ensemble many times and use those profiles to extract likelihoods and then estimate the average likelihood of the rescaled matter profile.

Following the likelihood for a single cluster matter profile in Eq.~\ref{eq:single_cluster_likelihood}, we write the WL mass calibration likelihood for an ensemble of $n$ clusters with associated observables $\hat\lambda_i$, $\hat\zeta_i$ and $z_i$ as
\begin{equation}
    \begin{split}
    \mathcal{L_{\mathrm{bin}}} 
    = \langle P\left(\widetilde{\Delta\Sigma}(\vec{\hzeta,\hlambda,z}, \vec p)| \widetilde{\Delta\Sigma}_{\mathrm{mod}}, \vec z\right) \rangle \prod_{i=1}^n P(\hlambda_i|\hzeta_i, z_i, \vec p), 
    \end{split}
    \label{eq:stacked_like}
\end{equation}
where $\langle P\left(\widetilde{\Delta\Sigma}(\vec{\hzeta,\hlambda,z}, \vec p)| \widetilde{\Delta\Sigma}_{\mathrm{mod}}, \vec z\right) \rangle$ is the average lensing likelihood of the average rescaled matter profile built from the ensemble.  The observable vectors $\vec{\hzeta}$, $\vec{\hlambda}$, and $\vec z$ each contain the measurements for the $n$ clusters in the ensemble.
For an $n$ cluster ensemble, it takes the form
\begin{equation}
    \begin{split}
   \langle P&\left(\widetilde{\Delta\Sigma}\left(\vec{\hzeta,\hlambda,z}, \vec p\right)| \widetilde{\Delta\Sigma}_{\mathrm{mod}}, \vec z\right) \rangle =\\
    &\idotsint \diff M_{\mathrm{WL}_1}...\diff M_{\mathrm{WL}_n} P(M_{\mathrm{WL}_1}|\hzeta_1, \hlambda_1, z_1, \vec p)\times...\\
    &\ \ \ \ \ \ \ \ \ \ \ \ \ \ \ \ P(M_{\mathrm{WL}_n}|\hzeta_n, \hlambda_n, z_n, \vec p) \\
    &\ \ \ \ \ \ \ \ \ \ \ \ \ \ \ \ P\left(\widetilde{\Delta \Sigma}(M_{\mathrm{WL}_1},...,M_{\mathrm{WL}_n},\vec p)|\widetilde{\Delta\Sigma}_{\mathrm{mod}}, \vec z\right),\\
    \end{split}
\end{equation}
where we note that the ${M_{\mathrm{WL}}}$ is needed to build the rescaled matter profile (using Eq.~\ref{eq:obs_rescaledDeltaSigma} with ${M_{\mathrm{WL}}}$ instead of ${M_{\mathrm{200c}}}$) and is calculated from observables ($\hat\lambda$, $\hat\zeta$ and $z$) and the $M_{\mathrm{WL}}$-mass relation using Eq.~\ref{eq:WL_mass_rel}.

The final likelihood for $m$ $\hzeta-\hlambda-z$ bins can be written as the product of the likelihood of individual bins
\begin{equation}
    \mathcal{L} = \prod_{\mathrm{bin}=1}^m\mathcal{L_{\mathrm{bin}}}.
\end{equation}

\subsection{Modeling and correcting for systematic effects}
\label{sec:systematics}
\subsubsection{Cluster mis-centering distribution}
\label{sec:mis-centering}
For each of the clusters in our sample, we have two measurements of the cluster center.  The first is the tSZE center as measured by the SPT and the second is the optical center extracted using the MCMF algorithm. MCMF adopts the BCG as the center if it is within 250~kpc of the cluster position determined by SPT; otherwise, the position of the peak of the galaxy density map is used. We only make use of MCMF centers for our analysis. As the observationally determined center is not a perfect tracer of the true halo center, the effect of this mis-centering must be taken into account when modeling the cluster matter profile. We adopt the mis-centering model and the parameters from the recent work by \cite{BocquetI2024PhRvD.110h3509B}.
The mis-centering distribution for the tSZE and optical centers is modeled using the double Rayleigh distribution.
\begin{equation}
    \begin{split}
    P_{\mathrm{offset}}(r) &= \rho \text{Rayl}(r,\sigma_0) + (1-\rho) \text{Rayl}(r,\sigma_1),\\
    \sigma_i &= \sigma_{i,0} \left(\frac{\lambda}{60}\right)^{1/3} \text{for } i\in \{0,1\}.
    \label{eq:mis_dist}
    \end{split}
\end{equation}
The double Rayleigh distribution is a good description of the mis-centering of the optical center with respect to the true halo center. The constraints on the mis-centering parameters ($\rho, \sigma_0, \sigma_1$) are obtained by simultaneously fitting for SPT and optical centers. A large fraction of clusters ($\rho \approx 0.89$) are well centered, and the two scatter parameters are $\sigma_0 \approx 0.007~h^{-1} \mathrm{Mpc}$ and $\sigma_1 \approx 0.18~h^{-1} \mathrm{Mpc}$ \citep[for additional details, see][]{BocquetI2024PhRvD.110h3509B}.  

Crucial for our mass calibration analysis is to include the effects of uncertainties on the mis-centering distribution.  We do this as part of the $M_\mathrm{WL}$-mass relation calibration. However, given the radial range we adopt for mass calibration ($R>500~h^{-1}$~kpc), the miscentering itself and the uncertainties on the mis-centering have little impact on our results.

\subsubsection{Cluster member contamination}
\label{sec:cmc}
Cluster galaxies are generally included in the WL source galaxy sample in the case where the cluster redshift lies within the source galaxy redshift distribution associated with a tomographic bin (see Fig~\ref{fig:source_distribution}). These cluster galaxies are not sheared by their host cluster halo, and therefore their inclusion biases the lensing signal toward being low. To correct for this shear bias, we follow the methodology described in \cite{paulus_thesis}, developed for the DES Y1 WL dataset, and in \cite{BocquetI2024PhRvD.110h3509B}, where the method was extended for application to the DES Y3 WL tomographic bin based dataset. The method follows notionally the work by \cite{Varga+19} but more explicitly accounts for varying cluster redshift $z_\mathrm{cl}$. Here we extend this method again by modeling the contamination within each tomographic bin separately. 

In our analysis, the fractional contamination by cluster members $f_{\mathrm{cl},b}$ in tomographic source bin $b$ is extracted by modeling the probability density function of galaxies in redshift along the line of sight toward the cluster as the weighted sum of the cluster member distribution $P_{\mathrm{cl},b}(z)$ and the average source galaxy field distribution in tomographic bin $b$ as
\begin{equation}
    P_b(z)=f_{\mathrm{cl},b}P_{\mathrm{cl},b}(z)+(1-f_{\mathrm{cl},b})P_{\mathrm{field},b}(z),
    \label{eq:boost_model1}
\end{equation}
where $P_{\mathrm{field},b}(z)$ corresponds to the field component, which has been divided into three tomographic bins $b\in{2, 3, 4}$. For this analysis the individual source galaxy DNZ photo-z's \citep{devicente16} are employed.

To determine both the cluster and field components as described, the source galaxies from the DES~Y3 shear catalog associated with each cluster are divided into nine logarithmically spaced bins in projected radius, ranging from 0.7$h^{-1}$\,Mpc to 10$h^{-1}$\,Mpc from each cluster center. Because the projected cluster galaxy population falls off rapidly with radial distance from the cluster, the outermost two bins are dominated by the field source distribution.
As previously reported in \cite{paulus_thesis}, the depth inhomogeneities and masking variations in the DES WL source galaxy catalog lead to a field component surface density associated with each tomographic bin being relatively homogeneous on the scale of a cluster but varying significantly over the survey.  Thus, we used a local measure of the field surface density around each cluster when modeling the contamination.

The redshift distribution of the cluster component $P_{\mathrm{cl},b}(z)$ is modeled as a Gaussian distribution in the space of $z-z_\mathrm{cl}$ with a redshift dependent characteristic width of $\sigma_{\mathrm{z},b}(z)$ and redshift offset parameter $z_\mathrm{{off},b}(z)$ as  
\begin{equation}
P_{\mathrm{cl},b}(z)={1\over{\sqrt{2\pi}\sigma_{\mathrm{z},b}(z)}} 
e^{-\frac{(z-z_\mathrm{cl}-z_{\mathrm{off},b}(z))^2}{2\sigma_{\mathrm{z},b}^2(z)}}.
\label{eq:boostredshift}
\end{equation}
The characteristic width of the cluster members in redshift reflects the typical photo-z uncertainties, and the offset in redshift away from the cluster redshift can result from, for example, the selection applied in dividing source galaxies into tomographic bins.

The spatial distribution of the cluster component is modeled as a projected NFW profile $f_\mathrm{NFW}(R,c_\lambda)$ whose amplitude varies with observed cluster richness as $\hat\lambda^\mathrm{B_\lambda}$ and whose redshift variation is extracted directly through measurements within independent redshift bins \citep{paulus_thesis}.
For convenience, this projected NFW spatial model is normalized to a value of 1.0 at a projected radius of $1\,h^{-1}$\,Mpc, and
the NFW concentration $10^{c_{\lambda}}$ is modeled as a function of richness as 
\begin{equation}
    r_s = \frac{(\hat\lambda/60)^{1/3}}{10^{c_{\lambda}}} h^{-1}\mathrm{Mpc}.
\end{equation}

This approach allows adequate freedom to describe the galaxy populations around tSZE selected clusters over our redshift range of interest \citep{Hennig2017MNRAS.467.4015H}.  We note that we measure the contamination around the same cluster centers used for the matter profile analysis, and so mis-centering effects are automatically included.
In our analysis, we exclude the core region of the cluster where blending is more common.
Moreover, we apply no correction for WL magnification bias \citep[e.g.,][]{Chiu2016MNRAS.457.3050C}, which is strongest in the excluded cluster core region.

The fractional cluster member contamination extracted is used to apply a radially dependent correction $1/(1-f_{\mathrm{cl},b}(R))$ to the amplitude of each cluster matter profile $\Delta\Sigma(R)$ derived from each tomographic bin $b$ before these matter profiles are averaged.  It can be expressed as

\begin{eqnarray}
1\over{1-f_{\mathrm{cl},b}(R)} &=& 1+
e^{A_{\mathrm{eff},b}(z_\mathrm{cl})} 
\left({\hat\lambda}\over {60}\right)^{B_{\lambda,b}} f_\mathrm{NFW}(R,c_{\lambda,b}),
\label{eq:clustercontamination}
\end{eqnarray}
where 
\begin{equation}
    A_{\mathrm{eff},b}(z_\mathrm{cl}) = A_{\infty} + \sum_i A_i e^{ -\frac{1}{2} \frac{(z_\mathrm{cl}-z_i)^2}{\rho^2_\mathrm{corr} }},
\end{equation} 
is a normalization factor that is dependent on the cluster redshift and the extracted amplitudes $A_i$ that are extracted within redshift bins $z_i$, where the redshift centers of the bins are
$
    z_i\in \{0.2, 0.28, 0.36, 0.44, 0.52, 0.6 , 0.68, 0.76, 0.84, 0.92, 1.0\}
$. 
In this expression we have integrated over the cluster member redshift distribution.
For each tomographic bin $b$, the parameters $A_{\infty}$, $A_i$, $\rho_{\mathrm{corr}}$, $B_{\lambda,b}$, $c_{\lambda,b}$, $z_{\mathrm{off},b}(z)$ and $\sigma_{\mathrm{z},b}(z)$ are obtained by fitting to the entire cluster population.
The redshift dependence of the redshift width and offset in Eq.~\ref{eq:boostredshift} are assumed to be simple linear functions in redshift around a pivot redshift $z=0.5$ as $\sigma_\mathrm{z}(z)= \sigma_\mathrm{z_0}+\sigma_\mathrm{z_z}(z-0.5)$ and $z_\mathrm{off}(z)=z_\mathrm{off_0}+z_\mathrm{off_z}(z-0.5)$, respectively (again, within each tomographic bin $b$).

To solve for the parameters of the cluster member contamination model, we iterate over all clusters in the sample comparing our model (Eq.~\ref{eq:boost_model1}) to the observed surface density of source galaxies as a function of redshift and projected separation from the cluster.  Following \cite{paulus_thesis}, we apply a regularization term to the likelihood with a correlation length in redshift $\rho_\mathrm{corr-z}$ to the pairs of neighboring amplitudes $e^{A_{i,b}}$, which then prefers a solution with smooth variation in redshift \citep[as noted explicitly in Eqns. 24 through 27 in][]{BocquetI2024PhRvD.110h3509B}.  With larger cluster samples these regularization terms would no longer be important. The parameter constraints for each tomographic bin are given in Table \ref{tab:clmemcont}.
Because the method applied here is similar to that in \cite{BocquetI2024PhRvD.110h3509B}, we direct the reader to Fig. 9 in \cite{BocquetI2024PhRvD.110h3509B} for validation tests of the model.

\begin{figure}
\centering
\includegraphics[width=\columnwidth]{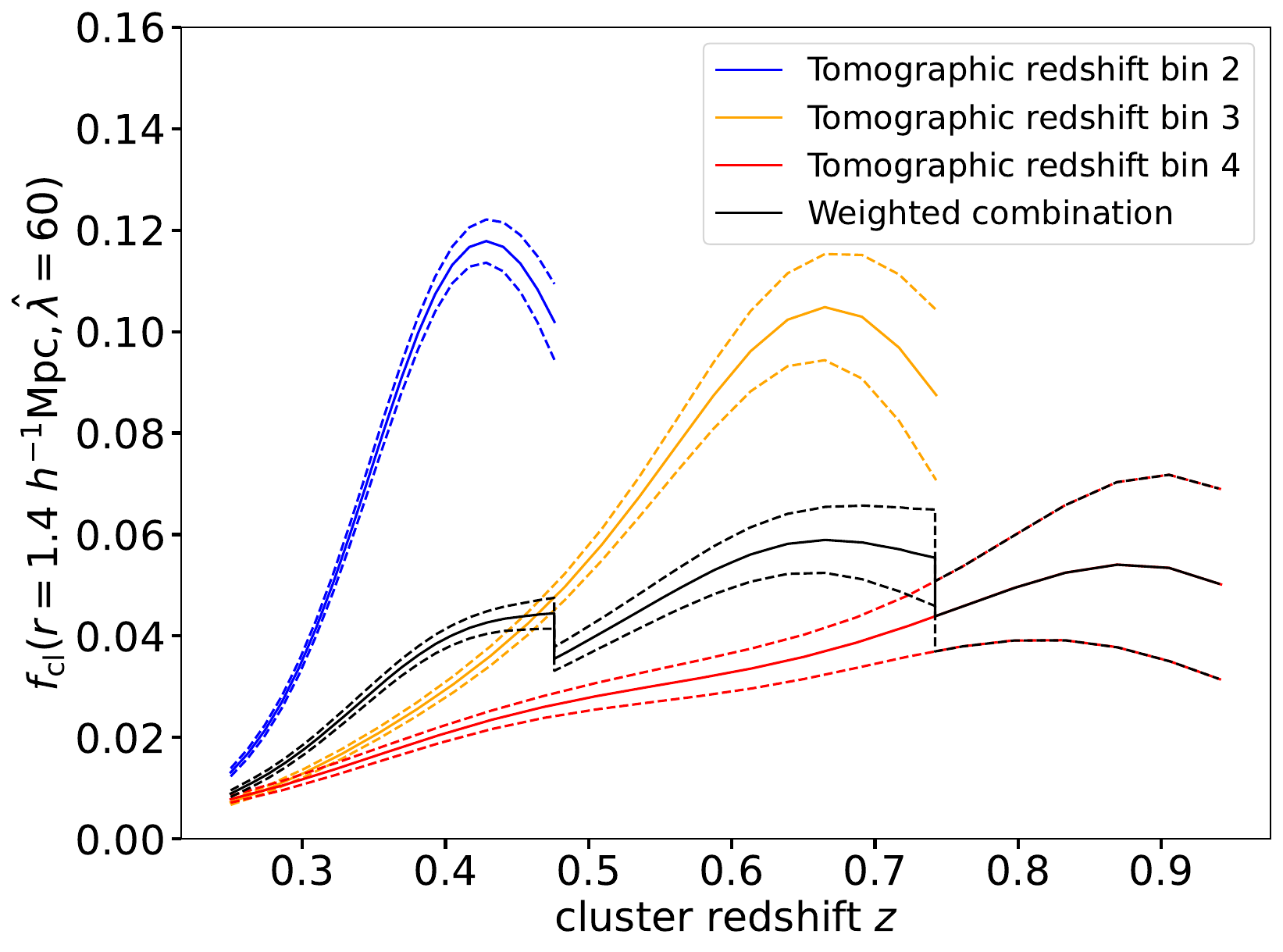}
  \vskip-0.10in
  \caption{Fractional cluster member contamination for a cluster with $\hat\lambda$=60 as a function of redshift for the three source galaxy tomographic bins (see Fig.~\ref{fig:source_distribution}) and their weighted combination. The solid (dashed) lines show the mean (68\% credible interval) of the contamination.  The colored lines extend over the redshift ranges for which each tomographic bin is employed in constructing the cluster matter profiles.  The 1.4$h^{-1}$\,Mpc radius is chosen because that is a characteristic radius from which the WL constraints are coming, given the adopted radial fitting range and the increasing number of source galaxies with radius.}
  \label{fig:f_cl_z_tomo}
\end{figure}

Figure~\ref{fig:f_cl_z_tomo} shows the fractional cluster member contamination at a projected radius of 1.4$h^{-1}$\,Mpc as a function of cluster redshift for the three source galaxy tomographic bins. The mean model is shown in color coded solid lines corresponding to the model parameters as listed in Table~\ref{tab:clmemcont} over the redshift range for which each tomographic bin is used. The dashed lines show the 68\% credible intervals. For a cluster at a given redshift, the total fractional contamination would be a weighted sum of the fractional contamination within each tomographic bin, where we apply the weight $\Sigma^{-1}_{\mathrm{crit}}$ appropriate for each bin.  This weighted contamination is shown as the black black line with associated dashed lines corresponding to the 68\% credible interval.  The radius 1.4$h^{-1}$\,Mpc is the characteristic radius for the cluster fitting when considering our radial fitting range and the increase of the number of source galaxies with radius. Thus, the typical contamination for a $\hlambda$=60 cluster varies from $\sim$1\% at the lowest cluster redshifts to $\sim$6\% at the highest, and the contamination varies as $\hlambda^{B_\lambda}$ where $B_\lambda$=0.78, 0.60 and 0.53 for tomographic bins 2, 3 and 4, respectively.

\begin{figure}
        \includegraphics[width=\columnwidth]{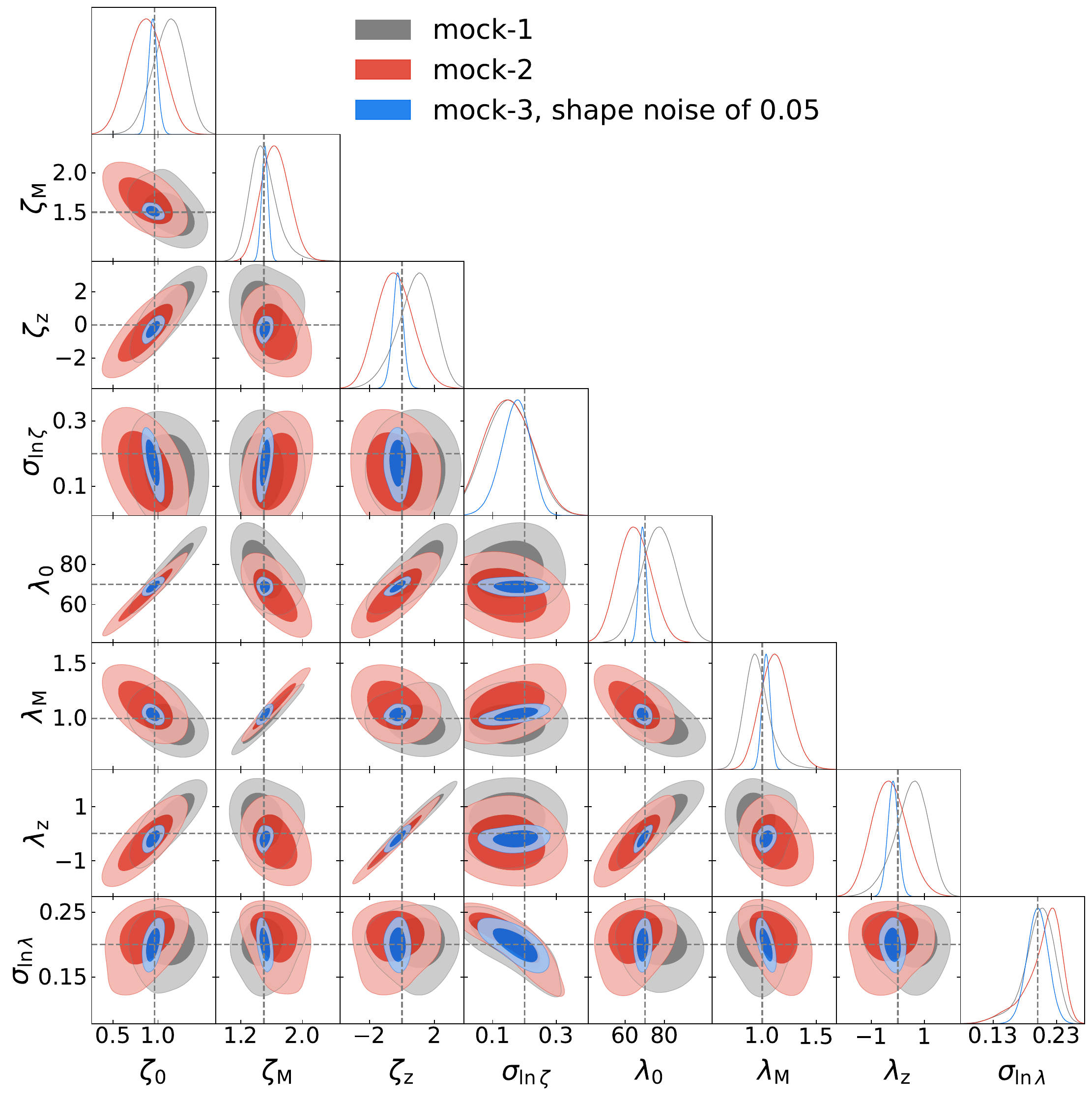}
        \vskip-0.10in
    \caption{Posterior constraints for average cluster mass calibration of three different mock SPT samples in a fiducial flat $\Lambda$CDM cosmology. The gray dashed line shows the input parameters used to generate the mock samples, which are recovered within the uncertainties. }
    \label{fig:mock_validation}
\end{figure}

\section{Results and discussion}
\label{sec:results}
In this section we first present validation tests of our new calibration method applied to mock data in Sect.~\ref{sec:mock_results}. In Sect.~\ref{sec:SPT_results} we show the resulting constraints from the real data, discuss the choices made in this analysis, such as which systematics are included, and present a validation of our adopted form for the observable-mass relations using the real data. We then present the average cluster matter profile out to larger radii (including the two-halo dominated region) and compare it with the simulation in Sect.~\ref{sec:outer_profile}.  Section~\ref{sec:comp_to_other_works} contains a comparison of our results with those reported in the recent literature.

\subsection{Validation of method using mock dataset}
\label{sec:mock_results}
We validate our analysis method using a realistic SPT+MCMF mock catalog with DES-like mock WL data. For this purpose, we create mock clusters and lensing data following the approach taken in the recent SPT$\times$DES analysis \citep{BocquetI2024PhRvD.110h3509B}. 

The first step in creating a mock SPT+MCMF cluster catalog within a fiducial cosmological model is to calculate the expected number of halos as a function of redshift and mass; for this we adopt the \cite{Tinker_2008} halo mass function scaled by the surveyed volume as a function of redshift,  imposing a mass range $10^{13} M_{\odot}< M_\mathrm{200c} < 10^{16} M_{\odot}$ and a redshift range 0.25 < $z$ < 0.94. We draw a Poisson realization of this sample, and then for each halo we assign cluster observables using the observable-mass relations presented in Sect.~\ref{sec:observable-mass}.  We then apply survey selection cuts in tSZE detection significance $\hat\zeta$ and optical richness $\hat\lambda$ consistent with those used to produce the real sample.  These realistic mocks follow the SPT and MCMF related survey depth geometry and produce a mock sample similar to the real dataset (see Fig.~\ref{fig:zeta_lambda_z}) that is fully consistent with the form of the observable-mass relations that we use to analyze the real dataset. 

To generate mock WL data for each selected cluster in our mock catalog, we first estimate the total source galaxies associated with the cluster by calculating its area on the sky (corresponding to the radial distance to cluster center $R=10h^{-1}$Mpc) and assuming a source galaxy density of 6\,arcmin$^{-2}$. We assume the same source redshift distribution as the DES~Y3 data for each tomographic bin for the mock source galaxies. For each source, we assign a weight by randomly drawing source weights from real DES data. We divide the total source galaxies equally among the three tomographic bins. We then assign a radial distance to each source galaxy by randomly drawing distance $\propto$ $R$. 
We then sample the amplitude of the rescaled matter profile $\widetilde{\Delta\Sigma}$ at the distance $R_i/R_{200\mathrm{c}}$ for each galaxy $i$, given the cluster radius $R_{200\mathrm{c}}$ and using our rescaled matter profile model (see Sect.~\ref{sec:hydro-model}).
We add tomographic bin-dependent cluster member contamination, consistent with our measurement using the DES data (see Sect.~\ref{sec:cmc}) and then convert the $\widetilde{\Delta\Sigma}$ amplitude for each galaxy to $g_t$ 
using $\left<\Sigma_{\mathrm{crit},b}\right>$ for the appropriate tomographic bin $b$.
We apply scatter to each $g_t$ measurement by drawing $g_t$ from a normal distribution with $\sigma_{\mathrm{eff}}=0.3$ (effective shape noise for DES data). This process produces realistically noisy and biased tangential shear data for each cluster. Specifically, these shear data include all the known systematic and stochastic effects needed to model cluster shear profiles in DES data.

\begin{table}
    \caption{ Observable-mass relation and cosmology parameter priors.
    }
    \label{tab:parameter_priors}
    \centering
    \begin{tabular}{ c c c }
    \hline
    \hline
    Parameter & Description & Prior \\
    \hline
    \multicolumn{3}{l}{tSZE  detection significance $\zeta$-mass relation}\\
    \hline
        $\azeta$ & amplitude & $\mathcal{U}(0.01, 1.5)$ \\
        $\bzeta$ & mass trend & $\mathcal{U}(0.5, 3)$ \\
        $\czeta$ & redshift trend & $\mathcal{U}(-5, 5)$ \\
        $\dzeta$ & intrinsic scatter & $\mathcal{U}(0.01, 0.5)$ \\
    \hline
    \multicolumn{3}{l}{Optical richness $\lambda$-mass relation}\\
    \hline
        $\alambda$ & amplitude & $\mathcal{U}(10, 70)$ \\
        $\blambda$ & mass trend & $\mathcal{U}(0.2, 2)$ \\
        $\clambda$ & redshift trend & $\mathcal{U}(-5, 5)$ \\
        $\dlambda$ & intrinsic scatter & $\mathcal{U}(0.01, 0.5)$ \\
    \hline
    \multicolumn{3}{l}{$M_{\mathrm{WL}}-M_{\mathrm{200c}}$ relation}\\
    \hline
      $\ln M_{\mathrm{WL}_{0}}$ & amplitude of bias & $0$ \\
      $\sigma_{\ln M_{\mathrm{WL}_{0,1}}}$ & error on amplitude & $\mathcal N(0, 1)$ \\
      $\sigma_{\ln M_{\mathrm{WL}_{0,2}}}$ & error on amplitude & $\mathcal N(0, 1)$ \\
      $\bWL$ & mass trend of bias & $\mathcal N(1.000, 0.006^2)$ \\
      $\ln\sigma^2_{\ln\mathrm{WL}_0}(z_0)$ & amplitude of scatter & $\mathcal N(-3.11, 0.04)$\\
      $\ln\sigma^2_{\ln\mathrm{WL}_0}(z_1)$ & amplitude of scatter & $\mathcal N(-3.07, 0.05)$\\
      $\ln\sigma^2_{\ln\mathrm{WL}_0}(z_2)$ & amplitude of scatter & $\mathcal N(-2.84, 0.06)$\\
      $\ln\sigma^2_{\ln\mathrm{WL}_0}(z_3)$ & amplitude of scatter & $\mathcal N(-1.94, 0.10)$\\
      $\bsigmaWL$ & mass trend of scatter & $\mathcal N(-0.23,0.04^2)$ \\
    \hline
    \multicolumn{3}{l}{Cosmology}\\
    \hline
        $\Omega_{\mathrm{m}}$ & matter density & $\mathcal{N}(0.315, 0.007)$ \\
        $\log_{10} A_s$ & amplitude of $P(k)$ & Fixed to -8.696 \\
        $H_0$ & Hubble parameter & Fixed to 70 \\
        $\Omega_{\textrm{b}, 0}$ & baryon density & Fixed to 0.0493 \\
        $n_s$ & scalar spectral index & Fixed to 0.96 \\
        $w_0$, $w_\mathrm{a}$ & EoS parameters & Fixed to -1, 0 \\
        $\sum m_\nu$ & sum of neutrino masses & Fixed to 0.06\\
        $\Omega_{\textrm{k}, 0}$ & curvature density & Fixed to 0 \\
    \hline
    \hline
    \end{tabular}
\end{table}

We create several statistically independent mock catalogs to assess the performance of our likelihood model and the software. For the analysis, we divide our data into $3\times3\times3$ $\hzeta-\hlambda-z$ observable bins, and the likelihood calculation is performed following the formalism outlined in Sect.~\ref{sec:likelihood_model}. Figure~\ref{fig:mock_validation} shows the posteriors for three mock catalogs with the same set of input observable-mass relation and cosmological parameters but with different uniform random deviate seeds. As expected, all the mock catalogs contain a number of clusters that is similar to the real SPT sample. The corresponding sets of lensing data are also generated with different random number seeds. Mock-1 and mock-2 are generated with a shape noise ($\sigma_\mathrm{eff}=0.3$), similar to DES~Y3, while mock-3 is created with a shape noise value of 0.05, which is equivalent to scaling up the lensing source galaxy density by a factor of 36, therefore providing a more stringent test of the software. 

To effectively and efficiently sample the high dimensional parameter space, we used the Markov Chain Monte Carlo algorithm $\textsc{MultiNest}$ \citep{Feroz_2009,Feroz_2019} for our likelihood analysis.
As is clear in Fig.~\ref{fig:mock_validation}, the posteriors are in good agreement with the input parameters (plotted as dashed lines).  These validation tests show no signs of biases.

Our $\widetilde{\Delta\Sigma}$ profile analyses of both mock and real datasets typically converge in a factor of five less time on similar computing resources than in the case of the cluster-by-cluster analysis.  Interestingly, the time required for a single iteration of the $\widetilde{\Delta\Sigma}$ likelihood is similar to that for the cluster-by-cluster analysis, but the number of iterations required for convergence is typically five times less.  This faster convergence seems to be due to the difference in SNR of the average profiles as compared to the individual cluster matter profiles, which influences the stability of the likelihood far away from the best fit parameter values.  In future analyses using the average profile method, we plan to present further efficiency improvements.  Ongoing testing indicates that these approximate methods reduce the time required for a single likelihood evaluation by more than an order of magnitude.

\subsection{SPT$\times$DES analysis}
\label{sec:SPT_results}
With the validation of the code, we move on to apply the analysis method outlined above to the SPT$\times$DES sample. Following \cite{Grandis21}, we restrict our analysis to the radial range $0.5<R/(h^{-1}\mathrm{Mpc})<3.2/(1+z_{\mathrm{cl}})$. This radial cut allows us to restrict the analysis to the one-halo region while simultaneously avoiding the central region of the cluster, which is most affected by cluster member contamination, mis-centering, and baryonic processes. Throughout the analysis, we use DNF redshifts \citep{devicente16}, which are used to calibrate cluster member contamination. The shape noise per tomographic 
bin for the DES~Y3 data is taken from \cite{Amon_2022}, which is in good agreement with our bootstrap error estimates
\[
\sigma_{\mathrm{eff},b} =
\begin{cases}
0.262 & \text{ $b=2$} \\
0.259 & \text{ $b=3$} \\
0.301 & \text{ $b=4$}. \\
\end{cases}
\]
\begin{figure}
        \includegraphics[width=\columnwidth]{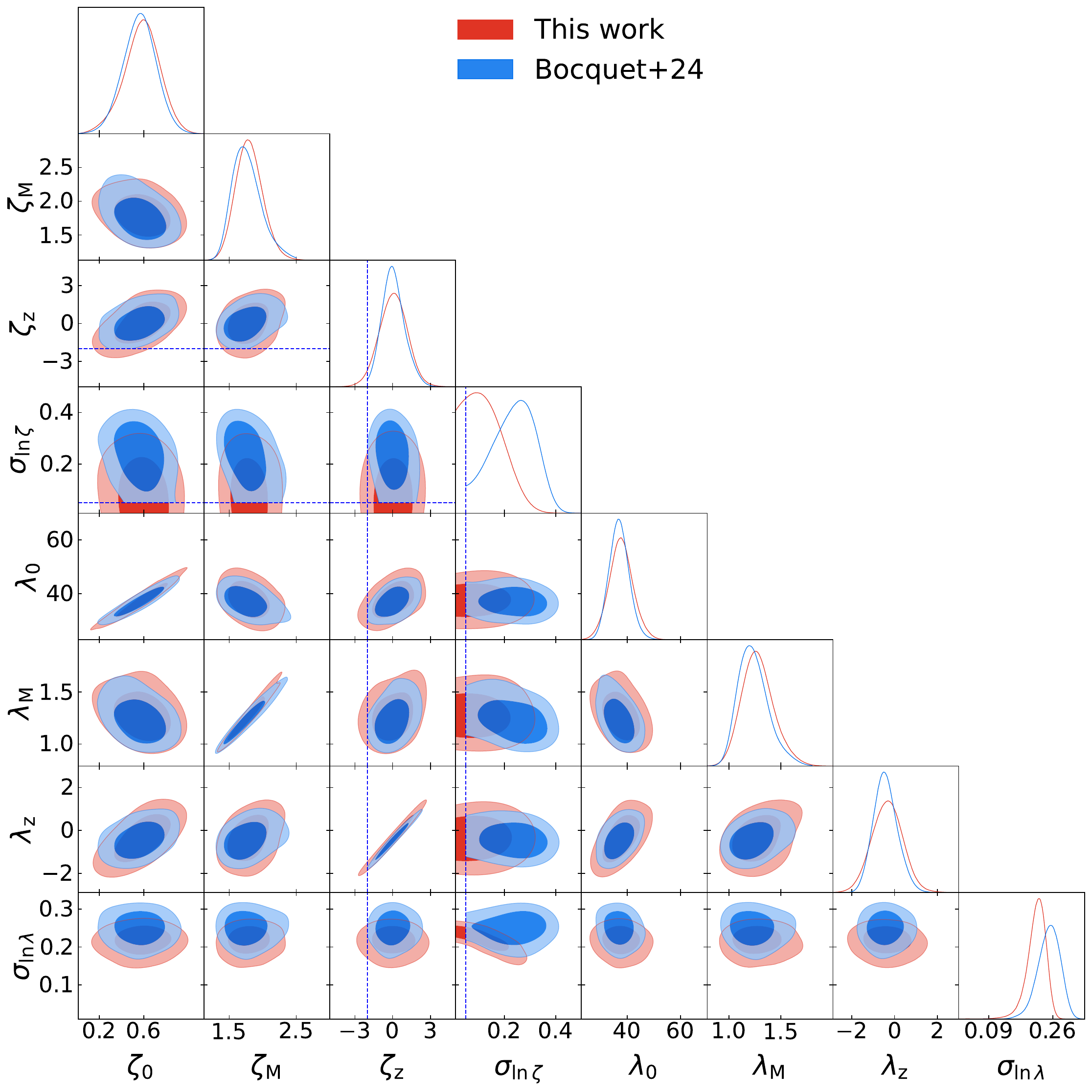}
        \vskip-0.10in
    \caption{ Contour plot showing the posterior constraints for cluster mass calibration using average matter profiles (in red) of the SPT cluster sample in a $\Lambda$CDM model. The blue contour shows cluster-by-cluster WL-only mass calibration results from \citet{Bocquet2024IIPhRvD.110h3510B} for the same SPT sample. The dashed blue line shows the prior boundary used in \citet{Bocquet2024IIPhRvD.110h3510B} for two parameters which is smaller than ours. }
    \label{fig:real_SPT_stacked}
\end{figure}
\begin{figure*}
        \includegraphics[width=\textwidth]{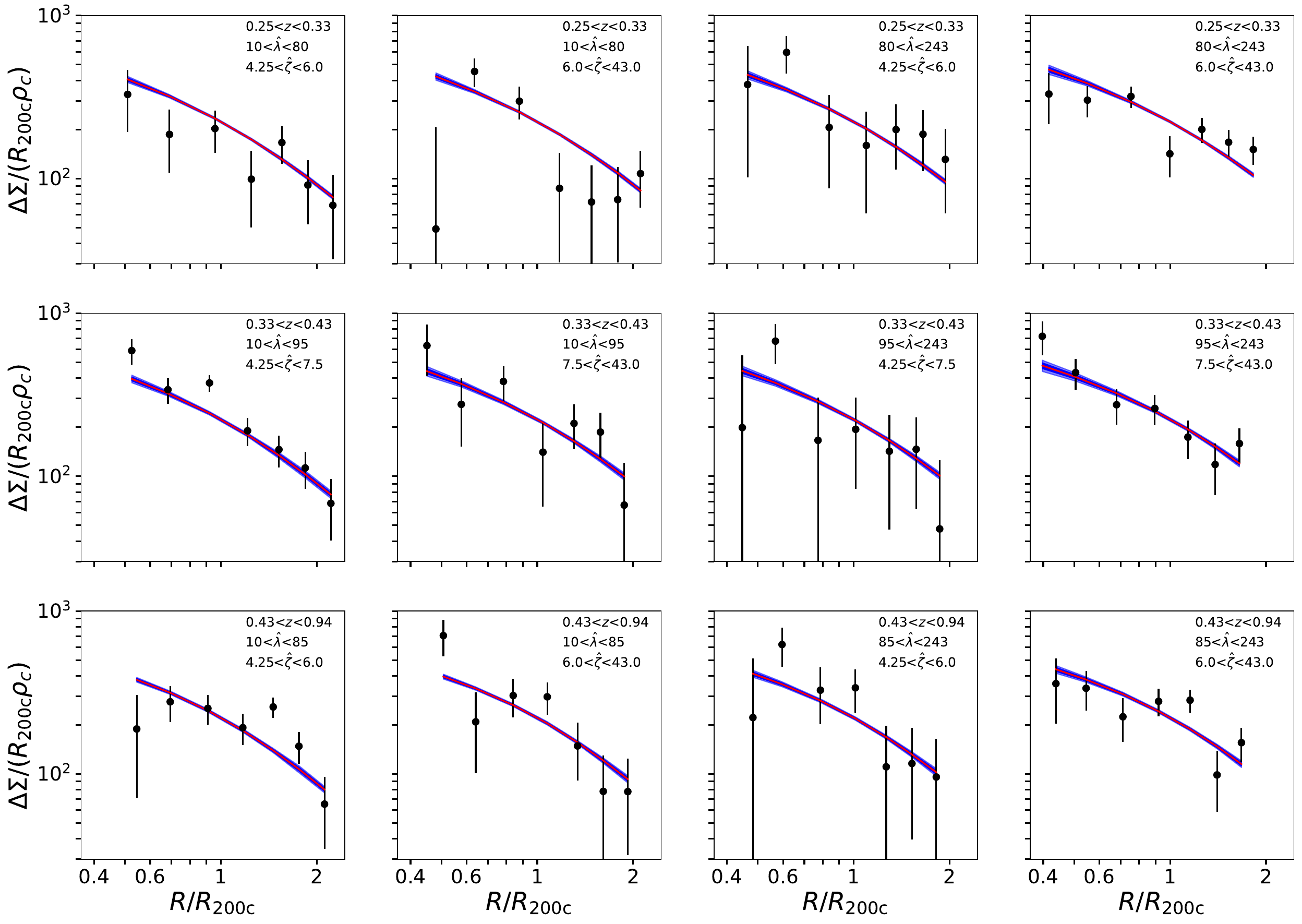}
        \vskip-0.10in
    \caption{Average SPT cluster matter profiles corresponding to the mean recovered parameters in twelve $\hzeta-\hlambda-z$ bins are shown with black data points with 1$\sigma$ error bars. The WL model is shown in the red line and the shaded blue region represents 2$\sigma$ error region on the model.}
    \label{fig:SPT_profile_validation}
\end{figure*}
We analyzed the SPT clusters in the redshift range $0.25\leq z<0.94$, which contains 698 clusters with DES WL data.  As with the mock validation, we divide the data into $3\times 3\times 3$ $\hzeta-\hlambda-z$ bins, leading to 27 independent average rescaled matter profiles $\widetilde{\Delta\Sigma}$. Because the WL signal in our sample has a higher SNR for lower redshift clusters, the highest redshift bin is chosen to be wider so that it has a sufficient SNR to approximately equal the SNR in the two lower redshift bins. Each redshift bin is further divided into 9 bins (3$\times$3 $\hzeta-\hlambda$).  Here, also, we choose bin boundary values such that each bin has a similar SNR. The observable bins for each redshift range are as follows
\begin{equation}
\begin{split}
   0.25\leq  z & < 0.33\nonumber  \\
    &0 \leq \hlambda < 60,\ \ 60 \leq \hlambda < 120,\ \ 120 \leq \hlambda < 250\nonumber \\
    &4.25 \leq \hzeta < 5,\ \  5 \leq \hzeta < 7,\ \  7 \leq \hzeta < 50\nonumber \\
    0.33\leq  z & < 0.43  \\
    &0 \leq \hlambda < 75,\ \  75 \leq \hlambda < 120, 120 \leq \hlambda < 250 \\
    &4.25 \leq \hzeta < 6.5,\ \  6.5 \leq \hzeta < 8.5,\ \  8.5 \leq \hzeta < 50 \\
    0.43\leq  z & < 0.94 \\
    &0 \leq \hlambda < 65,\ \  65 \leq \hlambda < 110,\ \  110 \leq \hlambda < 250 \\
    &4.25 \leq \hzeta < 5,\ \  5 \leq \hzeta < 7,\ \  7 \leq \hzeta < 50.
\end{split}
\end{equation}

Parameter priors for our run are listed in Table~\ref{tab:parameter_priors}.  In summary, we fixed the sum of neutrino masses to the minimum allowed value of 0.06\,eV. Nearly all other cosmological parameters were fixed to their mean Planck values \citep{Planck2020}, except for $\Omega_\mathrm{m}$, which has a Gaussian prior (our results are unaffected when using a wide flat prior on $\Omega_\mathrm{m}$). The observable-mass relation parameters are assigned a wide flat prior. Moreover, we assume no correlated scatter between $\zeta$ and $\lambda$. 

Rather than precisely following the form of the $M_{\mathrm{WL}} - M_{200\mathrm{c}}$ relation presented in Eqns~\ref{eq:WL_mass_rel} and \ref{eq:WL_mass_var}, our analysis follows the approach adopted in \cite{BocquetI2024PhRvD.110h3509B} where they analyze the same SPT$\times$DES sample.  We follow the redshift variation by interpolating between the relations determined through mock observations of specific simulation outputs at redshifts $z\in\{0.252, 0.470, 0.783, 0.963\}$. We set the amplitude and mass trend of the bias to 0 and 1, respectively, because we have adopted the simulation matter profiles as our model. 
We adopt the scatter parameter priors from the mock analysis of the simulations at four different redshifts, and we interpolate linearly to obtain the expectation at a given redshift. Additionally, we include the redshift dependent uncertainty on the amplitude of the bias with Gaussian random deviates whose values are scaled using the parameters $\sigma_{\ln M_{\mathrm{WL}_{0,1}}}$ and $\sigma_{\ln M_{\mathrm{WL}_{0,2}}}$.

\begin{table}
        \centering
        \caption{Mean parameter posteriors and 1-$\sigma$ uncertainties from our mass calibration analysis.}
        \label{tab:SPT_posterior}
        \begin{tabular}{l c} 
                \hline
            \hline
                Parameter & Posterior \\
                \hline
    \multicolumn{2}{l}{tSZE  detection significance $\zeta$-mass relation}\\
    \hline
                $\azeta$ &  0.586 $\pm$ 0.158 \\
                $\bzeta$ &  1.797 $\pm$ 0.195 \\ 
                $\czeta$ & 0.045 $\pm$ 1.054 \\
                $\dzeta$ & 0.127 $\pm$ 0.068 \\
    \hline
    \multicolumn{2}{l}{Optical richness $\lambda$-mass relation}\\
    \hline
                $\alambda$ &  37.69 $\pm$ 4.37 \\
                $\blambda$ &  1.275 $\pm$ 0.150 \\ 
                $\clambda$ & $-0.349$ $\pm$ 0.690 \\
                $\dlambda$ & 0.216 $\pm$ 0.025 \\   
    \hline
    \multicolumn{2}{l}{$M_{\mathrm{WL}}-M_{\mathrm{200c}}$ relation}\\
    \hline
                $\sigma_{\ln M_{\mathrm{WL}_{0,1}}}$ & $-0.35 \pm 0.80$ \\
                $\sigma_{\ln M_{\mathrm{WL}_{0,2}}}$ & $0.20 \pm 1.00$ \\
                $\bWL$ &  $1.000$ $\pm$ $0.004$ \\   
                $\ln\sigma^2_{\ln\mathrm{WL}_0}(z_0)$ & $-3.08 \pm 0.02$ \\
                $\ln\sigma^2_{\ln\mathrm{WL}_0}(z_1)$ & $-3.04 \pm 0.03$ \\
                $\ln\sigma^2_{\ln\mathrm{WL}_0}(z_2)$ & $-2.80 \pm 0.03$ \\
                $\ln\sigma^2_{\ln\mathrm{WL}_0}(z_3)$ & $-1.88 \pm 0.05$ \\
                $\bsigmaWL$ & $-0.22$ $\pm$ $0.04$\\
    \hline
    \multicolumn{2}{l}{Cosmology}\\
    \hline
                $\Omega_{\mathrm{m}}$ &  0.315 $\pm$ 0.006 \\
            \hline
                \hline
        \end{tabular}
\end{table}

The resulting posteriors inferred from applying the new mass calibration software to the SPT$\times$DES data are shown in Fig.~\ref{fig:real_SPT_stacked}. The mean posterior, along with the corresponding 68\% credible intervals, are listed in Table \ref{tab:SPT_posterior}. 
We find that the mass trend $\bzeta$ of the $\zeta$-mass relation has a value of 1.797$\pm$0.195, which is close to the $5/3$ scaling one would expect for the tSZE measured within the cluster virial region (e.g., $r<r_\mathrm{200c}$). However, given that the angular filtering in the tSZE cluster detection removes more flux from larger, more massive clusters, this measured mass trend is likely evidence for a steeper than self-similar relation. 
The redshift trend $\czeta$ is consistent with 0. We find that the redshift trend $\clambda$  of the $\lambda$-mass relation and the mass trend $\blambda$, are statistically consistent with 0 and 1, respectively.

We note that we marginalize over all the crucial systematic errors in our analysis by adopting the $M_{\mathrm{WL}} - M_{200\mathrm{c}}$ relation. We have repeated the analysis without marginalizing over the systematics, and we do not notice any significant difference.
This is expected, because our analysis is shape-noise dominated. \cite{Bocquet2024IIPhRvD.110h3510B} find the same as shown in their Fig. 3.

The blue contours in Fig.~\ref{fig:real_SPT_stacked} show the constraints from \cite{Bocquet2024IIPhRvD.110h3510B} WL analysis of the same SPT$\times$DES sample with MCMF center and DNF redshifts. They perform the analysis on a cluster-by-cluster basis compared to our average profile approach and they use the same radial fitting range as ours. Our analysis is done with a prior on $\Omega_{\mathrm{m}}$ unlike in \cite{Bocquet2024IIPhRvD.110h3510B}, where they use a flat prior. We use a wider prior on $\dzeta$ and $\czeta$ compared to their work as can be seen in Fig.~\ref{fig:real_SPT_stacked}. The slight differences in intrinsic scatter posteriors can be attributed to the different modeling of the observational Poisson noise on the richness  in this and the \cite{Bocquet2024IIPhRvD.110h3510B} analyses. In general, the results from our analysis show very good agreement with their single cluster analysis method and every parameter agrees on average at the $\approx 0.4\sigma$ level.

\begin{figure}
        \includegraphics[width=\columnwidth]{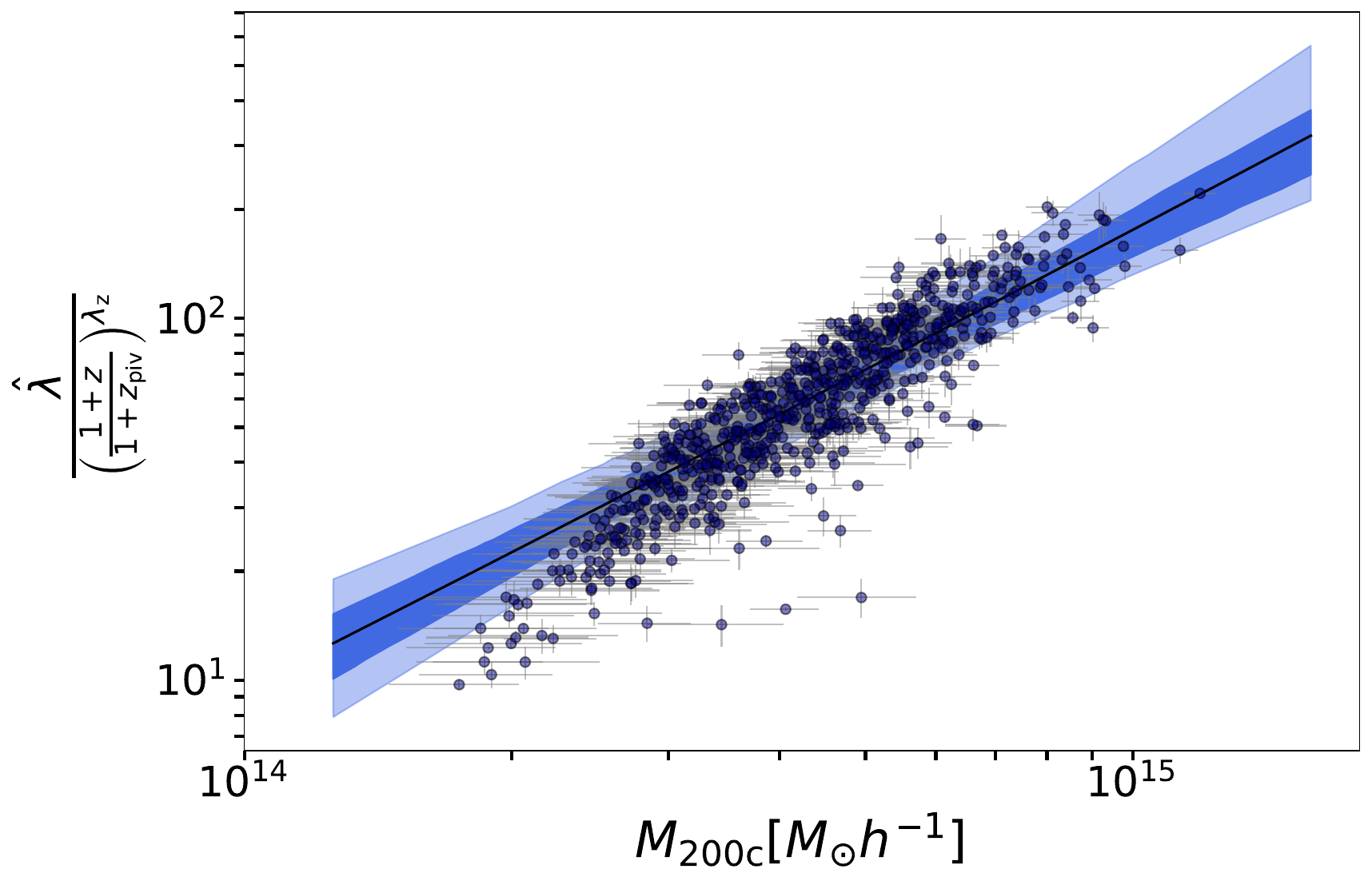}
        \includegraphics[width=\columnwidth]{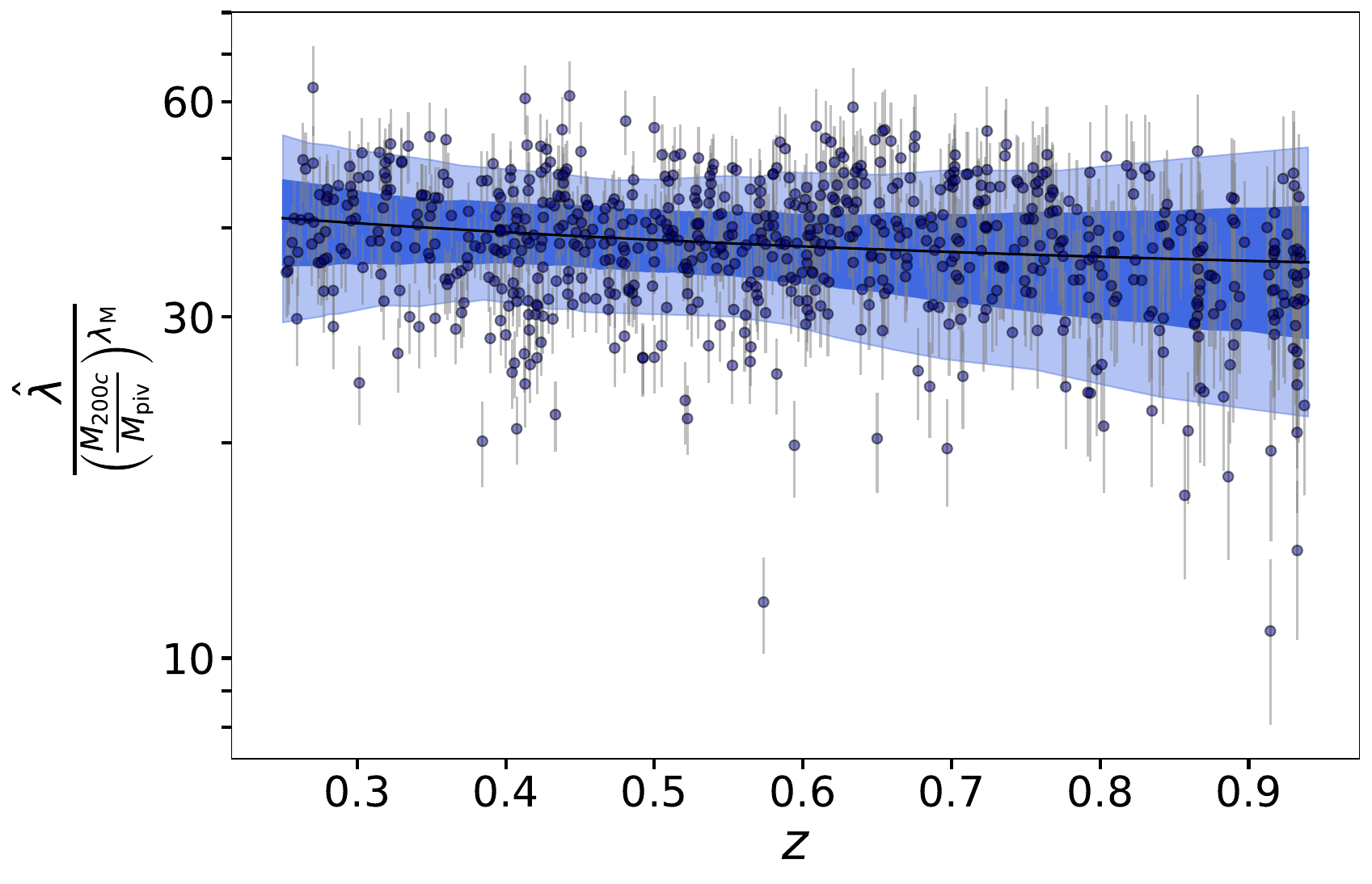}
        \vskip-0.10in
    \caption{Observed optical richness $\hat\lambda$ of SPT clusters as a function of the cluster halo mass (top) and redshift (bottom). The clusters are shown with filled circles, where the error bars also capture the error in the observable-mass relations and the estimated cluster halo mass.  The top and bottom plots shows the richness $\hat\lambda$ normalized at the pivot redshift $z_{\mathrm{piv}}$ = 0.6 and the pivot mass $M_{\mathrm{piv}} = 3\times10^{14} h^{-1} \mathrm{Mpc}$, respectively. The intrinsic model (Eq.~\ref{eq:lambdamass}) is shown in blue.
   The light and dark-shaded regions in both the panels represent 68$\%$ and 95$\%$ credible intervals of the mean model respectively.  }\label{fig:lambda_mass_redshift_validation}
\end{figure}

\begin{figure}
        \includegraphics[width=\columnwidth]{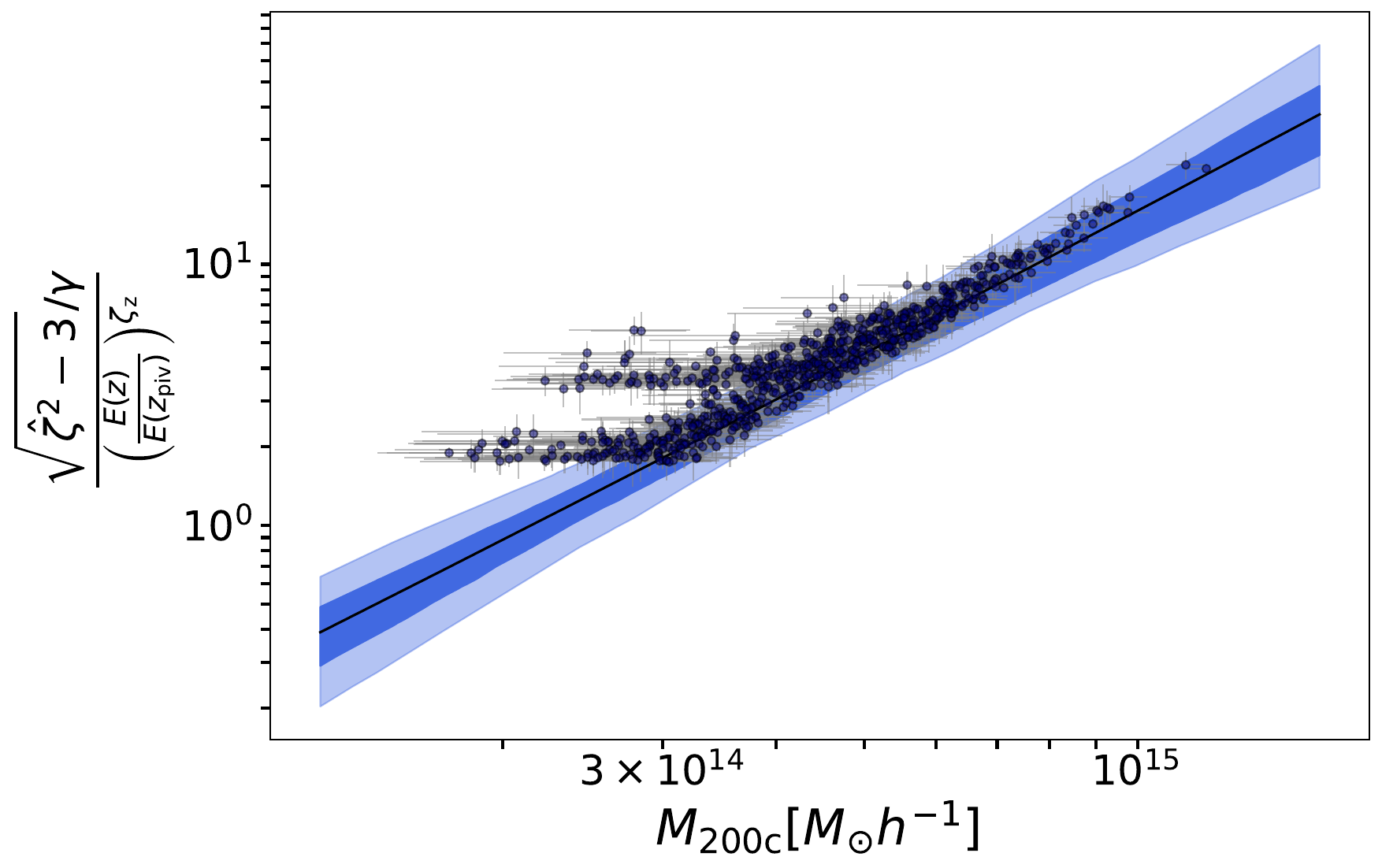}
        \includegraphics[width=\columnwidth]{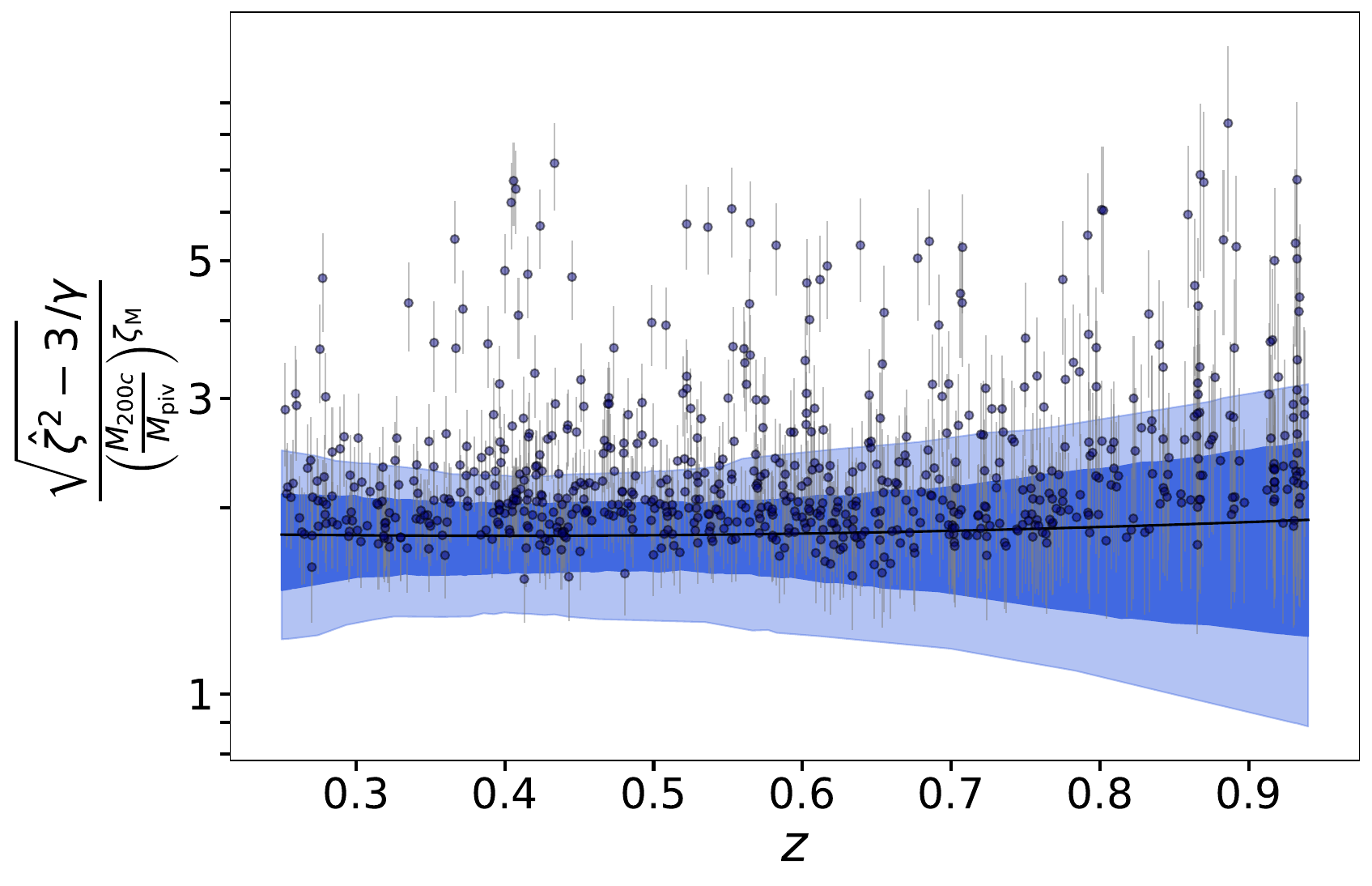}
        \vskip-0.10in
    \caption{Observed debiased detection significance $\hat\zeta$ of SPT clusters as a function of the cluster halo mass (top) and redshift (bottom).  The intrinsic model (Eq.~\ref{eq:zetaM}) is shown in blue. The plotting scheme is the same as Fig.~\ref{fig:lambda_mass_redshift_validation}. The effects of Eddington bias and selection on \hzeta can be seen in the above plots.}\label{fig:zeta_mass_redshift_validation}
\end{figure}
\begin{figure}
        \includegraphics[width=0.49\columnwidth]{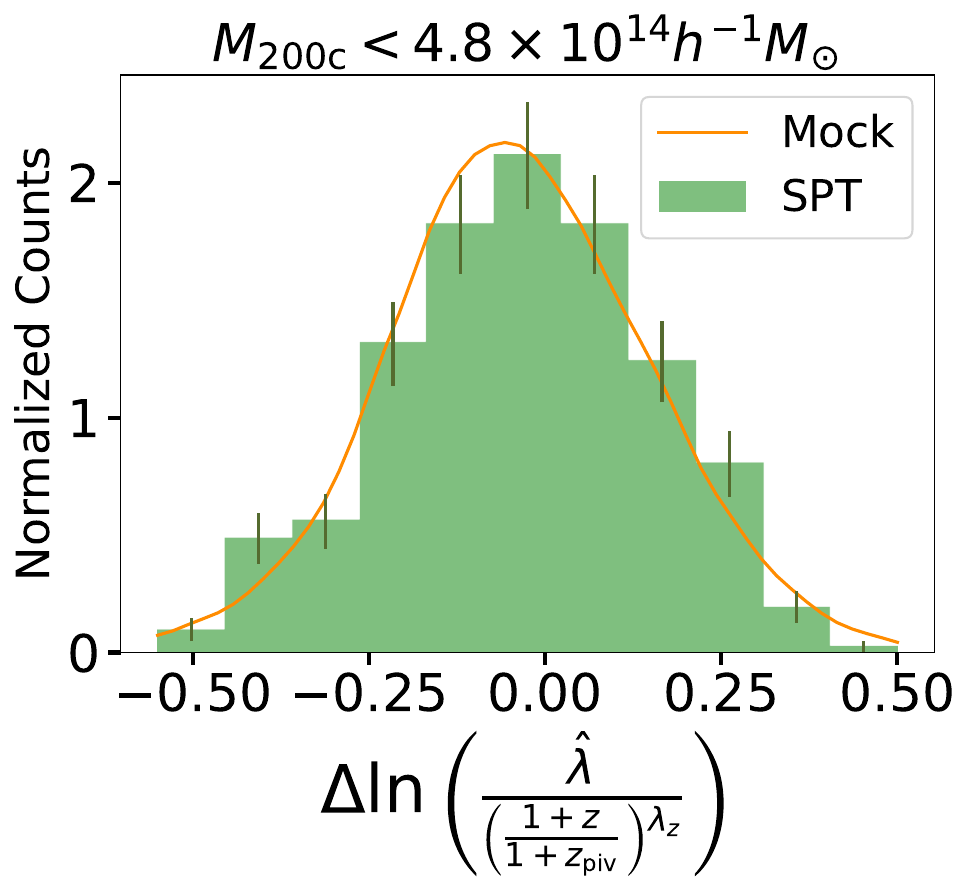}
        \includegraphics[width=0.49\columnwidth]{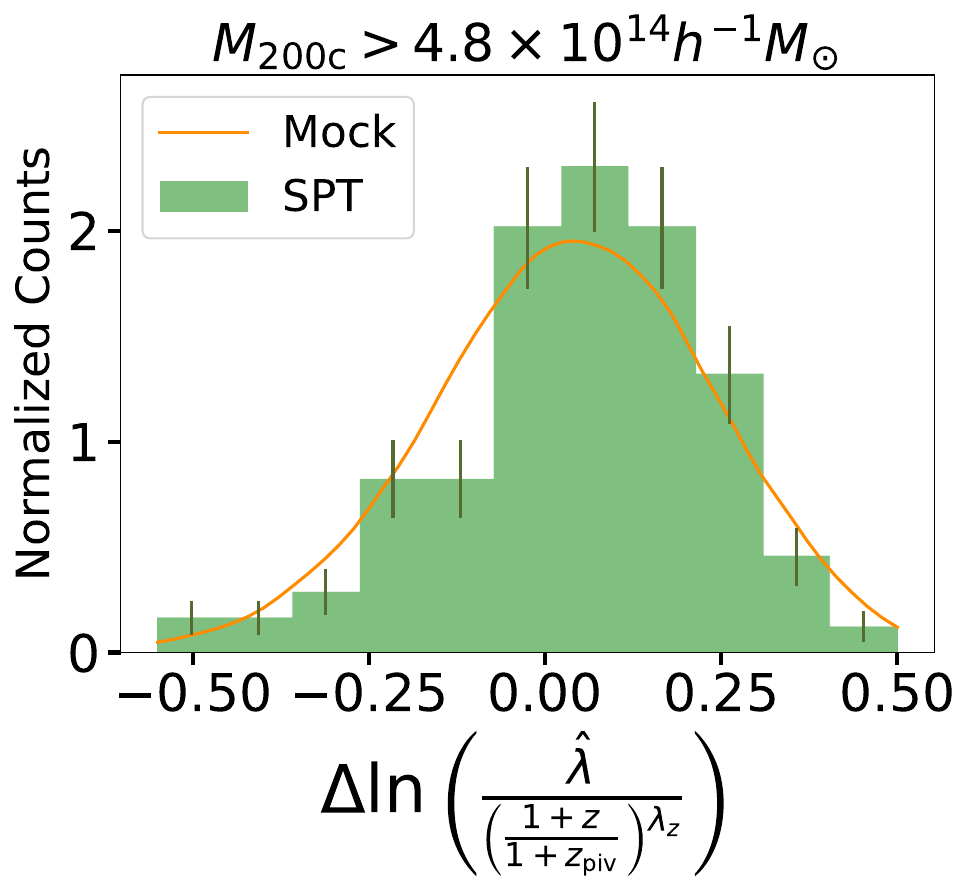}
        \includegraphics[width=0.49\columnwidth]{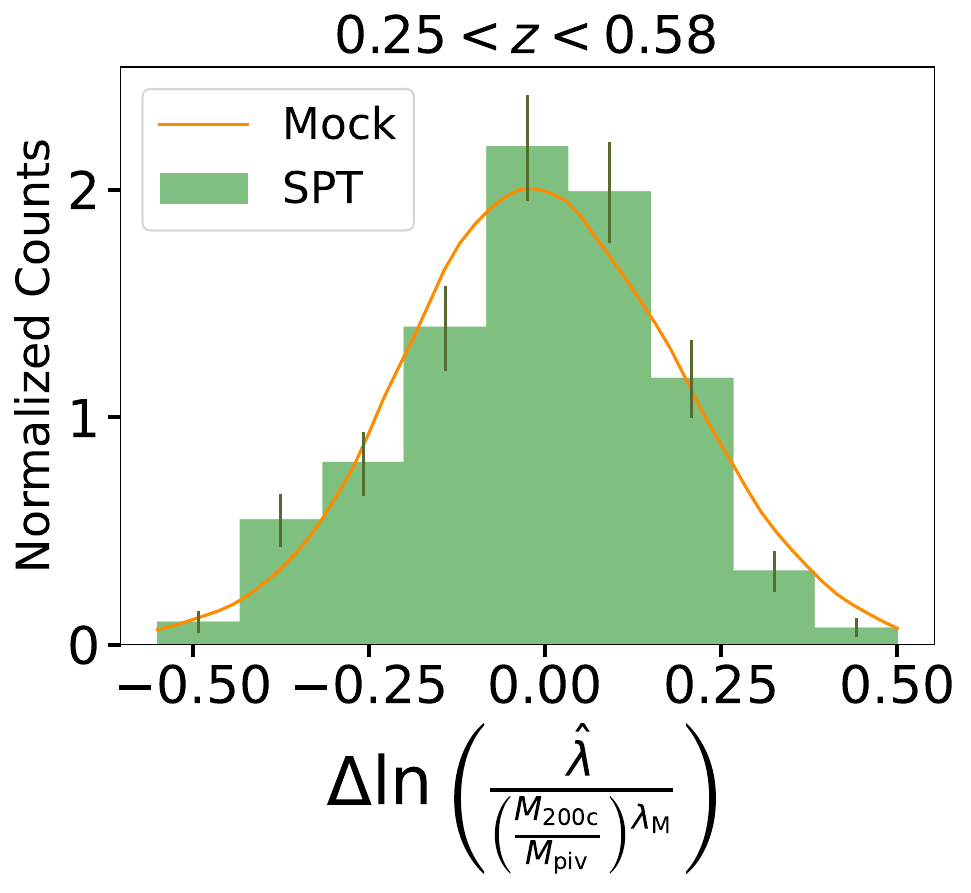}
        \includegraphics[width=0.49\columnwidth]{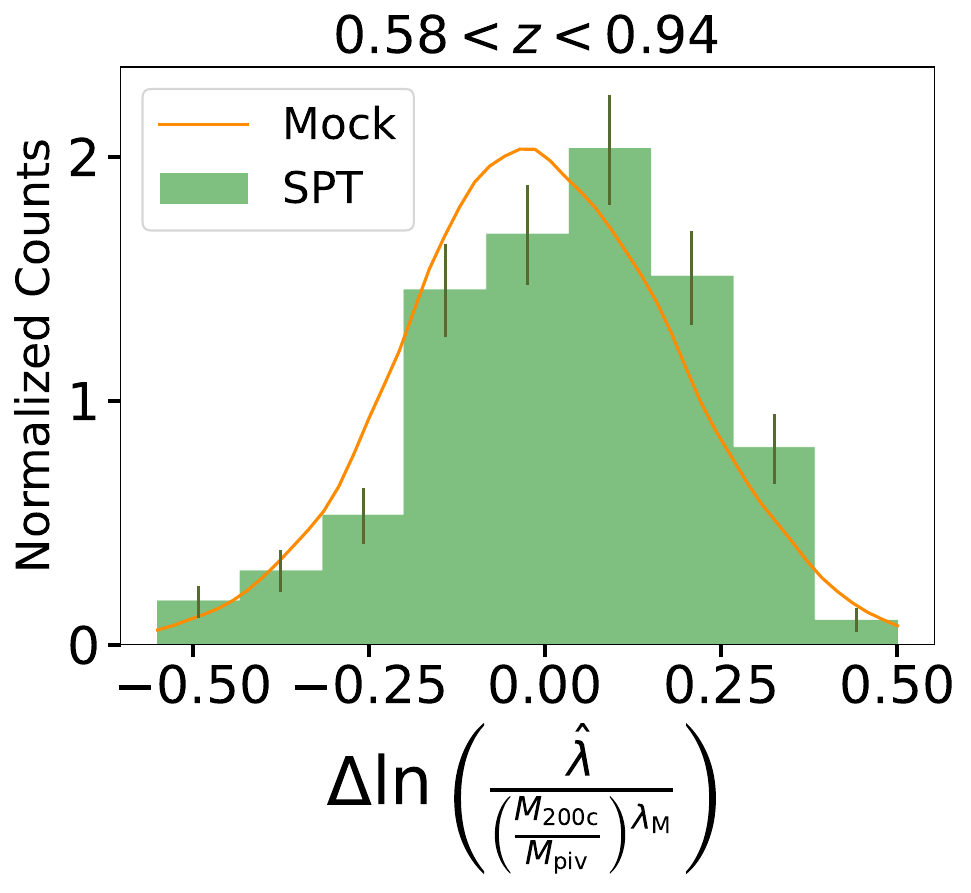}
        \vskip-0.10in
    \caption{Distributions of deviations in $\hlambda$ about the intrinsic observable-mass relation shown for the SPT sample (green histogram) and a mock dataset (100 times larger) drawn from that same intrinsic relation (orange line).  The top (bottom) panels show the deviations in low and high mass (redshift) with the measured redshift (mass) trend removed as in Fig.~\ref{fig:lambda_mass_redshift_validation}.
    }
    \label{fig:lambda_scatter_val_mass_redshift}
\end{figure}
\begin{figure}
        \includegraphics[width=0.49\columnwidth]{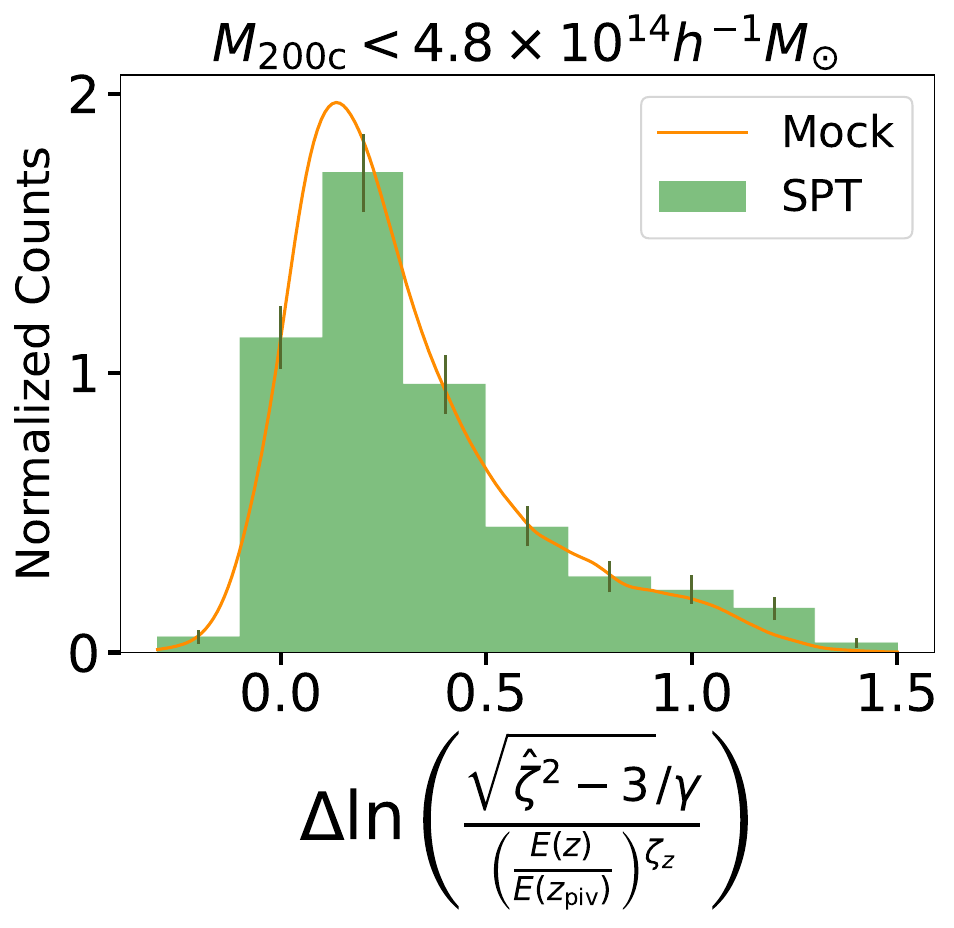}
        \includegraphics[width=0.49\columnwidth]{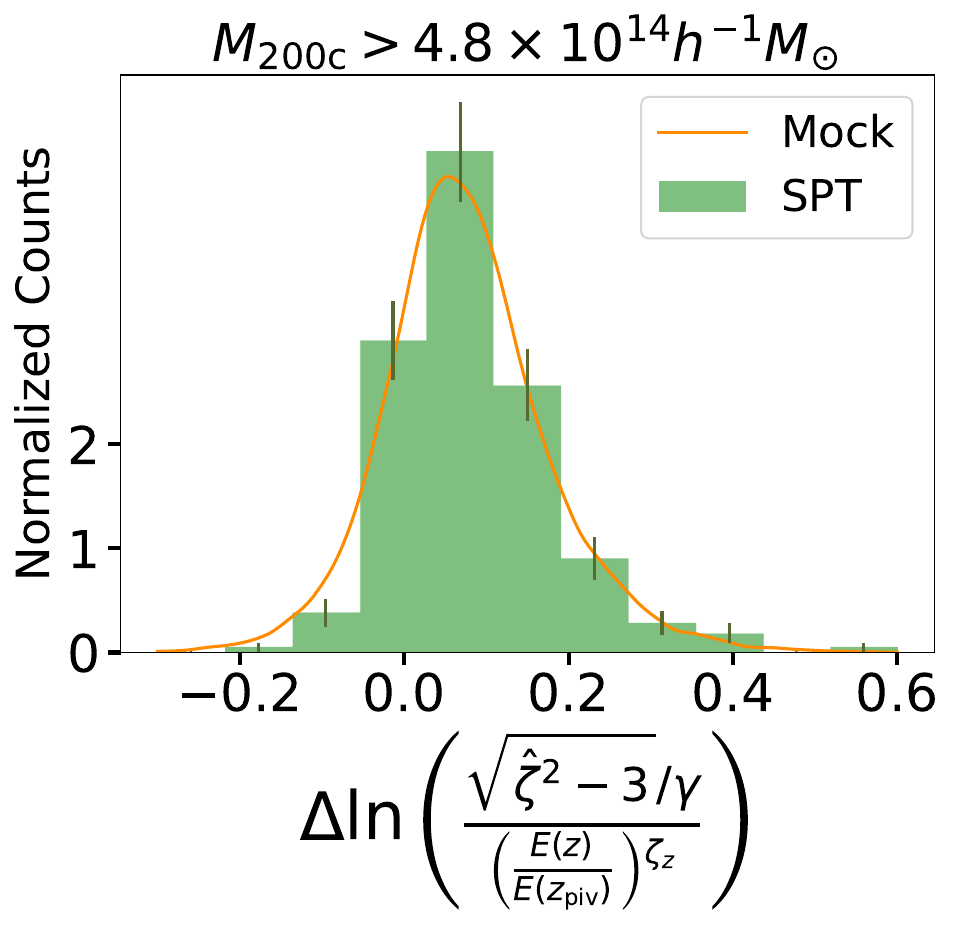}
        \includegraphics[width=0.49\columnwidth]{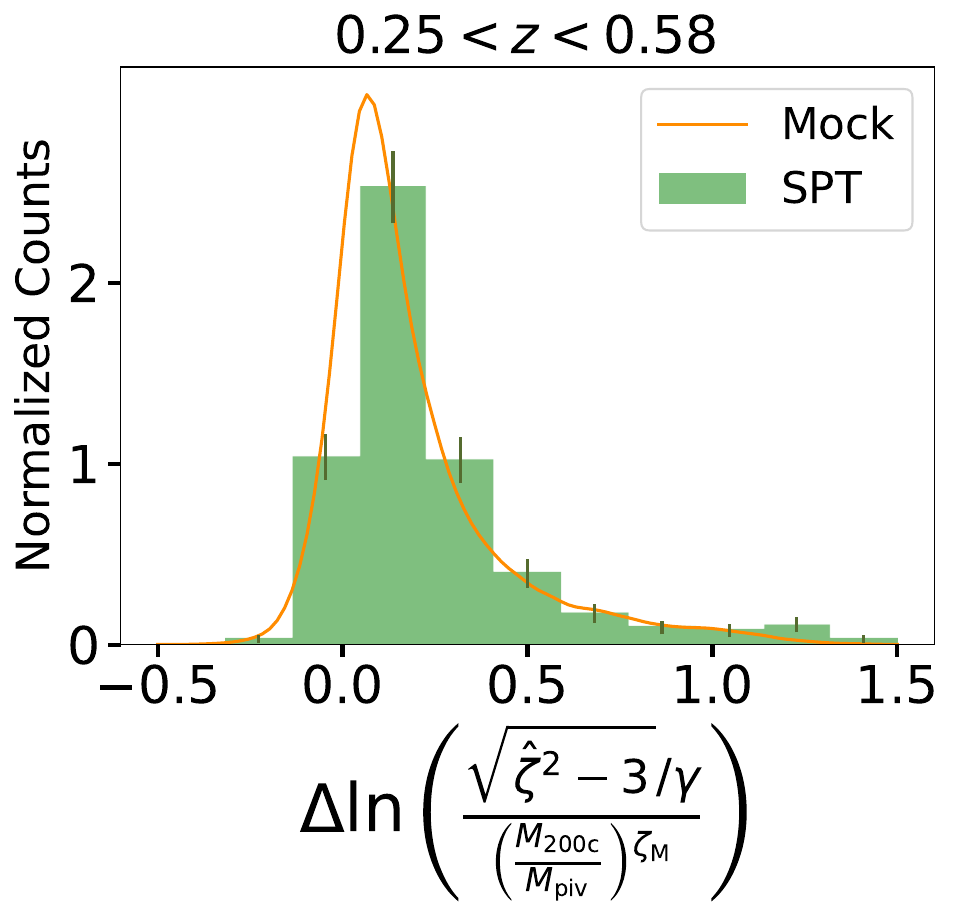}
        \includegraphics[width=0.49\columnwidth]{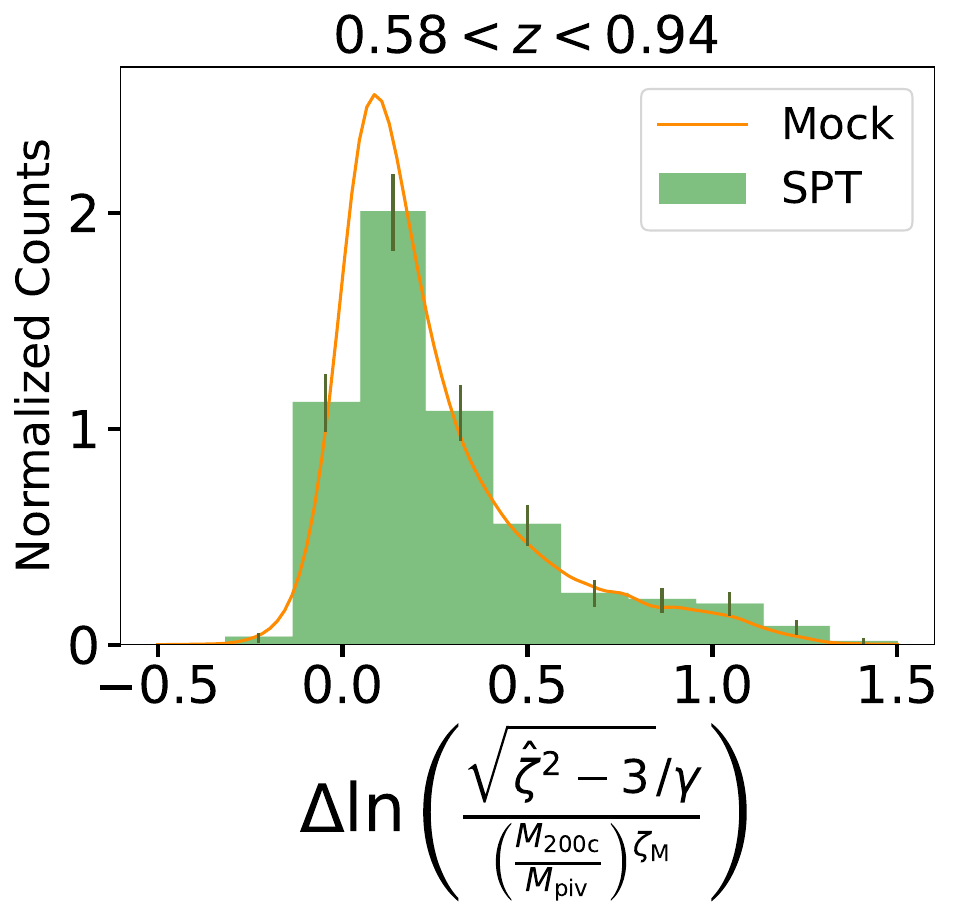}
        \vskip-0.10in
    \caption{Same as Fig.~\ref{fig:lambda_scatter_val_mass_redshift} but for normalized $\hat\zeta$.  The top (bottom) panels show the deviations in low and high mass (redshift) with the measured redshift (mass) trend removed as in Fig.~\ref{fig:zeta_mass_redshift_validation}.
    }
    \label{fig:zeta_scatter_val_mass_redshift}
\end{figure}

\subsubsection{Goodness of fit}
\label{sec:goodness_fit}
To assess the goodness of fit of the data to our model,  we compare
the average matter profile model to the average cluster matter profile in $3 \times 2 \times 2$ $\hzeta-\hlambda-z$ bins and perform a $\chi^2$ fit to all bins. Figure~\ref{fig:SPT_profile_validation} shows 12 average matter profiles (observations represented with black data points), each for a given $\hzeta-\hlambda-z$ bin along with the model and its corresponding 2$\sigma$ region (shown in shaded blue region). The profiles are extracted corresponding to the mean parameter obtained from our posterior. We obtain a chi-squared value of $\chi^2 = 97.83$ from 84 data points, effectively constraining 8 parameters. This corresponds to a reduced chi-squared value of $\chi_{\mathrm{red}}^2 = 1.27$ and a probability of exceeding the observed $\chi^2$ of $p$=0.047.

\subsubsection{Observable-mass relation validation}
\label{sec:obs_mass_validation}
To further validate our mass calibration results, we perform a series of tests on the richness and tSZE observable-mass relations.  In these tests we are determining whether the data are consistent with our model description of the observable-mass relations. In Fig.~\ref{fig:lambda_mass_redshift_validation} we show the $\hlambda$-mass (top) and $\hlambda$-redshift (bottom) relations. To analyze the $\hlambda$-mass relation, we calculate the richnesses at the pivot redshift by simply dividing $\hlambda$ by $\left(({1+z})/({1+z_{\mathrm{piv}}})\right)^{\clambda}$. This factor removes the measured redshift trend and allows us to study only the mass trend. On the x-axis, we show mean mass posteriors obtained using dual-observables, given by the following equation: 
\begin{equation}
\begin{split}
     & P(M_{\mathrm{200c}}|\hzeta, \hlambda, z, \vec p) = \\
     & \ \ 
     \frac{\iint \diff \lambda \diff \zeta P(\hlambda|\lambda) P(\hzeta|\zeta) P(\zeta, \lambda|M_{\mathrm{200c}}, z, \vec p) P(M_{\mathrm{200c}}|z, \vec p)}{\iiint \diff M \diff \lambda \diff \zeta P(\hlambda|\lambda) P(\hzeta|\zeta) P(\zeta, \lambda|M_{\mathrm{200c}}, z, \vec p) P(M_{\mathrm{200c}}|z, \vec p)}.
\label{eq: m200c_cal}    
\end{split}
\end{equation}
where of course the parameters of the observable-mass relations have been constrained using the WL mass calibration described above.

Similarly, we study the richness-redshift relation by normalizing the richnesses with $\left({\mathrm{M}_\mathrm{200c}}/{\mathrm{M}_\mathrm{{piv}}}\right)^{\blambda}$. In both panels the solid black line shows the 
intrinsic observable-mass relation (Eq.~\ref{eq:lambdamass}) corresponding to the mean posterior values. The dark and light-shaded blue bands represent 68\% and 95\% credible intervals on the intrinsic observable-mass relation model. The gray error bar on the data points represents the statistical errors and uncertainties in the observable-mass relations and the estimated cluster halo masses.

In Fig.~\ref{fig:zeta_mass_redshift_validation}
we show $\hzeta$-mass and $\hzeta$-redshift relation. On the y-axis, we plot $(\sqrt{{\hzeta^2-3}})/\gamma$,
where $\gamma$ is a scale factor that is used to correct for the different depths of fields in the SPT survey. Again, we normalize the y-axis with a factor of $\left({E(z)}/{E(z_{\mathrm{piv}})}\right)^{\czeta}$ while analyzing the relation with mass and with a factor of $\left({\mathrm{M}_\mathrm{200c}}/{\mathrm{M}_\mathrm{{piv}}}\right)^{\bzeta}$ while analyzing its relation with redshift. The intrinsic observable-mass relation model (Eq.~\ref{eq:zetaM}) with mean posterior parameters is shown with a black line and the error region is shown in blue bands as in the previous figure.

Compared to the intrinsic observable-mass relations (defined by blue bands) the observables in our analysis suffer from selection effects, which can be seen in Figs.~\ref{fig:lambda_mass_redshift_validation} and \ref{fig:zeta_mass_redshift_validation} (we further dicuss this in Appendix~\ref{sec:mock_SPT_obs_mass}). The Eddington bias is clearly visible in the left side (low mass portion) of the top panels in   Figs.~\ref{fig:lambda_mass_redshift_validation} and \ref{fig:zeta_mass_redshift_validation} from the points lying above the best-fitting line in $\hzeta$-mass and below the relation in $\hlambda$-mass.  These points are preferentially scattered away from the mean expected observables $\hzeta$ and $\hlambda$. A particularly strong feature is the two extensions of the observable $\hzeta$ to lower mass away from the mean relation.  These two features are created by the $\hzeta$ selection thresholds in the two SPT survey fields, and these are often referred to as Malmquist bias.

To understand whether the data points are behaving consistently with expectation, we create a mock sample 100 times larger than SPT sample using the mean parameter values from the measured posteriors and apply the same selection in $\hzeta$ and $\hlambda$ that was applied in creating the real cluster sample. The comparison of the mock and the real data then allows for robust validation.

In Fig.~\ref{fig:lambda_scatter_val_mass_redshift} we plot the difference between the observed $\hlambda$ and the mean intrinsic observable-mass relation model in log space, within two different mass bins (top panels) and two different different redshift bins (bottom panels). The orange line represents the 100 times larger mock sample and the green histogram is the real SPT sample. Poisson uncertainties are shown as error bars for each real data bin. We see good agreement between the mocks and the real sample in bins of high ($\chi^2 = 20.68$, $p$=0.04) and low mass ($\chi^2 = 15.15$, $p$=0.17) and high ($\chi^2 = 26.92$, $p$=0.002) and low redshift ($\chi^2 = 16.86$, $p$=0.05), indicating that our observable-mass relation model (including intrinsic and observed scatter components) provides a reasonably good description of the real dataset.  The $p$ value in the high redshift deviation plot indicates a $\sim$3$\sigma$ tension, which is notable. This tension could likely be reduced  by introducing a redshift dependent intrinsic scatter component in the $\lambda$-mass relation, but we leave that for a future discussion.  We note that the deviations in the top panels have the redshift trend scaled out, and the deviations in the bottom panels have the mass trend scaled out, similar to what we show in Fig.~\ref{fig:lambda_mass_redshift_validation}.

Similarly, in Fig.~\ref{fig:zeta_scatter_val_mass_redshift} we show the distribution of deviations in observed debiased $\hzeta$ around the mean intrinsic observable-mass relation for the SPT sample (green histogram) and the 100 times larger mock sample (orange line). The top panels present data within two mass ranges, and the bottom panels show two redshift ranges. We obtain quite reasonable $\chi^2$ in all four cases, with p values of 0.21 and 0.35 for the high- and low-mass bins, respectively, while the p values corresponding to the high- and low-redshift bins are 0.48 and 0.11, respectively.  Overall, the scatter of the real and mock clusters around the intrinsic observable-mass relation is similar, indicating that the $\zeta$-mass relation we adopt (together with its intrinsic scatter and measurement noise) provides a good description of the real dataset. The shift in the histogram peak away from zero is reflective of the Eddington bias introduced by the $\hzeta$ selection, which is more pronounced when compared to richness. As with the previous plot, the deviations in the top panels have the redshift trend scaled out, and the deviations in the bottom panels have the mass trend scaled out, similar to what we show in Fig.~\ref{fig:zeta_mass_redshift_validation}.

In the appendix we present in Fig.~\ref{fig:zeta_mass_redshift_mocks} a version of the $\hzeta$-mass figure shown for real data in Fig.~\ref{fig:zeta_mass_redshift_validation} that is created by down-sampling the 100 times larger mock sample to a similar number of clusters as for the real data.  The mock and real data behave similarly, showing the same selection related scatter of the data around the best fits intrinsic relation.

\begin{figure}
        \includegraphics[width=\columnwidth]{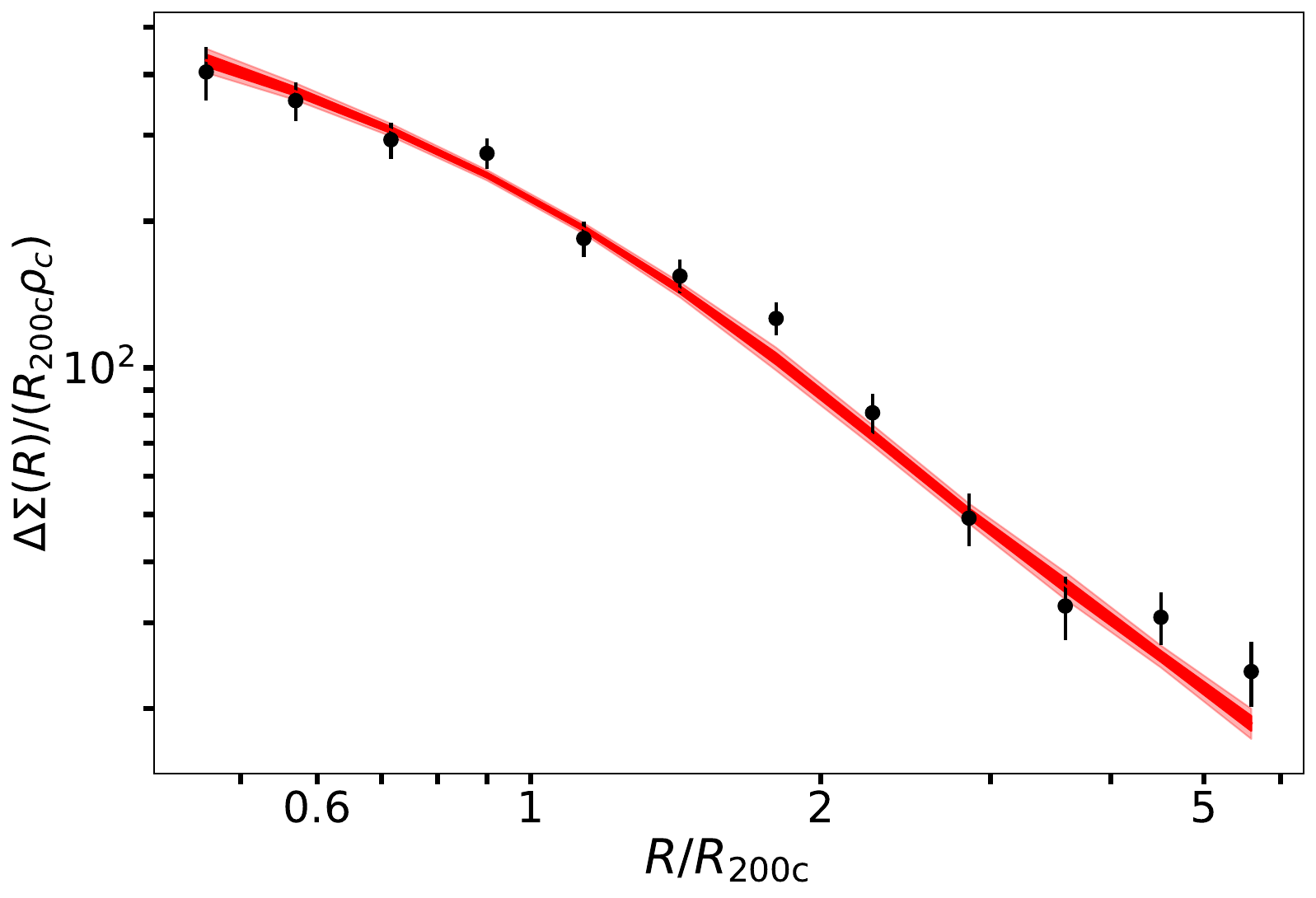}
        \vskip-0.10in
 \caption{Average rescaled matter profile $\widetilde{\Delta\Sigma}({R}/{R_\mathrm{200c}})$ of the full sample of 698 clusters over the full mass and redshift ranges estimated using the mean mass posterior $P(M_{\mathrm{200c}}|\hzeta, \hlambda, z, \vec p)$ for each cluster (see Eq.~\ref{eq: m200c_cal}).
 The black points represent the measured mean profile, and the error bars include not only shape noise but also marginalization over the cluster mass uncertainties. The dark and light-shaded red bands show 68\% and 95\% credible intervals, respectively, on the model profile extracted from hydrodynamical simulations.}\label{fig:full_profile}
\end{figure}
\subsection{Matter profile extending to cluster outskirts  }
\label{sec:outer_profile}
In our mass calibration analysis, we restricted our radial range to $R<3.2/(1+z) h^{-1} \mathrm{Mpc}$ to avoid the impact of the two-halo regime. 
Using our mass calibration results (Table \ref{tab:SPT_posterior}), we combine the whole SPT sample with redshift, $0.25<z<0.94$ and all masses to create an average rescaled matter profile  $\widetilde{\Delta\Sigma}$ that includes regions beyond the one-halo region. We create 200 realizations of the average matter profile using the parameter posteriors to marginalize over the observable-mass relation parameter uncertainty in the profile. The SNR of the full average matter profile is $\sim$36 out to 6 $R/R_\mathrm{200c}$. Figure~\ref{fig:full_profile} shows the full profile compared to the mean model calculated at the mean redshift of the sample. The light and dark-shaded red bands represent the 68\% and 95\% credible intervals,respectively. We obtain $\chi^2 = 15.51$ for 12 degrees of freedom, which corresponds to a probability to exceed observed $\chi^2$ of $p$=0.21. This suggests that the model and data are in good agreement, even in the cluster outskirts. 

The agreement between simulations (red) and observations (points) in this high SNR measured profile over this radial range is quite interesting.  It suggests that structure formation modeled using CDM and baryonic physics within a $\Lambda$CDM context provides quite a good description of not only cluster halo profiles but also cluster infall regions.  Extending the profile inward to smaller radii should enable interesting tests of baryonic feedback and the CDM scenario in the limit that cluster mis-centering and cluster member contamination can be sufficiently controlled.  This topic lies beyond the scope of the current analysis.

\subsection{Comparison to previous work}
\label{sec:comp_to_other_works}
\begin{figure}
        \includegraphics[width=0.49\textwidth]{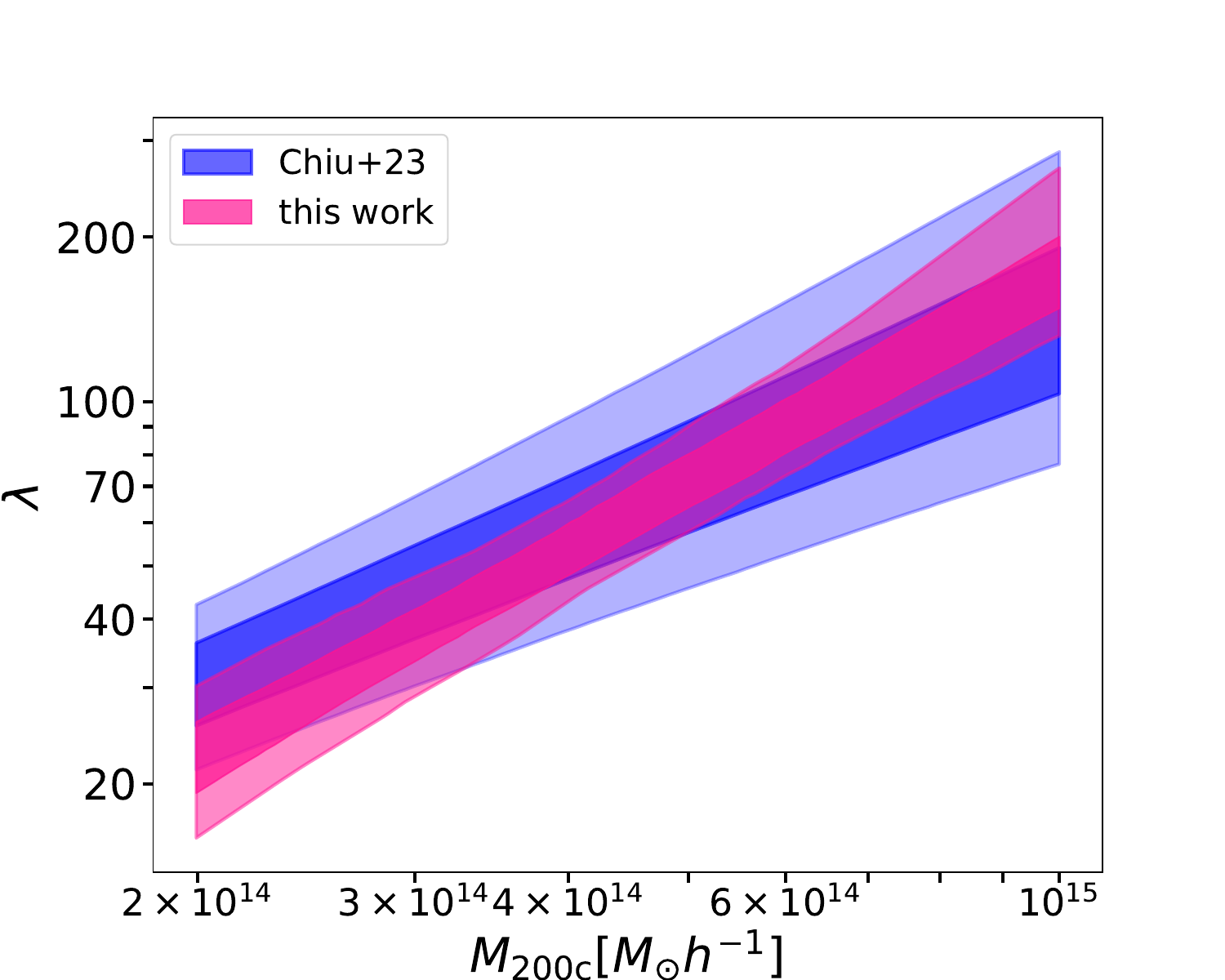}
        \includegraphics[width=0.49\textwidth]{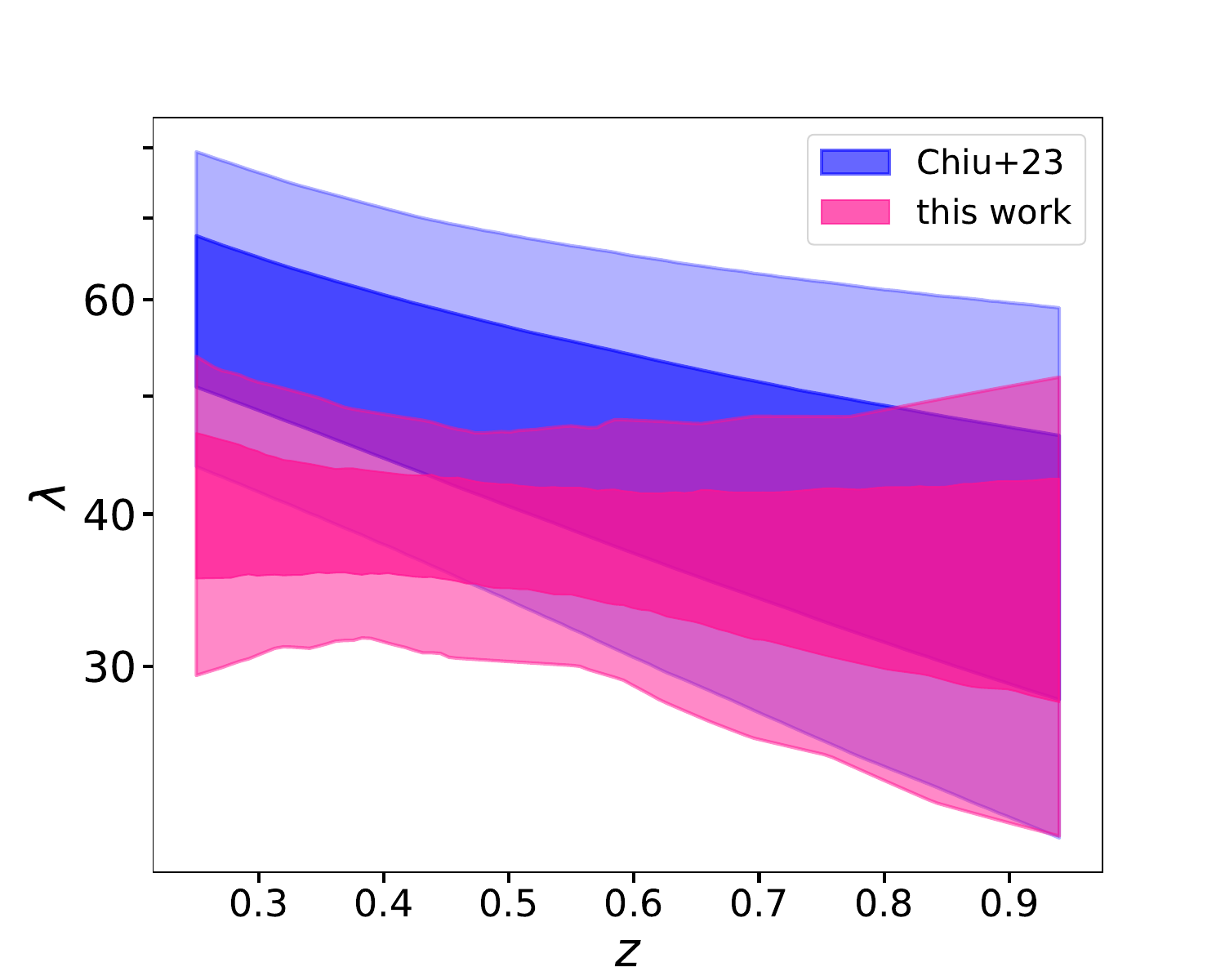}
        \vskip-0.10in
    \caption{Best fit $\lambda$-mass (top) and $\lambda$-redshift (bottom) relations evaluated at our pivot mass and redshift compared to \protect\cite{Chiu23} work (in blue). Dark and light-shaded regions represent 68\% and 95\% credible intervals, respectively. Our analysis of the tSZE selected and MCMF confirmed sample (in pink) shows good agreement with the analysis of an X-ray extent-selected sample (blue). }\label{fig:Chi_Adi_lam_comp}
\end{figure}

In this section we compare our tSZE and richness observable-mass relation results to those from previous studies. \cite{Chiu23} analyzed the eROSITA Final Equatorial
Depth Survey (eFEDS) cluster sample with Hyper Suprime-Cam (HSC) WL data along with MCMF richness and calibrated the richness-mass-redshift relation. Their results are shown in shaded blue color in Fig.~\ref{fig:Chi_Adi_lam_comp} plotted at our pivot redshift of 0.6 and pivot mass of $M_\mathrm{200c} = 3 \times 10^{14} h^{-1}\mathrm{Mpc}$. Because their analysis is performed with $M_{500c}$, we used the conversion relation provided in \cite{Ragagnin_2020} to convert $M_{500c}$ to $M_\mathrm{200c}$. Our work (in pink) results in tighter constraints than those presented in \cite{Chiu23}.  Our mass trend is slightly steeper and redshift trend is slightly weaker than those from \cite{Chiu23}, but given the uncertainties there is no tension between the two analyses.

\section{Summary}
\label{sec:summary}
In this work, we have studied galaxy cluster matter profiles in observations and simulations. We have developed a new mass calibration technique that employs average matter profiles and applied it to the SPT+MCMF tSZE selected and optically confirmed cluster sample with associated DES weak gravitational lensing measurements.

Our analysis began with a comparison of the scaling properties, the so-called self-similarity, of the simulated and observed matter profiles $\Delta\Sigma(R)$ of the cluster samples. An examination of \textit{Magneticum} and IllustrisTNG simulations indicated remarkable self-similarity among galaxy clusters with varying redshift and masses. In particular, we analyzed the rescaled matter profiles $\widetilde{\Delta\Sigma}({R}/{R_\mathrm{200c}})$, which is $\Delta\Sigma(R/R_\mathrm{200c}) /(R_\mathrm{200c}\rho_{\mathrm{crit}})$. Rescaling individual profiles by their corresponding $R_\mathrm{200c}$ and associated critical density $\rho_{\mathrm{crit}}(z)$ significantly reduces the mass and redshift dependences, respectively. We quantified the observed self-similarity by computing fractional variation with redshift and mass in the rescaled space and comparing it with the fractional variation obtained with the original matter profiles $\Delta\Sigma(R)$. The fractional variation with redshift is roughly six times lower in the rescaled matter profile. We observed a remarkably low fractional variation with mass of $\approx 1\%$, which is $\approx 23$ times lower than the fractional variation obtained in $\Delta\Sigma(R)$. This self-similar behavior is ideal for analyzing the average cluster matter profiles because it minimizes the cluster-to-cluster variation among the profiles and allows one to combine clusters with a wide range of redshift and mass, enabling studies of cluster matter profiles in a high S/N regime.

We exploited the self-similarity of the average matter profiles in the rescaled space in order to develop a new mass calibration method that relies on average rescaled matter profiles, and we employed that method to calibrate the masses of SPT tSZE-selected and MCMF optically confirmed clusters. For this method, we used hydrodynamical simulations to construct a model average rescaled matter profile $\widetilde{\Delta\Sigma}({R}/{R_\mathrm{200c}})$ while accounting for small residual redshift trends and assuming perfect self-similarity with mass. We created average SPT $\times$ DES-WL cluster matter profiles with their appropriate weights and accounted for all crucial systematic errors through the $M_{\mathrm{WL}}-M_{200\mathrm{c}}$ relation. Our likelihood constrains the $\lambda-M-z$ and $\zeta-M-z$ relation parameters and takes into account the observational and intrinsic scatter on the observables. Additionally, we accounted for the Eddington and Malmquist biases that arise from the selection applied to the observables in defining the cluster sample.

We calibrated the $\lambda-M-z$ and $\zeta-M-z$ observable-mass relations using the average rescaled matter profile calibration method described in Sect.~\ref{sec:method}, which simultaneously constrains the amplitude, mass trend, redshift trend, and intrinsic scatter for both observable-mass relations. Our constraints on the observable-mass relation parameters show mass trends that are steeper than but statistically consistent with the self-similar expectations for both the richness and tSZE observable-mass relations. Moreover, we found no statistically significant evidence for a redshift trend in richness, $\clambda = -0.349 \pm 0.690 $, or tSZE $\zeta$, $\czeta =  0.045 \pm 1.054$. Our constraints on $\lambda-M-z$ parameters are in good agreement with \cite{Chiu23}, who has analyzed the eFEDS X-ray extent-selected cluster sample with HSC WL.
 
In addition, \cite{Bocquet2024IIPhRvD.110h3510B} have previously analyzed the SPT selected and MCMF confirmed cluster sample with DES WL using a cluster-by-cluster WL calibration method together with a simultaneous fit to the tSZE halo observable function. Our results from the average rescaled matter profile analysis for both the observable-mass relations are in good agreement with their work.  Because our analysis uses no information from the underlying halo mass function, which is strongly cosmologically dependent, the agreement between the two methods is an interesting indication that with the DES WL dataset, the direct mass constraints from WL are not significantly biased by the inclusion of the cluster abundance information.

We have presented a new validation of the observable-mass relations that examines the scatter of the cluster sample observables around the mean intrinsic relations.  Using a large sample of mock clusters drawn from the best-fit relation, we found no significant differences between the real and mock samples for the $\zeta$-mass relation. This indicates that the power-law form of the observable-mass relations, including our modeling of intrinsic scatter and measurement noise, provides a good description of the data. For the $\lambda$-mass relation, however, we find that the high-redshift bin indicates a $\sim3\sigma$ tension, which could suggest a potential redshift dependence of the intrinsic scatter in the $\lambda$-mass relation.

The validation of the form of the observable-mass relations, along with the evidence showing that the mass calibration reported using the \cite{BocquetI2024PhRvD.110h3509B} method is unaffected by potential biases from the underlying halo mass function, lend further weight to recent findings. These findings include standard cosmological results on interacting dark matter, modified gravity, and combined probe results reported using variations of that method to analyze the SPTxDES dataset.
\citep{Bocquet2024IIPhRvD.110h3510B,Mazoun2024arXiv241119911M,Vogt2024arXiv240913556V,BocquetCombined2024arXiv241207765B}.

Employing the rescaled matter profile method, we produced a high S/N average rescaled matter profile from the entire SPT and MCMF selected sample of 698 clusters that extends to 6$R/R_{\mathrm{200c}}$ (see Fig.~\ref{fig:full_profile}).  This profile deviates from a null profile with an S/N$\sim$36, providing the most precise measurement of the average ICM selected cluster matter profile to date.  A comparison of this profile to hydrodynamical simulations carried out within the $\Lambda$CDM paradigm over the same radial range shows good agreement within the current uncertainties.

The upcoming stage IV WL surveys will offer a vast amount of much higher quality lensing data. The average rescaled matter profile mass calibration method, which we have successfully demonstrated here, provides a promising new tool for analyzing these future datasets. Moreover, this new method will enable efficient analyses of much larger cluster samples for cosmology and structure formation constraints. Analyzing the average cluster matter profiles in rescaled space will further help create high S/N profiles by combining cluster measurements over wide redshift and mass ranges. The shape of the average cluster matter profile can then be used to study baryonic feedback, different modified gravity models, and the collisional nature of dark matter.

\begin{acknowledgements}
We acknowledge financial support from the MPG Faculty Fellowship program and the Ludwig-Maximilians-Universit\"at (LMU-Munich).

The South Pole Telescope program is supported by the National Science Foundation (NSF) through the Grant No. OPP-1852617. Partial support is also provided by the Kavli Institute of Cosmological Physics at the University of Chicago.
PISCO observations were supported by US NSF grant AST-0126090. 
Work at Argonne National Laboratory was supported by the U.S. Department of Energy, Office of High Energy Physics, under Contract No. DE-AC02-06CH11357.

Funding for the DES Projects has been provided by the U.S. Department of Energy, the U.S. National Science Foundation, the Ministry of Science and Education of Spain, 
the Science and Technology Facilities Council of the United Kingdom, the Higher Education Funding Council for England, the National Center for Supercomputing 
Applications at the University of Illinois at Urbana-Champaign, the Kavli Institute of Cosmological Physics at the University of Chicago, 
the Center for Cosmology and Astro-Particle Physics at the Ohio State University,
the Mitchell Institute for Fundamental Physics and Astronomy at Texas A\&M University, Financiadora de Estudos e Projetos, 
Funda{\c c}{\~a}o Carlos Chagas Filho de Amparo {\`a} Pesquisa do Estado do Rio de Janeiro, Conselho Nacional de Desenvolvimento Cient{\'i}fico e Tecnol{\'o}gico and 
the Minist{\'e}rio da Ci{\^e}ncia, Tecnologia e Inova{\c c}{\~a}o, the Deutsche Forschungsgemeinschaft and the Collaborating Institutions in the Dark Energy Survey. 

The Collaborating Institutions are Argonne National Laboratory, the University of California at Santa Cruz, the University of Cambridge, Centro de Investigaciones Energ{\'e}ticas, 
Medioambientales y Tecnol{\'o}gicas-Madrid, the University of Chicago, University College London, the DES-Brazil Consortium, the University of Edinburgh, 
the Eidgen{\"o}ssische Technische Hochschule (ETH) Z{\"u}rich, 
Fermi National Accelerator Laboratory, the University of Illinois at Urbana-Champaign, the Institut de Ci{\`e}ncies de l'Espai (IEEC/CSIC), 
the Institut de F{\'i}sica d'Altes Energies, Lawrence Berkeley National Laboratory, the Ludwig-Maximilians Universit{\"a}t M{\"u}nchen and the associated Excellence Cluster Universe, 
the University of Michigan, NSF's NOIRLab, the University of Nottingham, The Ohio State University, the University of Pennsylvania, the University of Portsmouth, 
SLAC National Accelerator Laboratory, Stanford University, the University of Sussex, Texas A\&M University, and the OzDES Membership Consortium.

Based in part on observations at Cerro Tololo Inter-American Observatory at NSF's NOIRLab (NOIRLab Prop. ID 2012B-0001; PI: J. Frieman), which is managed by the Association of Universities for Research in Astronomy (AURA) under a cooperative agreement with the National Science Foundation.

The DES data management system is supported by the National Science Foundation under Grant Numbers AST-1138766 and AST-1536171.
The DES participants from Spanish institutions are partially supported by MICINN under grants ESP2017-89838, PGC2018-094773, PGC2018-102021, SEV-2016-0588, SEV-2016-0597, and MDM-2015-0509, some of which include ERDF funds from the European Union. IFAE is partially funded by the CERCA program of the Generalitat de Catalunya.
Research leading to these results has received funding from the European Research
Council under the European Union's Seventh Framework Program (FP7/2007-2013) including ERC grant agreements 240672, 291329, and 306478.
We  acknowledge support from the Brazilian Instituto Nacional de Ci\^encia
e Tecnologia (INCT) do e-Universo (CNPq grant 465376/2014-2).

This manuscript has been authored by Fermi Research Alliance, LLC under Contract No. DE-AC02-07CH11359 with the U.S. Department of Energy, Office of Science, Office of High Energy Physics.
\end{acknowledgements}
\section*{Data availability}
The data underlying this article will be shared upon reasonable request to the corresponding author.


\bibliographystyle{aa}
\bibliography{bibliography}


\section*{Affiliations}
\begin{enumerate}
\item University Observatory, Faculty of Physics, Ludwig-Maximilians-Universitat, Scheinerstr. 1, 81679 Munich, Germany 
\item Max Planck Institute for Extraterrestrial Physics, Giessenbachstr. 1, 85748 Garching, Germany 
\item Universitat Innsbruck, Institut fur Astro- und Teilchenphysik, Technikerstr. 25/8, 6020 Innsbruck, Austria 
\item Laborat\'orio Interinstitucional de e-Astronomia - LIneA, Rua Gal. Jos\'e Cristino 77, Rio de Janeiro, RJ - 20921-400, Brazil 
\item Fermi National Accelerator Laboratory, P. O. Box 500, Batavia, IL 60510, USA 
\item Department of Physics, University of Michigan, Ann Arbor, MI 48109, USA 
\item Instituto de F\'isica Te\'orica, Universidade Estadual Paulista, S$\tilde{\mathrm{a}}$o Paulo, Brazil 
\item Institute of Cosmology and Gravitation, University of Portsmouth, Portsmouth, PO1 3FX, UK 
\item Department of Physics and Astronomy, Pevensey Building, University of Sussex, Brighton, BN1 9QH, UK 
\item Department of Physics \& Astronomy, University College London, Gower Street, London, WC1E 6BT, UK 
\item Instituto de Astrofisica de Canarias, E-38205 La Laguna, Tenerife, Spain 
\item Institut de F\'{\i}sica d'Altes Energies (IFAE), The Barcelona Institute of Science and Technology, Campus UAB, 08193 Bellaterra (Barcelona) Spain 
\item Astronomy Unit, Department of Physics, University of Trieste, via Tiepolo 11, I-34131 Trieste, Italy 
\item INAF-Osservatorio Astronomico di Trieste, via G. B. Tiepolo 11, I-34143 Trieste, Italy 
\item Institute for Fundamental Physics of the Universe, Via Beirut 2, 34014 Trieste, Italy 
\item Hamburger Sternwarte, Universit\"{a}t Hamburg, Gojenbergsweg 112, 21029 Hamburg, Germany 
\item Department of Physics, IIT Hyderabad, Kandi, Telangana 502285, India 
\item Jet Propulsion Laboratory, California Institute of Technology, 4800 Oak Grove Dr., Pasadena, CA 91109, USA 
\item Kavli Institute for Cosmological Physics, University of Chicago, Chicago, IL 60637, USA 
\item Instituto de Fisica Teorica UAM/CSIC, Universidad Autonoma de Madrid, 28049 Madrid, Spain 
\item Institut d'Estudis Espacials de Catalunya (IEEC), 08034 Barcelona, Spain 
\item Institute of Space Sciences (ICE, CSIC),  Campus UAB, Carrer de Can Magrans, s/n,  08193 Barcelona, Spain 
\item Center for Astrophysical Surveys, National Center for Supercomputing Applications, 1205 West Clark St., Urbana, IL 61801, USA 
\item Department of Astronomy, University of Illinois at Urbana-Champaign, 1002 W. Green Street, Urbana, IL 61801, USA 
\item Santa Cruz Institute for Particle Physics, Santa Cruz, CA 95064, USA 
\item Center for Cosmology and Astro-Particle Physics, The Ohio State University, Columbus, OH 43210, USA 
\item Department of Physics, The Ohio State University, Columbus, OH 43210, USA 
\item Center for Astrophysics $\vert$ Harvard \& Smithsonian, 60 Garden Street, Cambridge, MA 02138, USA 
\item Australian Astronomical Optics, Macquarie University, North Ryde, NSW 2113, Australia 
\item Lowell Observatory, 1400 Mars Hill Rd, Flagstaff, AZ 86001, USA 
\item Departamento de F\'isica Matem\'atica, Instituto de F\'isica, Universidade de S\~ao Paulo, CP 66318, S\~ao Paulo, SP, 05314-970, Brazil 
\item George P. and Cynthia Woods Mitchell Institute for Fundamental Physics and Astronomy, and Department of Physics and Astronomy, Texas A\&M University, College Station, TX 77843,  USA 
\item LPSC Grenoble - 53, Avenue des Martyrs 38026 Grenoble, France 
\item Instituci\'o Catalana de Recerca i Estudis Avan\c{c}ats, E-08010 Barcelona, Spain 
\item Department of Astrophysical Sciences, Princeton University, Peyton Hall, Princeton, NJ 08544, USA 
\item Observat\'orio Nacional, Rua Gal. Jos\'e Cristino 77, Rio de Janeiro, RJ - 20921-400, Brazil 
\item Department of Physics, Northeastern University, Boston, MA 02115, USA 
\item Centro de Investigaciones Energ\'eticas, Medioambientales y Tecnol\'ogicas (CIEMAT), Madrid, Spain 
\item School of Physics and Astronomy, University of Southampton,  Southampton, SO17 1BJ, UK 
\item Computer Science and Mathematics Division, Oak Ridge National Laboratory, Oak Ridge, TN 37831 
\item Argonne National Laboratory, 9700 S Cass Ave, Lemont, IL 60439, USA 
\item Department of Astronomy, University of California, Berkeley,  501 Campbell Hall, Berkeley, CA 94720, USA 
\item Lawrence Berkeley National Laboratory, 1 Cyclotron Road, Berkeley, CA 94720, USA
\end{enumerate}

\institute{University Observatory, Faculty of Physics, Ludwig-Maximilians-Universitat, Scheinerstr. 1, 81679 Munich, Germany 
        \and Max Planck Institute for Extraterrestrial Physics, Giessenbachstr. 1, 85748 Garching, Germany 
        \and Universitat Innsbruck, Institut fur Astro- und Teilchenphysik, Technikerstr. 25/8, 6020 Innsbruck, Austria 
        \and Laborat\'orio Interinstitucional de e-Astronomia - LIneA, Rua Gal. Jos\'e Cristino 77, Rio de Janeiro, RJ - 20921-400, Brazil 
        \and Fermi National Accelerator Laboratory, P. O. Box 500, Batavia, IL 60510, USA 
        \and Department of Physics, University of Michigan, Ann Arbor, MI 48109, USA 
        \and Instituto de F\'isica Te\'orica, Universidade Estadual Paulista, S$\tilde{\mathrm{a}}$o Paulo, Brazil 
        \and Institute of Cosmology and Gravitation, University of Portsmouth, Portsmouth, PO1 3FX, UK 
        \and Department of Physics and Astronomy, Pevensey Building, University of Sussex, Brighton, BN1 9QH, UK 
        \and Department of Physics \& Astronomy, University College London, Gower Street, London, WC1E 6BT, UK 
        \and Instituto de Astrofisica de Canarias, E-38205 La Laguna, Tenerife, Spain 
        \and Institut de F\'{\i}sica d'Altes Energies (IFAE), The Barcelona Institute of Science and Technology, Campus UAB, 08193 Bellaterra (Barcelona) Spain 
        \and Astronomy Unit, Department of Physics, University of Trieste, via Tiepolo 11, I-34131 Trieste, Italy 
        \and INAF-Osservatorio Astronomico di Trieste, via G. B. Tiepolo 11, I-34143 Trieste, Italy 
        \and Institute for Fundamental Physics of the Universe, Via Beirut 2, 34014 Trieste, Italy 
        \and Hamburger Sternwarte, Universit\"{a}t Hamburg, Gojenbergsweg 112, 21029 Hamburg, Germany 
        \and Department of Physics, IIT Hyderabad, Kandi, Telangana 502285, India 
        \and Jet Propulsion Laboratory, California Institute of Technology, 4800 Oak Grove Dr., Pasadena, CA 91109, USA 
        \and Kavli Institute for Cosmological Physics, University of Chicago, Chicago, IL 60637, USA 
        \and Instituto de Fisica Teorica UAM/CSIC, Universidad Autonoma de Madrid, 28049 Madrid, Spain 
        \and Institut d'Estudis Espacials de Catalunya (IEEC), 08034 Barcelona, Spain 
        \and Institute of Space Sciences (ICE, CSIC),  Campus UAB, Carrer de Can Magrans, s/n,  08193 Barcelona, Spain 
        \and Center for Astrophysical Surveys, National Center for Supercomputing Applications, 1205 West Clark St., Urbana, IL 61801, USA 
        \and Department of Astronomy, University of Illinois at Urbana-Champaign, 1002 W. Green Street, Urbana, IL 61801, USA 
        \and Santa Cruz Institute for Particle Physics, Santa Cruz, CA 95064, USA 
        \and Center for Cosmology and Astro-Particle Physics, The Ohio State University, Columbus, OH 43210, USA 
        \and Department of Physics, The Ohio State University, Columbus, OH 43210, USA 
        \and Center for Astrophysics $\vert$ Harvard \& Smithsonian, 60 Garden Street, Cambridge, MA 02138, USA 
        \and Australian Astronomical Optics, Macquarie University, North Ryde, NSW 2113, Australia 
        \and Lowell Observatory, 1400 Mars Hill Rd, Flagstaff, AZ 86001, USA 
        \and Departamento de F\'isica Matem\'atica, Instituto de F\'isica, Universidade de S\~ao Paulo, CP 66318, S\~ao Paulo, SP, 05314-970, Brazil 
        \and George P. and Cynthia Woods Mitchell Institute for Fundamental Physics and Astronomy, and Department of Physics and Astronomy, Texas A\&M University, College Station, TX 77843,  USA 
        \and LPSC Grenoble - 53, Avenue des Martyrs 38026 Grenoble, France 
        \and Instituci\'o Catalana de Recerca i Estudis Avan\c{c}ats, E-08010 Barcelona, Spain 
        \and Department of Astrophysical Sciences, Princeton University, Peyton Hall, Princeton, NJ 08544, USA 
        \and Observat\'orio Nacional, Rua Gal. Jos\'e Cristino 77, Rio de Janeiro, RJ - 20921-400, Brazil 
        \and Department of Physics, Northeastern University, Boston, MA 02115, USA 
        \and Centro de Investigaciones Energ\'eticas, Medioambientales y Tecnol\'ogicas (CIEMAT), Madrid, Spain 
        \and School of Physics and Astronomy, University of Southampton,  Southampton, SO17 1BJ, UK 
        \and Computer Science and Mathematics Division, Oak Ridge National Laboratory, Oak Ridge, TN 37831 
        \and Argonne National Laboratory, 9700 S Cass Ave, Lemont, IL 60439, USA 
        \and Department of Astronomy, University of California, Berkeley,  501 Campbell Hall, Berkeley, CA 94720, USA 
        \and Lawrence Berkeley National Laboratory, 1 Cyclotron Road, Berkeley, CA 94720, USA 
        }

\begin{appendix}
\section{Cluster matter profile interpolation}
\begin{figure}[h]
        \includegraphics[width=0.44\textwidth]{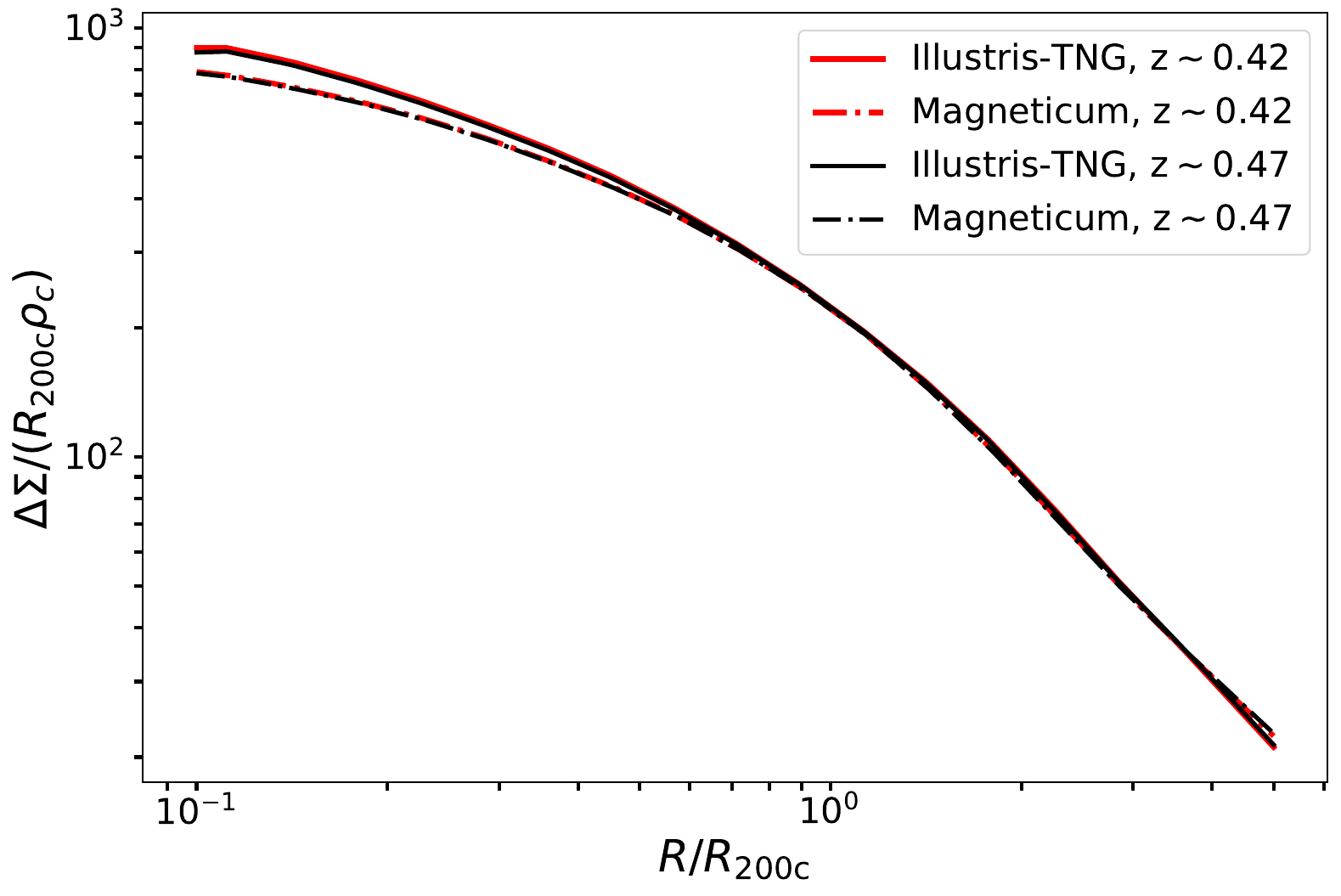}
        \includegraphics[width=0.44\textwidth]{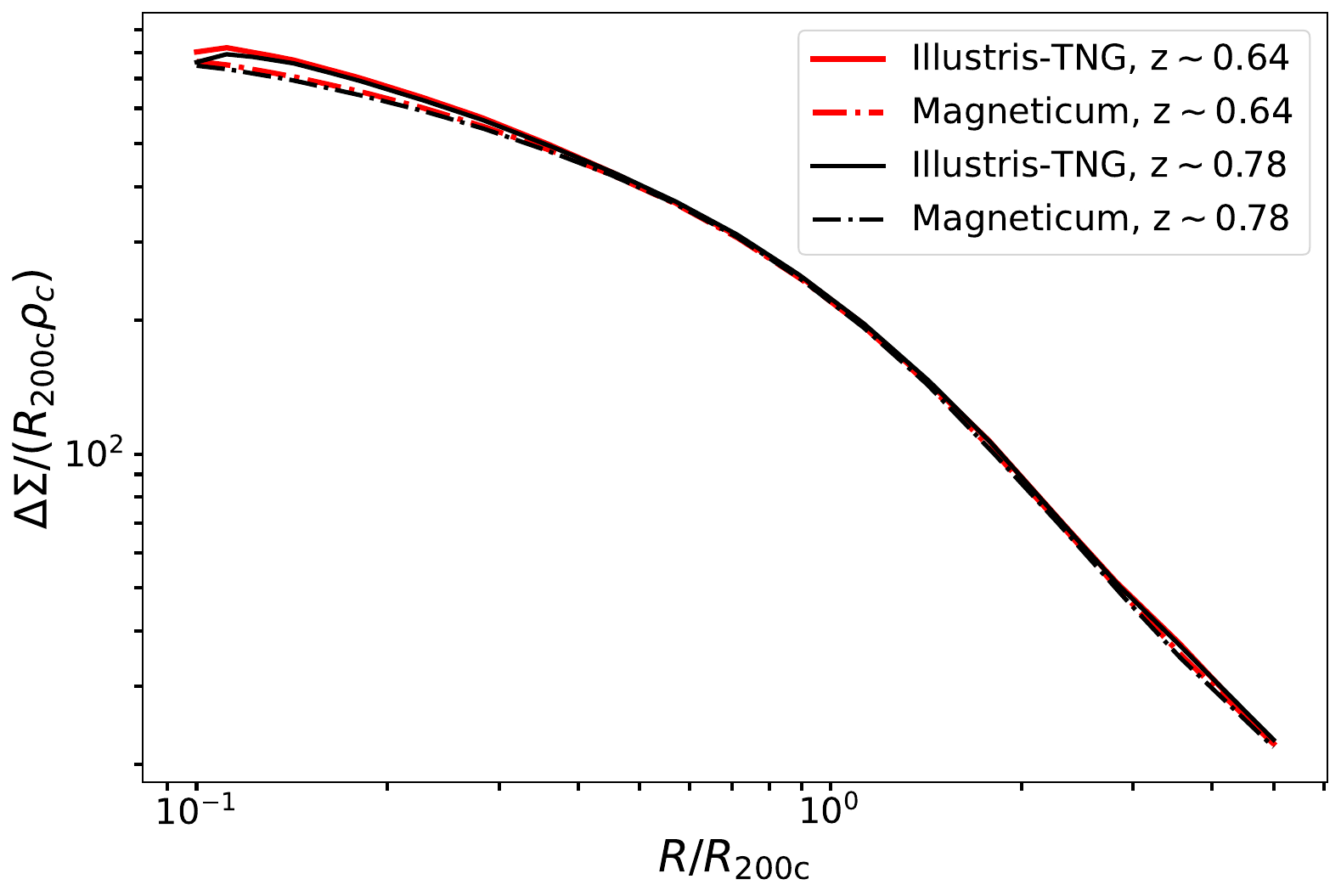}
        \vskip-0.10in
    \caption{Comparison of average matter profiles from IllustrisTNG and \textit{Magneticum} at the redshift of 0.42 and 0.47 (top figure) and at the redshift of 0.64 and 0.78 (bottom figure). The profiles exhibit very small differences with a small redshift change for a given simulation throughout the radial range $R/R_\mathrm{200c}>0.5$}\label{fig:Mag_TNG_redshift_differences}
\end{figure}

In Fig.~\ref{fig:Mag_TNG_redshift_differences} (top panel) we show the differences in the IllustrisTNG average cluster matter profiles at 0.42 vs. 0.47 in solid red and solid black lines respectively. In dashed red and black lines we show a average cluster matter profile in \textit{Magneticum} at the redshift of 0.42 and 0.47 respectively. The profiles show very small differences with slight changes in redshift. Similarly in the bottom figure, we compare the profiles at the redshift of 0.64 vs. 0.78 and find that both simulation profiles exhibit minor differences. The purpose of this plot is not to compare the different simulations themselves, but, the differences in the simulations in the inner radial region at the same redshift are due to different baryonic effects.  That region is avoided in the mass calibration method presented in Sect.~\ref{sec:method}.
\section{Observable mass relation}
\label{sec:mock_SPT_obs_mass}
\begin{figure}
        \includegraphics[width=0.44\textwidth]{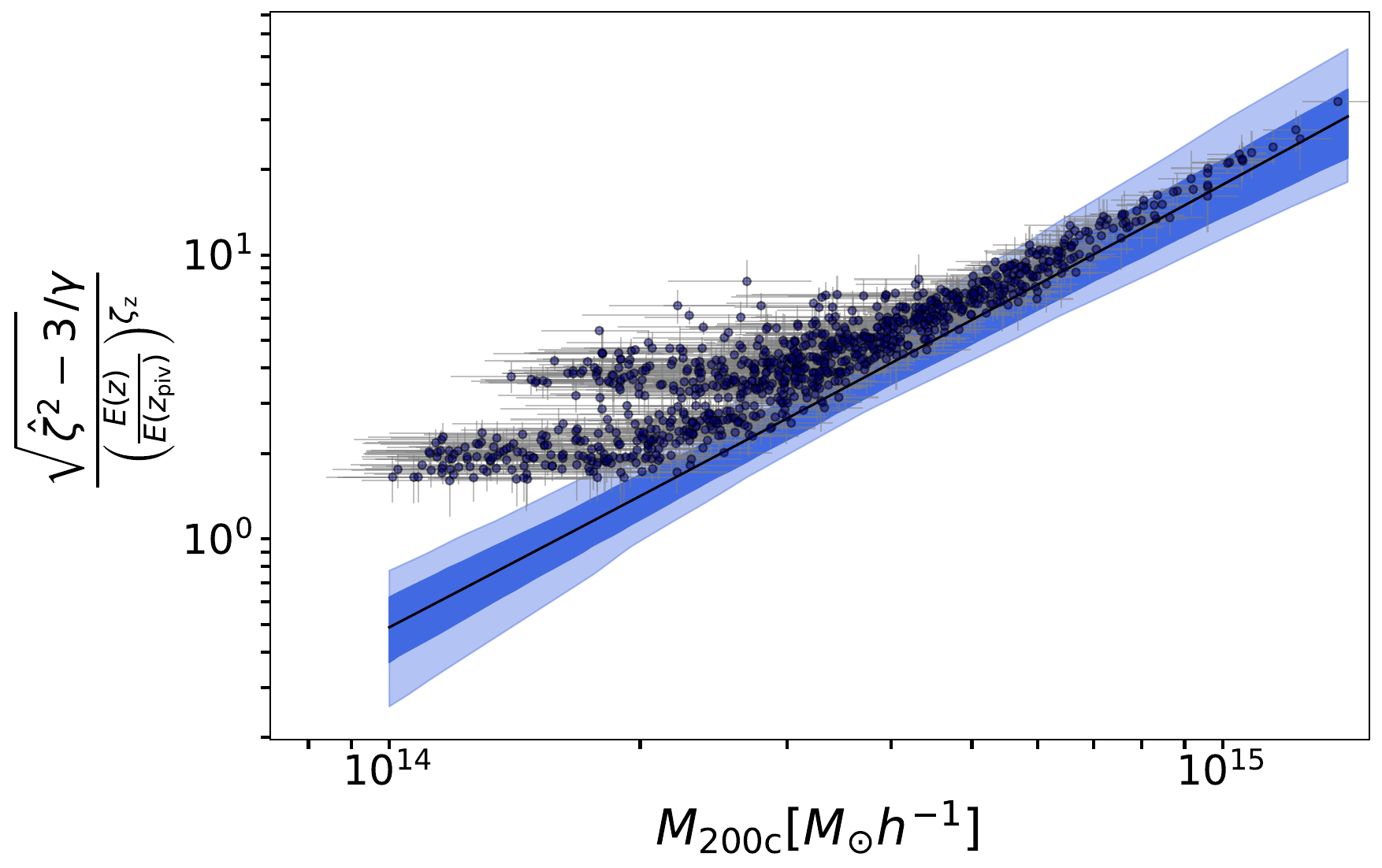}
        \vskip-0.10in
    \caption{The observed debiased detection significance, $\hzeta$, as a function of cluster halo mass for a mock SPT catalog (black dots). The solid black line indicates the intrinsic mean relation, and the shaded regions denote the 68\% and 95\% credible intervals.  The similarities between the deviations of the mock clusters and of the real clusters (Fig.~\ref{fig:zeta_mass_redshift_validation}) about the mean relation are striking.}
    \label{fig:zeta_mass_redshift_mocks}
\end{figure}

To confirm that the features observed in Fig.~\ref{fig:zeta_mass_redshift_validation} (top) is a real effect caused by $\hzeta$ tSZE selection, we analyze a mock SPT catalog along with mock DES~Y3 lensing data. We perform a mass calibration analysis on this mock data and plot the debiased $\hzeta$ as a function of the calibrated $M_{\mathrm{200c}}$, just as we have done with the real data in Sect.~\ref{sec:obs_mass_validation}. Figure~\ref{fig:zeta_mass_redshift_mocks} illustrates this relation for the mock sample, where the mock data is represented by black dots, the mean relation is shown as a black line, and the shaded regions denote the 68\% and 95\% credible intervals on that model. We observe the same features at low mass and low debiased $\hzeta$ as seen in Fig.~\ref{fig:zeta_mass_redshift_validation} (top). This analysis confirms that the feature is introduced by the tSZE selection, where only clusters exceeding the selection threshold make it into our sample.  At low masses, only those clusters that are scattered up in $\hzeta$ due to the combination of intrinsic scatter and measurement noise are selected.  Additionally, we observe the same qualitative features in the mock analysis as those seen in Fig.\ref{fig:lambda_mass_redshift_validation} and the bottom plot of Fig.\ref{fig:zeta_mass_redshift_validation}.

\section{Robustness of the mass calibration}
\begin{figure*}
        \includegraphics[width=0.49\textwidth]{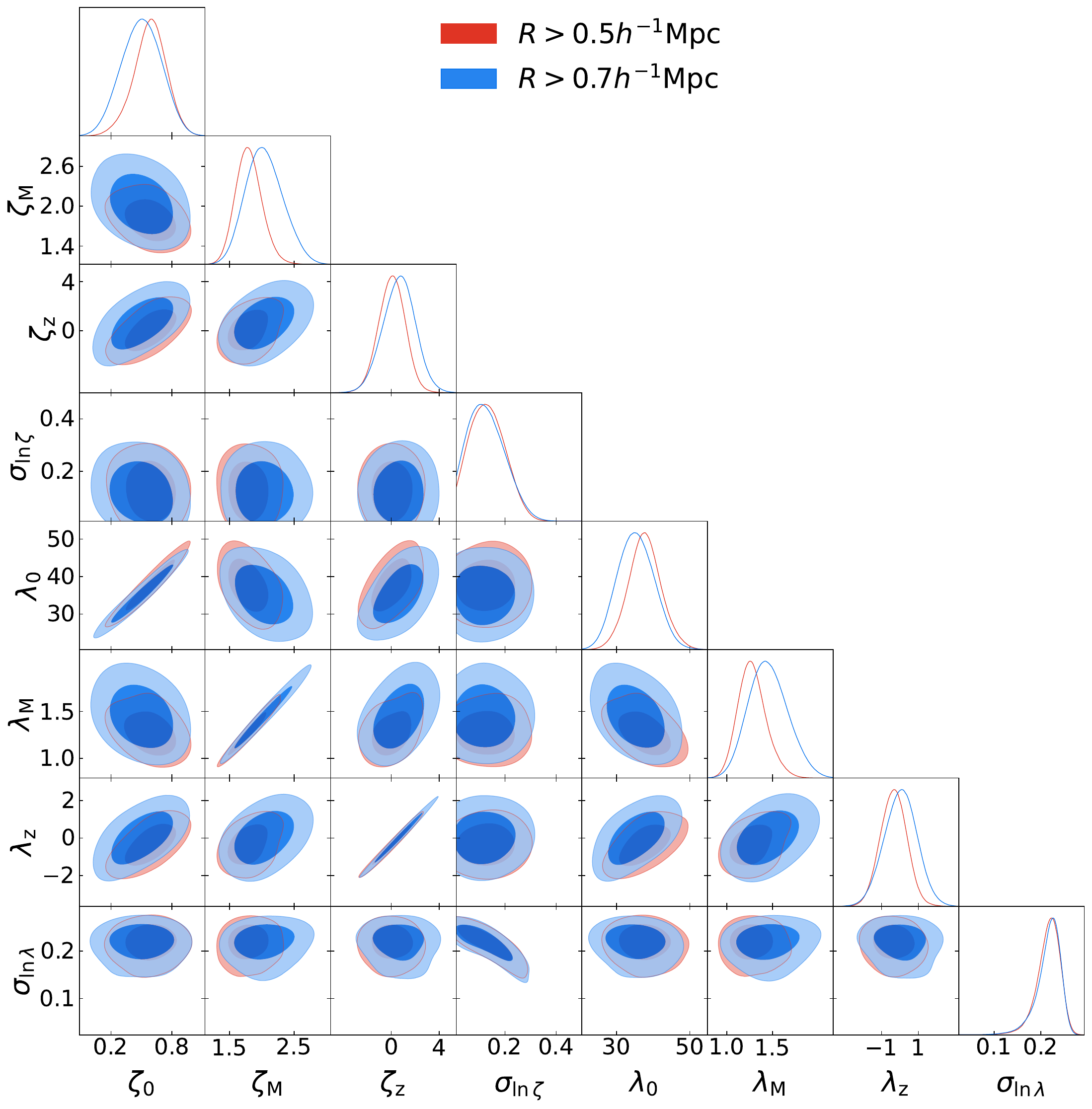}
        \includegraphics[width=0.49\textwidth]{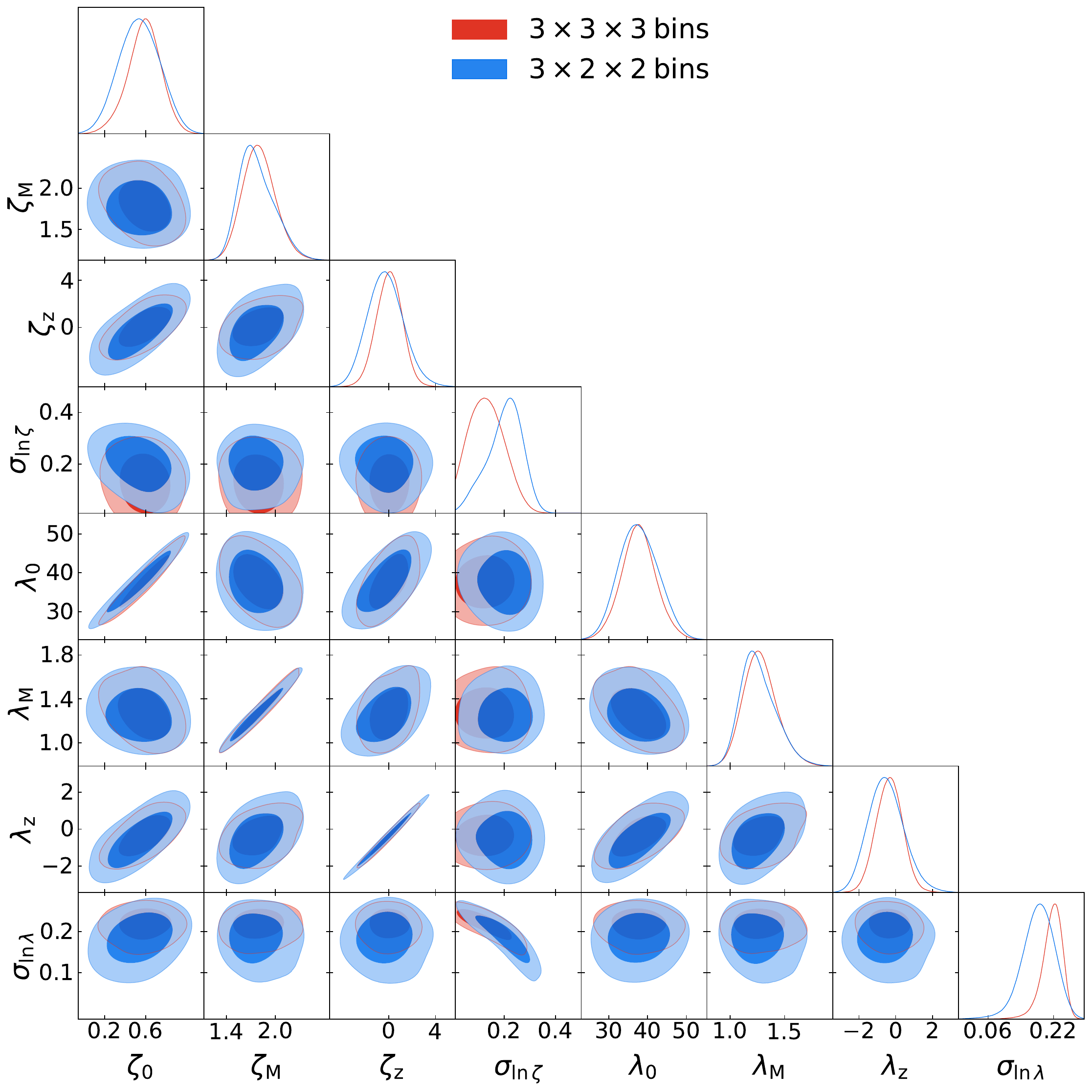}
        \vskip-0.10in
    \caption{Comparison of the posterior for the SPT sample with the same binning but different inner radius fit (left figure) of $0.5h^{-1}\mathrm{Mpc}$ (red) and $0.7h^{-1}\mathrm{Mpc}$ (blue). The blue posterior is in good agreement with the red posterior with a slightly larger error region. In the right figure, we compare the results from two different observable binning schemes while keeping the inner-fitting radii fixed at $0.5h^{-1}\mathrm{Mpc}$. We see good agreement between the two results suggesting our results are not heavily affected by the choice of observable binning.}\label{fig:robustness_plot}
\end{figure*}

In this section, we assess the robustness of our analysis method by changing the inner fitting region of the cluster matter profile and also by changing the binning of the sample in $\hzeta-\hlambda-z$. In the top figure of Fig.~\ref{fig:robustness_plot}, we compare the posterior of the observable-mass relation parameter for inner fitting radii of $0.5h^{-1}\mathrm{Mpc}$ and $0.7h^{-1}\mathrm{Mpc}$ (we note that for this analysis we have binned our observables in $3\times 3\times 3$ bins ). The blue posterior ($R>0.7h^{-1}\mathrm{Mpc}$) results in a larger error compared to the red posterior, which is expected given that we have fewer source galaxies as we restrict our fitting range. 

In the bottom figure of Fig. \ref{fig:robustness_plot}, we compare the observable-mass relation parameters posterior resulting from different binning of observables for the SPT sample compared to our fiducial $3\times 3\times 3$ binning. For the new binning, we divide our observables in $3\times 2\times 2$ bins as follows
\begin{equation}
\begin{split}
    0.25\leq z & < 0.33 \nonumber \\
    &0 \leq \hlambda < 80,\ \  80 \leq \hlambda < 243\nonumber \\
    &4.25 \leq \hzeta < 6,\ \ 6 \leq \hzeta < 50\nonumber \\
    0.33\leq z & < 0.43 \nonumber \\
    &0 \leq \hlambda < 95,\ \  95 \leq \hlambda <243\nonumber \\
    &4.25 \leq \hzeta < 7.5,\ \ 7.5 \leq \hzeta <50\nonumber \\
    0.43\leq z & < 0.94 \nonumber\\
    &0 \leq \hlambda < 85,\ \ 85 \leq \hlambda < 243\nonumber\\
    &4.25 \leq \hzeta < 6,\ \ 6 \leq \hzeta < 50.
\end{split}
\end{equation}
The above binning is such that each redshift bin is further divided into observable bins with roughly similar SNR. The blue and red posteriors show good agreement, indicating that the choice of $\hzeta-\hlambda-z$ binning does not strongly impact parameter values. 
\begin{table*}[t]
\centering
\caption{\label{tab:clmemcont}
Parameters of the cluster member contamination model as described in Sect.~\ref{sec:cmc}. For each parameter the mean and 68\% credible region of the posterior are given.}
\begin{tabular*}{\textwidth}{@{\extracolsep{\fill}} cccc }
\toprule
Parameter & \multicolumn{3}{c@{}}{DNF photo-$z$ and MCMF center}\\ 
\midrule
 &  tomographic bin 2 & tomographic bin 3 & tomographic bin 4 \\
\midrule          
$z_\mathrm{off_0}$ & $-0.009\pm0.002$ & $0.050\pm0.002$ & $0.164\pm0.007$\\
$z_\mathrm{off_z}$ & $-0.22\pm0.02$ & $-0.23\pm0.02$ & $-0.45\pm0.02$ \\
$\sigma_\mathrm{z_0}$ & $0.047\pm0.002$ & $0.076\pm0.002$ & $0.130\pm0.006$ \\
$\sigma_\mathrm{z_z}$ & $-0.018\pm0.021$ & $-0.086\pm0.017$ & $-0.16\pm0.02$ \\
$\log(c_{\lambda})$ & $0.44\pm0.03$ & $0.51\pm0.04$ & $0.26\pm0.06$ \\
$B_\lambda$ & $0.78\pm0.04$ & $0.60\pm0.04$ & $0.53\pm0.07$ \\
$\rho_\mathrm{corr-z}$ & $0.27\pm0.01$ & $0.18\pm0.01$ & $0.494\pm0.006$ \\
$A_0$ & $0.12\pm0.06$ & $-0.32\pm0.17$ & $0.16\pm0.03$ \\
$A_1$ & $0.19\pm0.04$ & $-0.22\pm0.13$ & $0.16\pm0.03$ \\
$A_2$ & $0.29\pm0.04$ & $-0.01\pm0.12$ & $0.17\pm0.03$ \\
$A_3$ & $0.42\pm0.05$ & $0.30\pm0.12$ & $0.18\pm0.02$ \\
$A_4$ & $0.53\pm0.06$ & $0.64\pm0.11$ & $0.19\pm0.03$ \\
$A_5$ & $0.62\pm0.06$ & $0.86\pm0.11$ & $0.19\pm0.02$ \\
$A_6$ & $0.68\pm0.06$ & $0.87\pm0.12$ & $0.20\pm0.02$\\
$A_7$ & $0.72\pm0.06$ & $0.69\pm0.12$  & $0.21\pm0.02$ \\
$A_8$ & $0.74\pm0.07$ & $0.44\pm0.13$ & $0.22\pm0.03$ \\
$A_9$ & $0.75\pm0.08$ & $0.24\pm0.17$ & $0.22\pm0.03$ \\
$A_{10}$ & $0.75\pm0.09$ & $0.13\pm0.21$ & $0.23\pm0.03$ \\
$A_\infty$ & $-4.94\pm0.06$ & $-4.61\pm0.38$ & $-4.76\pm0.21$ \\
\bottomrule
\end{tabular*}
\end{table*}
\end{appendix}
\end{document}